\documentclass[floats,floatfix,showpacs,amssymb,prd,twocolumn,superscriptaddress,nofootinbib,nolongbibliography,reprint]{revtex4-2}

\usepackage{amssymb,amsmath,verbatim,mathtools,needspace,enumitem,etoolbox,graphicx,physics,microtype,afterpage,bigints,gensymb,tabularx,mathtools,siunitx}

\usepackage[dvipsnames, usenames]{xcolor}
\definecolor{linkcolor}{rgb}{0.0,0.3,0.5}
\definecolor{dodgerblue}{HTML}{1E90FF}
\usepackage{hyperref}
\hypersetup{unicode, colorlinks=true, linkcolor=linkcolor, citecolor=linkcolor, filecolor=linkcolor,urlcolor=linkcolor, pdfusetitle}
\usepackage[all]{hypcap}
\usepackage[T1]{fontenc}
\usepackage[utf8]{inputenc}
\usepackage{orcidlink}

\usepackage [english]{babel}
\usepackage [autostyle, english = american]{csquotes}
\MakeOuterQuote{"}

\def\aap{{Astron.~Astrophys.}}

\def\mnras{{Mon.~Not.~R.~Astron.~Soc.}}

\def\pr{{Phys.~Rev.}}

\def\cqg{{Class. Quantum Gravity}}

\def\advss{{Adv. Space Res.}}     
\def\app{{Astropart. Phys.}}
\def\natrp{{Nat. Rev. Phys.}}

\newcommand{\ssim}{\mathchar"5218\relax\,}

\renewcommand{\emph}[1]{\textit{#1}}

\newcommand{\umani}{\affiliation{Department of Physics and Astronomy \& Winnipeg Institute for Theoretical Physics, University of Manitoba, Winnipeg, R3T 2N2, Canada}}
\newcommand{\milan}{\affiliation{Dipartimento di Fisica ``G. Occhialini'', Universit\'a degli Studi di Milano-Bicocca, Piazza della Scienza 3, 20126 Milano, Italy}}
\newcommand{\infn}{\affiliation{INFN, Sezione di Milano-Bicocca, Piazza della Scienza 3, 20126 Milano, Italy}}
\newcommand{\bham}{\affiliation{School of Physics and Astronomy \& Institute for Gravitational Wave Astronomy, \\University of Birmingham, Birmingham, B15 2TT, United Kingdom}}
\newcommand{\herts}{\affiliation{Centre for Astrophysics Research \& Department of Physics, Astronomy and Mathematics, University of Hertfordshire, College Lane, Hatfield AL10 9AB, United Kingdom}}

\newcommand{\Msol}{\rm \,M_{\odot}}
\newcommand{\Mscale}{\num{2e7}\,\Msol}

\begin{document}

\title{Probing AGN jet precession with LISA}

\author{Nathan Steinle$\,$\orcidlink{0000-0003-0658-402X}}
\email{nathan.steinle@umanitoba.ca}
\umani \bham

\author{Davide Gerosa$\,$\orcidlink{0000-0002-0933-3579}}
\milan
\infn
\bham

\author{Martin G.~H. Krause$\,$\orcidlink{0000-0002-9610-5629}}
\herts
\date{\today}

\begin{abstract}
The precession of astrophysical jets produced by active-galactic nuclei is likely related to the dynamics of the accretion disks surrounding the central supermassive black holes (BHs) from which jets are launched. Two main mechanisms that can drive jet precession arise from Lense-Thirring precession  and tidal torquing. These can explain direct and indirect observations of precessing jets; however, such explanations often utilize crude approximations of the disk evolution and observing jet precession can be challenging with electromagnetic facilities. Simultaneously, the Laser Interferometer Space Antenna (LISA) is expected to measure gravitational waves from the mergers of massive binary BHs with high accuracy and probe their progenitor evolution. In this paper, we connect the LISA detectability of binary BH mergers to the possible jet precession during their progenitor evolution. We make use of a semi-analytic model that self-consistently treats disk-driven BH alignment and binary inspiral and includes the possibility of disk breaking. We find that tidal torquing of the accretion disk provides a wide range of jet precession timescales depending on the binary separation and the spin direction of the BH from which the jet is launched. Efficient disk-driven BH alignment results in shorter  timescales of $\ssim 1$ yr which are correlated with higher LISA signal-to-noise ratios. Disk breaking results in the longest possible times of $\ssim 10^7$ yrs, suggesting a deep interplay between the disk critical obliquity (i.e. where the disk breaks) and jet precession. Studies such as ours will help to reveal the cosmic population of precessing jets that are detectable with gravitational waves.
\end{abstract}

\maketitle

\section{Introduction} \label{sec:Intro}

Astrophysical jets are bipolar outflows observed across astronomical scales and often originate from forming stars or accretion processes involving compact objects such as pulsars, stellar-mass black holes (BHs), and active galactic nuclei (AGN) hosting supermassive BHs \cite{2005AdSpR..35..908D,2017Natur.544..202M,2020NewAR..8801539H,2023A&ARv..31....3B,2019A&A...622A..17S,2021MNRAS.504..338P,2001ARA&A..39..403R,2020A&ARv..28....1L,1998Natur.392..673M,2023MNRAS.518.1243F,2014SSRv..183..323F,2014Natur.515..376G,2016Galax...4...29L,2018Natur.562..233V,2018MNRAS.478L.132G}. Multi-wavelength electromagnetic (EM) observations combined with the long history of modeling of jets have led to advancements in understanding the dynamical evolution of the jet components, interactions with their environments, and their connection to the evolution of their hosts \cite{2019ARA&A..57..467B,2019NewAR..8701541H}. 

Among the best studied jets are those associated with AGN \cite{2019ARA&A..57..467B,2020NewAR..8801539H}. These jets can have an observed range of morphologies \cite{2013AJ....146..120L,2019MNRAS.482..240K,2022MNRAS.511.3250M}, exhibit substantial variability \cite{2013AJ....146..120L,2019Galax...7...28R,2020MNRAS.496.1706S}, and extend across a wide range of spatial and angular scales \cite{2023A&A...672A.163O}. The components of a jet can be ejected with different apparent proper motions and directions on the sky plane, typically interpreted as arising from the precession of the jet about a rotation axis \cite{2013AJ....146..120L,2019MNRAS.482..240K,2021ApJ...908..178N,2023Natur.621..711C,2023ApJ...951..106B}. Observing this precession can be challenging and relating it to the kinematically driven evolution of the jet at $\ssim 0.1$ pc spatial and $\ssim 1$ mas angular scales is uncertain \cite{2022ApJ...933...71F}. 

The precession of a jet is most easily directly observable if its axis of rotation is sufficiently along the line of sight. A classic example is the ``Rosetta Stone'' blazar OJ287, whose jet has been observed for 100+ years and is precessing over a timescale of $\ssim 20$ yrs \cite{2018MNRAS.478.3199B,2023ApJ...951..106B}, possibly driven by the evolution of a BH binary \cite{2018ApJ...866...11D}. 
The high-redshift blazar J0017+8135 has a jet precessing over a timescale $\ssim 12$ yrs \cite{2016Galax...4...10R}. 
Alternatively, if the AGN is within sufficient proximity from us, the precession of an off-the-line-of-sight jet can also be directly observed; this is the case of the nearby galaxy M81, whose jet precesses over a timescale of $\ssim 7$--$12$ yrs \cite{2011A&A...533A.111M,2023Natur.621..711C,2023A&A...672L...5V}.
Other systems observed to host dynamically evolving outflows, such as misaligned bubbles and radio lobes, are thought to be tied to the motion of a jet precessing over timescales $\gtrsim$ Myr \cite{2019MNRAS.482..240K}. 
A good example is Hydra A, where 3D hydrodynamic modelling of the high-resolution radio observations suggest a precession timescale of $10^6$ yrs \cite{2016MNRAS.458..802N}. 
Jet precession is also a possible explanation for the locations of the bubbles in the Perseus cluster, NGC 1275 (3C 84) \cite{2010ApJ...713L..74F,2019Galax...7...72B}, where the jet precession timescale was measured to be $\ssim 10^7$ yrs by identifying the formation of four components: inner jets, outer lobes, ghost bubbles, and ancient bubbles \cite{2006MNRAS.366..758D}, though, ``cluster weather,'' i.e., gas motions due to substructure mergers, probably 
also plays a role.

Currently, EM observations of AGNs with jets indicate that many can display strong curvature, see e.g.'s \cite{1988A&A...189...45Z,2023arXiv230107751M}, some fraction of which may be due to jet precession \cite{2020MNRAS.499.5765H}. The fundamental cause(s) of jet precession are uncertain, but two main possibilities involve the precession of an accretion disk of a nuclear supermassive BH \cite{1996A&A...308..321F,2018MNRAS.474L..81L,2019ARA&A..57..467B} where either (i) the accretion disk precesses because of relativistic Lense-Thirring torques or (ii) the accretion disk precesses because of the tidal torque induced by a BH companion. 
The physics of BH accretion and binary BH disk evolution are themselves also uncertain, and connecting these processes with theoretical and observed jet precession properties is a nontrivial task, see e.g. Ref.~\cite{2022NewAR..9501661B} and references therein. 

As direct EM observations of jet precession are limited to systems that are sufficiently nearby, to jet precession timescales that are $\lesssim$ a human lifespan, or to environments that are sufficiently dense for the trace of the jet to be seen over large distances, and as it is challenging to accurately infer long jet precession timescales from indirect EM observations of outflows, there exists an opportunity in galactic astrophysics for complementary probes of jet precession to emerge \cite{2022NewAR..9501661B}. 

One possible new avenue comes from the gravitational waves (GWs) that are emitted by merging binary BHs. 
The Laser Interferometer Space Antenna (LISA)~\cite{2017arXiv170200786A,2024arXiv240207571C} will be able to detect mili-Hertz GWs from the mergers of supermassive binary BHs that result from galactic mergers across a large range of BH masses $M \ssim 10^{4-8} \Msol$ and redshifts $z \gtrsim 10$. A significant fraction of these mergers are expected to harbor sufficient gas for accretion to be relevant during the galactic/BH merger process \cite{2012MNRAS.423.2533B,2013CQGra..30x4008M,2022ApJ...933..104L}, implying that LISA might have the potential to probe the broad range of AGN jet precession timescales spanned by EM observations. 
However, to do so will ultimately require accurate models of supermassive BH binary evolution which is also currently an important open problem~\cite{2014SSRv..183..189C,2021NatRP...3..732V,2023arXiv231117144B}. 

The journey of two BHs from initial pairing to forming a Keplerian binary and evolving until GW emission guarantees merger depends on many astrophysical processes, such as dynamical friction, stellar loss-cone hardening, and viscous-drag in disk migration \cite{1980Natur.287..307B}. While much work has anticipated methods to constrain the complicated supermassive binary BH evolution with GWs and EM counterparts \cite{2022PhRvD.106j3017M,2024APh...15402892C}, for example accretion disk flares and jets associated with the BH binary merger or with post-merger accretion of the merger remnant, these require observational campaigns targeting rare EM transients coincident with GW signals providing accurate sky localization. Thus, a couple questions naturally arise: Can we use GWs to constrain the precession of jets for AGN systems that are challenging to observe with EM facilities? How would such constraints depend on the uncertainties related to the astrophysical evolution of binary BH binaries? 

In this work, we address these questions by connecting jet precession that can occur during the phase of disk migration at binary separations $\lesssim$ 0.1 pc with the LISA detectability of GWs from binary mergers. To compute the jet precession timescales that arise from Lense-Thirring and tidal torquing, the latter of which can drive the disk to break, we assume that jet precession is steered by the precession of an accretion disk about the more massive BH. We make use of a state-of-the-art semi-analytic model \cite{2020MNRAS.496.3060G} that solves the coupled evolution of the BH spin orientations and the binary inspiral through a circumbinary accretion disk.
We then evolve the binary to merger and compute its LISA signal-to-noise ratio. This approach simultaneously parameterizes the two jet precession timescales and correlates their dependencies on astrophysical uncertainties with expectations for LISA detections of GWs from the binary merger. 

While the jet precession timescale due to Lense-Thirring disk precession is generally of $\mathcal{O}({\rm Myr})$, we find that the orientation of the primary BH spin plays an important role in determining the bounds on the tidal jet-precession timescale, which can generically range from $\ssim 1$ to $10^7$ yrs subject to the parameters of the binary BH and the uncertainties of  disk migration. Accretion disk breaking is correlated with the longest possible tidal jet-precession timescales, since they share the common cause of perturbations of the accretion disk from the binary companion. 
Meanwhile, BH spin alignment is correlated with the shortest tidal jet-precession timescales. 
Coupling these predictions with LISA detectability of binary BH mergers, we show how aligned-spin systems have higher signal-to-noise than misaligned systems, implying a selection effect where LISA is more sensitive to systems that evolved from progenitors that experienced jet precession over a shorter timescale as compared to systems whose accretion disk broke.  
Additionally, we speculate how our model can help explain the observed jet precession timescales of a few AGN systems, and we discuss various future routes of exploration that are motivated by our study.

This paper is organized as follows. We present the adopted model of binary BH evolution and LISA detections in Sec.~\ref{sec:Meth}, which we explore in Sec.~\ref{sec:Results}. In Sec.~\ref{sec:Summary} we summarize our findings and in Sec.~\ref{sec:Discussion} we discuss their implications.

\section{Models of binary disk migration, jet precession, and LISA}
\label{sec:Meth}

We utilize semi-analytic models of three binary-BH evolution phases: disk migration, post-Newtonian inspiral, and merger. 
We contend that this encapsulates important phenomenological dependencies between the possible precession of a jet during disk migration, the evolution of the BH binary spin orientations, and the LISA detectability of binary mergers.

\subsection{Disk migration and jet precession}
\label{subsec:DiskMig}

To connect the jet precession timescales to the evolution of supermassive BH binaries, we use the semi-analytic model first presented in Ref. \cite{2020MNRAS.496.3060G}. 
This model assumes that the BH binary is initialized in a cavity carved from a circumbinary disk which feeds the minidisk of each BH  \cite{2018ApJ...853L..17B} in order to compute the coupled evolution of the disk-driven binary inspiral and the BH-spin alignment. 
These minidisks are modeled as collections of annuli whose evolution is governed by the classic disk evolution equations \cite{1983MNRAS.202.1181P,1996MNRAS.282..291S} that correspond to the conservation of mass and angular momentum. 
Simplifying assumptions are utilized to form a tractable boundary value problem without resorting to numerically solving the full set of evolution equations \cite{2014MNRAS.441.1408T,2020MNRAS.496.3060G}. 

This model is well-suited for connecting the precession of the accretion disk with the precession of a jet as it self-consistently incorporates three main aspects of the Bardeen-Petterson effect \cite{1975ApJ...195L..65B,1978Natur.275..516R,1985MNRAS.213..435K}: 
(i) the warping of the minidisk due to Lense-Thirring precession,
(ii) the warping of the minidisk due to the tidal torque from the binary companion, and
(iii) the possible breaking of the minidisk due to sufficiently strong tidal perturbations. 
Thus, it allows us to relate the properties of both the accretion disk and the BH with the precession of a jet during the BH-binary phase of disk migration. A full description of the model is provided in Refs.~\cite{2020MNRAS.496.3060G, 2023MNRAS.519.5031S}. Here, we briefly restate the ingredients that are essential for our analysis. 

We refer to the more massive BH in a binary as the primary denoted by a subscript 1 and the less massive BH as the secondary with a subscript 2. The masses of the two BHs are denoted by $m_{1,2}$, the spin dimensionless Kerr parameters are denoted by $\chi_{1,2}$, the binary separation is denoted by $r$ and the misalignment angles between the spins of the accreting BHs and the orbital angular momentum of the outer edge of the minidisk (which is assumed to be aligned with the angular momentum of the larger circumbinary disk \cite{1999MNRAS.307...79I}) are denoted by  $\theta_{1,2} \in[0, \pi]$. 
Before solving the coupled binary-inspiral and spin-alignment problem, one must first evaluate the warped disk profile. 
We use a locally isothermal nonlinear viscous warp theory \cite{2013MNRAS.433.2403O} to compute the viscosity coefficients that appear in the disk evolution equations, which we approximate via radially dependent power laws with spectral index $\beta = 3/2$ to recast the viscosities in terms of the dimensionless kinematic viscosity parameter $\alpha$ \cite{1973A&A....24..337S} and the warp amplitude $\psi$~\cite{2013MNRAS.433.2403O}. 

Once we have the warped-disk profile, we compute the time evolution $\dd \theta/\dd t$ by integrating the Lense-Thirring torque density in a quasi-adiabatic approximation such that the alignment timescale is much longer than the viscous timescale (the canonical Bardeen-Petterson effect~\cite{1975ApJ...195L..65B}) and much shorter than the mass-accretion timescale.
This implies that the mass and spin magnitude of the accreting BH remain constant. 
We assume that all gas funneled from the circumbinary disk reaches the binary, i.e., $f_1+f_2=f_{\rm T}$ where $f_1$ and $f_2$ are the relative Eddington fractions of the primary and secondary BHs, respectively, and $f_{\rm T}$ is the Eddington fraction of the circumbinary disk, and we assume differential accretion~\cite{2015MNRAS.451.3941G} that scales linearly with the BH masses $m_1$ and $m_2$, i.e.,  $f_2/f_1 = m_1/m_2$. In this multi-timescale framework, we solve the coupled evolution of $\dd\theta/\dd t$ and $\dd r/\dd t$ to obtain the inspiral dependence $\theta(r)$. 
The speed of the alignment of each accreting BH spin relative to the binary inspiral  enters the formalism of  Ref.~\cite{2020MNRAS.496.3060G} with a single dimensionless parameter $\omega$ that depends on the binary masses, the spin magnitude of the accreting BH, and on the disk and inspiral prescriptions~\cite{2023MNRAS.519.5031S}.

The tidal parameter $\kappa$ encodes the effect of the tidal torque from the companion on the   warped minidisk of the accreting BH~\cite{2020MNRAS.496.3060G,2014MNRAS.441.1408T}. For the primary BH one has
\begin{align}\label{E:kappa}
    \kappa_1 & \simeq 0.66 \left( \frac{m_1}{10^7\Msol} \right)^2 \left( \frac{\chi_1}{0.5} \right)^2 \left( \frac{m_2}{10^7\Msol} \right) \left( \frac{r}{0.1\,\rm pc} \right)^{-3} 
    \notag \\
    & \times \left( \frac{H/R}{0.002} \right)^{-6} \left( \frac{\alpha}{0.2} \right)^{-3} \left[ \frac{\zeta}{1/(2\times0.2^2)} \right]^{-3}\,,
\end{align}
where $H/R$ is the aspect ratio at the reference radius where the viscosities are quoted \citep{2007MNRAS.381.1617M,2009MNRAS.400..383M}, and $\zeta = \zeta(\alpha)$ is the ratio of the vertical to horizontal viscosity in the small-warp limit (with $\zeta\propto 1/2\alpha^2$ for $\alpha\to0$, \cite{1983MNRAS.202.1181P,1999MNRAS.304..557O}). 
For a generic binary, one has a $\kappa$ parameter for each BH, where for the secondary one should consider Eq.~(\ref{E:kappa}) with switched labels $1\longleftrightarrow 2$.

With this parameterization, the mass and momentum equations reduce to a one-parameter family of solutions -- that is, we solve $\dd\cos\theta/\dd\kappa$ according to $\kappa = (R_{\rm tid}/R_{\rm LT})^{-7/2}$, where $R_{\rm tid}$ and $R_{\rm LT}$ are the disk radii such that the companion tidal and Lense-Thirring torques, respectively, mostly affect the warp profile \cite{2020MNRAS.496.3060G}. 
As $\kappa \propto r^{-3}$, larger (smaller) $\kappa$ provide solutions that are more (less) perturbed by the binary companion. 

It is possible that the warped minidisk around each BH can break, see e.g. Refs.~\cite{2000MNRAS.315..570N,2014MNRAS.441.1408T,2022MNRAS.509.5608N}. In our semi-analytic model, this occurs for low viscosity $\alpha$, large tidal perturbations $\kappa$ from the binary companion, or a large in-plane component of the BH spin orientation $\theta$ \cite{2020MNRAS.496.3060G}. For given $\alpha$ and $\kappa$, the critical obliquity $\theta_{\rm crit}$ bounds the range of misalignments for which the disk breaks $\theta_{\rm crit}<\theta<\pi - \theta_{\rm crit}$. 
This implies that either the disk can be inititalized broken or can break as the BH spin aligns and the binary inspirals (i.e., as $\kappa$ increases). 
For simplicity, after a disk is broken we hold $\theta$ constant for the remainder of the disk-driven inspiral \cite{2023MNRAS.519.5031S}. 

Throughout this work, and without assumptions regarding the nature of its formation or kinematics, we consider a precessing jet from the primary BH that accretes from its minidisk. 
We assume that the precession of the minidisk is directly responsible for the precession of the jet, see e.g. Refs.~\cite{1988ApJ...334...95R,2007ApJ...671.1272L,2018MNRAS.474L..81L}. This implies two timescales over which this jet can precess corresponding to two main sources of disk precession in our model of binary BH disk evolution \cite{2023MNRAS.519.5031S}. 

The first is the classic Lense-Thirring precession timescale \cite{1998ApJ...506L..97N}, i.e. the time for the inner regions of the minidisk to align with the primary BH spin, given by~\cite{1999MNRAS.309..961N,1996MNRAS.282..291S,2013MNRAS.429L..30L,2020MNRAS.496.3060G}, 
\begin{align}\label{E:Tlt}
    t_{\rm LT} & \simeq 6 \left( \frac{\chi_1}{0.5} \right)^{2/3} \left( \frac{H/R}{0.002} \right)^{2/3}  \left( \frac{f_{\rm T}}{0.1} \right)^{-1} \notag \\
    & \times \left( \frac{\alpha}{0.2} \right)^{1/3} \left[ \frac{\zeta}{1/(2\times0.2^2)} \right]^{-2/3}\,{\rm Myr}\,.
\end{align}

\begin{figure}
\centering
\includegraphics[width=0.48\textwidth]{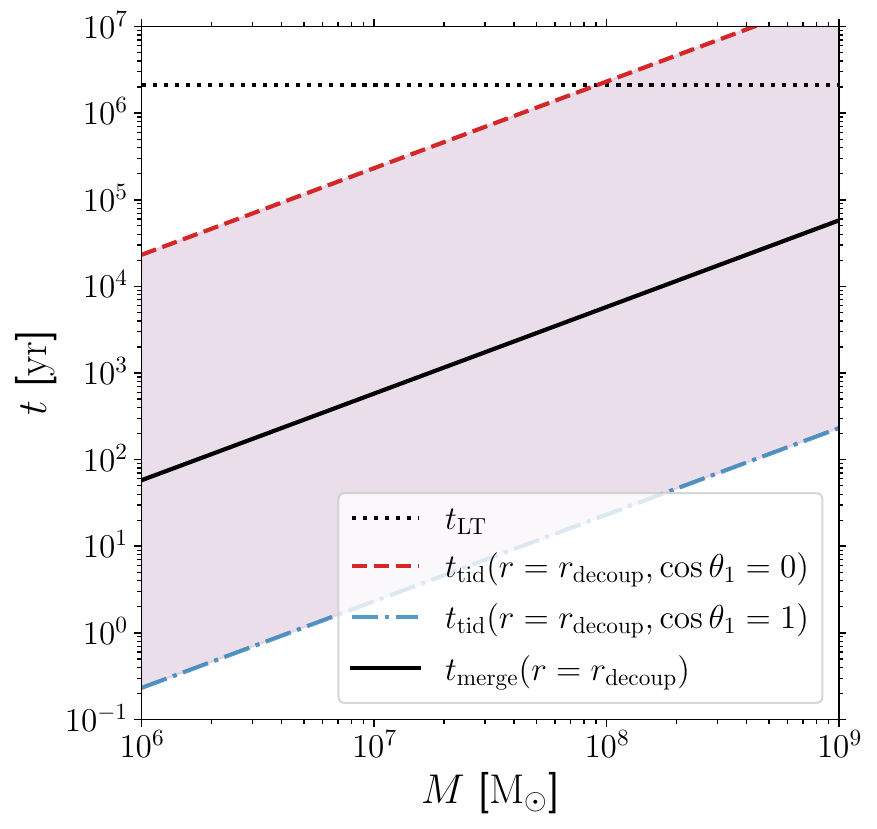}
\caption{The dependence of various timescales on the binary BH total mass, $M$. The dashed red and dot-dashed blue lines are the tidal disk precession timescale [Eq. (\ref{E:Ttid2})] for in-plane and aligned primary BH spin, respectively, which define the allowed range of values shown by the light-pink shaded patch. 
The dotted black line is the Lense-Thirring precession timescale $t_{\rm LT}$ [Eq. (\ref{E:Tlt})], and the black solid line is the GW-driven binary merger timescale $t_{\rm merge}$ [Eq. (\ref{E:Tmerge})]. We use our fiducial values for the remaining parameters as listed in Sec.~\ref{sec:Results}.
} \label{F:timescales}
\end{figure}

The second jet-precession timescale arises from the Newtonian tidal torque acting on the primary-BH minidisk from the binary companion. One has \cite{1997ApJ...478..527K}
\begin{align}\label{E:Ttid}
    t_{\rm tid} &\simeq \num{1.8e2}  \left( \frac{r}{0.1 \,\rm pc} \right)^{3}  \left( \frac{R_{\rm d}}{0.1 \,\rm pc} \right)^{-3/2}  \notag \\
    & \times \left( \frac{q \, m_2}{10^7\Msol} \right)^{-1/2} |\cos\theta_1|^{-1}\,\,\,\,{\rm yr},
\end{align}
where $q = m_2/m_1 \leq 1$ is the binary mass ratio and $R_{\rm d}$ is the radius of the primary minidisk. 
We estimate $R_{\rm d}$ directly from our model of disk migration, which arguably provides a more realistic approximation of the warping and precession of the accretion disk compared to typical analytic treatments wherein the disk is treated as a rigid object and $R_{\rm d}$ is taken as its outer radius~\cite{2007ApJ...671.1272L,2018MNRAS.478.3199B}. 
In Eq.~(\ref{E:Ttid}), we set $R_{\rm d} = R_{\rm tid} = \kappa^{-2/7} R_{\rm LT}$ for the disk radius since this is where the effect of the companion is strongest on the warp of the primary minidisk. 
The importance of $\kappa$ for $t_{\rm tid}$ in our parameterization illuminates the phenomenological dependence of $t_{\rm tid}$ on the binary disk evolution.
Hence, we can cast Eq.~(\ref{E:Ttid}) in terms of the disk and BH parameters by substituting Eq.~(\ref{E:kappa}) for $\kappa$ and Eq.~(21) of Ref.~\cite{2020MNRAS.496.3060G} for $R_{\rm LT}$; this yields
\begin{align}\label{E:Ttid2}
    t_{\rm tid} &\simeq \num{7.45e4} \, \frac{(1 + q)^{5/7}}{2^{5/7 }q^{4/7}} \left( \frac{M}{\Mscale} \right)^{-5/7} \notag \\
    & \times \left( \frac{r}{0.1 \,\rm pc} \right)^{12/7} \left( \frac{\chi_1}{0.5} \right)^{-1/7} \,\, |\cos\theta_1|^{-1} \notag \\
    & \times \left( \frac{H/R}{0.002} \right)^{-4/7} \left( \frac{\alpha}{0.2} \right)^{12/7} \left[ \frac{\zeta}{1/(2\times0.2^2)} \right]^{12/7}
    \,\,\,\,\,\,{\rm yr},
\end{align}
where $M = m_1 + m_2$ is the binary total mass. We note that the above formula for $t_{\rm tid}$ diverges at $\theta_1 = \pi/2$ which corresponds to the disk configuration where the minidisk is broken at the start of the disk migration phase for any set of initial parameters \cite{2020MNRAS.496.3060G}. Thus, we expect a correlation between binaries with broken primary minidisks and longer values of  $t_{\rm tid}$.

We initialize the phase of disk migration at the transition between binary hardening due to stellar loss-cone scattering and viscous disk migration. In our fiducial model where $f_{\rm T} = 0.1$ and the scaling parameters are $R_{\rm b} = 0.05$ pc, $t_{\rm b} = 1$ Myr, $t_{\rm s} = 10$ Myr, and $R_{\rm s} = 0.1$ pc~\cite{2023MNRAS.519.5031S},
this provides an initial binary separation $r_{\rm i} = 0.066$ pc. 
The phase of disk migration ends approximately when the angular momentum of the binary decouples from the angular momentum of the circumbinary disk
~\cite{2012PhRvL.109v1102F},
\begin{align}\label{E:DecoupSep}
\begin{aligned}
    r_{\rm decoup} &\simeq \num{3e-4} \left( \frac{M}{\Mscale} \right) \left[ \frac{4q}{(1 + q)^2} \right]^{2/5} \\
    & \quad \times \left( \frac{H/R}{0.002} \right)^{-4/5} \left( \frac{\alpha}{0.2} \right)^{-2/5} {\rm pc}\,.
\end{aligned}
\end{align}
The decoupling separation provides the smallest separation during the phase of disk migration and hence a lower limit for $t_{\rm tid}$.  
In the results of Sec.~\ref{sec:Results}, we explicitly specify the separation at which $t_{\rm tid}$ is calculated.
We also compute the merger timescale of the binary due to GW emission via \cite{1963PhRv..131..435P}, 
\begin{align}\label{E:Tmerge}
\begin{aligned}
    t_{\rm merge} \simeq \num{14.4} \frac{(1 + q)^2}{4 q} \left( \frac{r}{10^{-4} \, \rm pc} \right)^4 \left( \frac{M}{\Mscale} \right)^{-3} {\rm yr}\,. 
\end{aligned}
\end{align}

Figure \ref{F:timescales} shows the above timescales as the binary total mass $M$ is varied; the other parameters are set to fiducial values, see Sec.~\ref{sec:Results}. For the jet of the primary BH, the Lense-Thirring precession timescale $t_{\rm LT}$, shown as the black dotted line, is independent of $M$ in our model, and is generally larger than the tidal precession timescale $t_{\rm tid}$, which depends strongly on the spin orientation of the primary BH $\theta_1$ and the binary separation $r$. 
When evaluated at the decoupling separation from Eq.~(\ref{E:DecoupSep}), $t_{\rm tid}$ becomes linear in $M$,  cf. Eq.~(\ref{E:Ttid}).
As we shall explore further in Sec.~\ref{sec:Results}, a wide range of $t_{\rm tid}$ are possible, shown by the light-pink patch in Fig. \ref{F:timescales}, depending on the misalignment $\theta_1$. Understanding the phenomenological dependence of $\theta_1$ is important for predicting $t_{\rm tid}$ in AGN systems hosting binary BHs. 
The similar dependence on $r$ of $t_{\rm tid}$ and $t_{\rm merge}$ implies that both timescales should be positively correlated. 
Here and throughout this work, we fix the binary mass ratio to $q = 0.8$ for simplicity. Varying $q$ for a range of major mergers, i.e., $q \gtrsim 0.1$, causes $t_{\rm tid}$ and the GW merger timescale $t_{\rm merge}$, shown by the black solid line, to change only by a factor of $\lesssim 3$.

\begin{figure}
\centering
\includegraphics[width=0.48\textwidth]{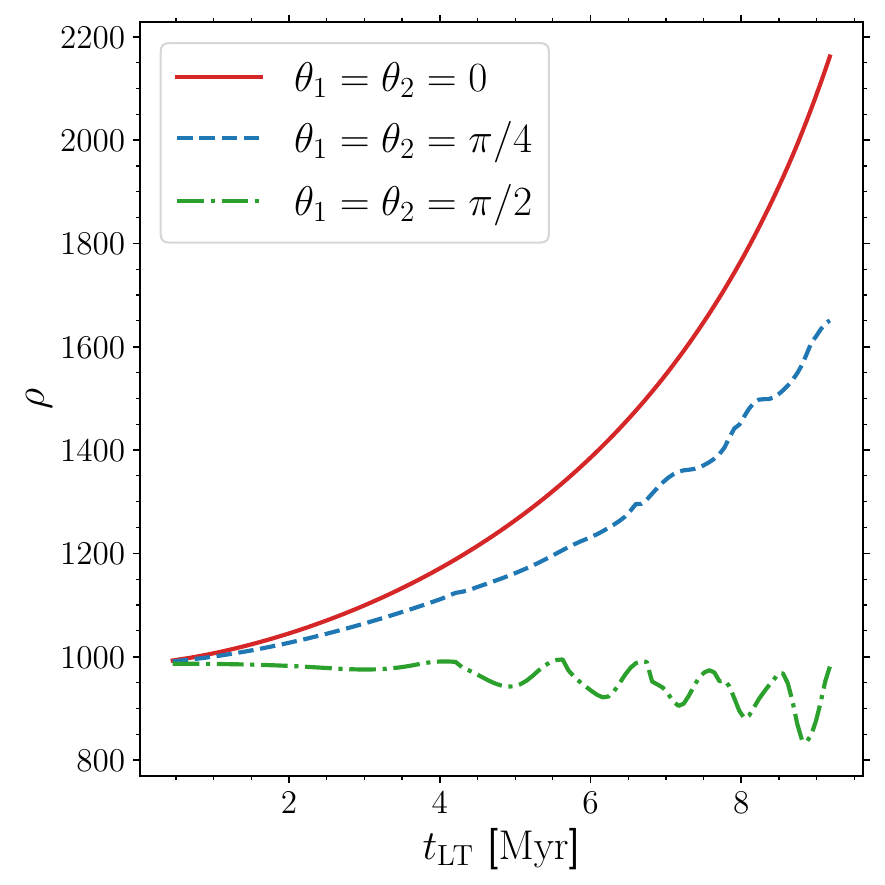}
\caption{
The Lense-Thirring jet precession timescale $t_{\rm LT}$ of the jet of the primary BH during the phase of disk migration and the LISA signal-to-noise ratio $\rho$ of the binary BH merger after the PN inspiral as we vary the dimensionless spin magnitude of the primary BH $\chi_1$ from 0.01 to 0.9 which translates to the Lense-Thirring period via Eq.~(\ref{E:Tlt}). Three initial binary BH spin orientations are indicated by the red (solid), blue (dashed), and green (dash-dot) lines. Note that each value of $\chi_1$ corresponds to a different system, as $\chi_1$ is constant in our model of binary disk migration (see Sec.~\ref{subsec:DiskMig}). 
} \label{F:singleBH}
\end{figure}

\begin{figure*}
\centering
\includegraphics[width=\textwidth]{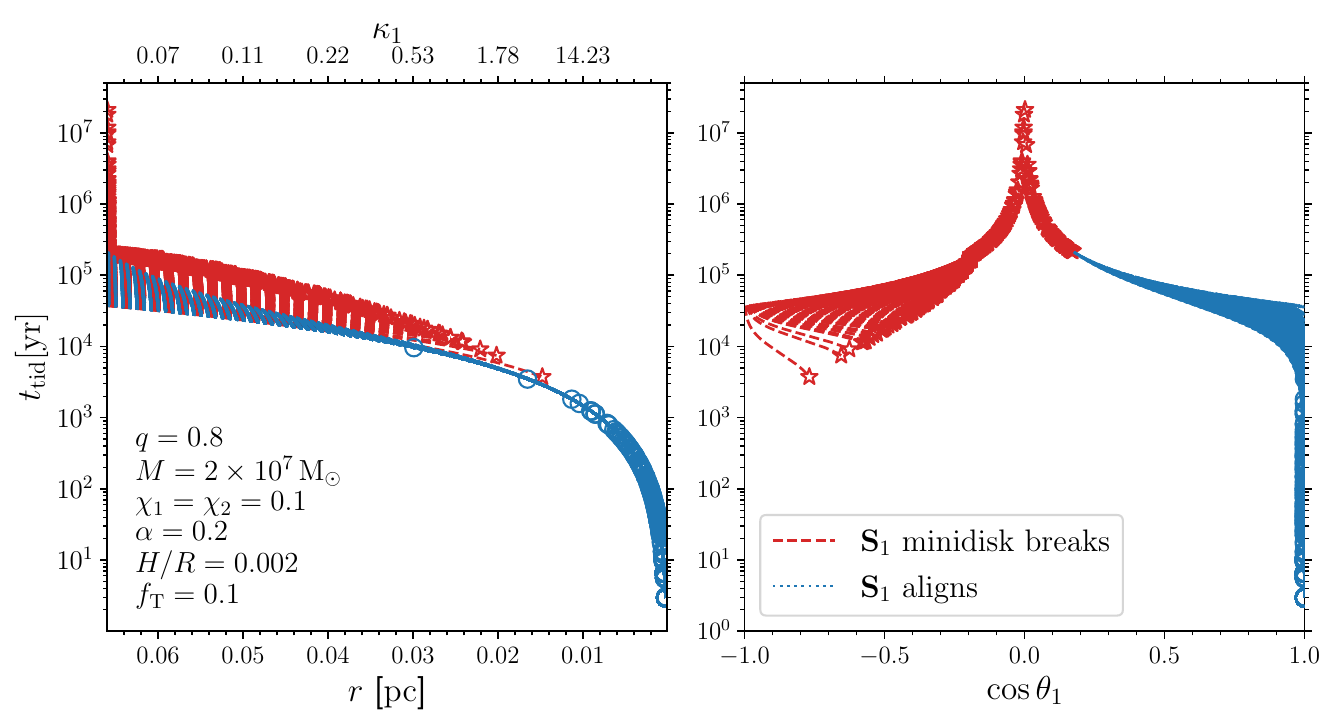}
\caption{The evolution of the tidal jet precession timescale $t_{\rm tid}$ during the phase of disk migration, which depends on both the binary separation $r$ (or equivalently $\kappa_1$) and the spin-orbit misalignment of the primary BH $\theta_1$. 
This distribution of binaries are initialized with our fiducial parameters (Sec.~\ref{sec:Results}) and differ only by their initial spin orientations which are isotropic.
Here we terminate the evolution of the binary when the spin of the primary BH becomes aligned, shown by the blue dotted lines and open circles, or ceases to align further due to encountering a broken accretion disk, as shown by the red dashed lines and stars. Note that $t_{\rm tid} \approx 3.6$ yrs at the decoupling separation, $r_{\rm decoup} \approx 0.006$ pc, for an aligned primary BH spin ($\cos\theta_1 = 1$).  
} \label{F:jet_disk}
\end{figure*}

\subsection{LISA detections of massive binary mergers}
\label{subsec:LISA}

LISA is expected to have tremendous detection capabilities for supermassive BH-binary mergers and is set to be the main experimental facility for the milli-Hz GW regime, with peak sensitivity in the frequency range $f \ssim 0.1$ mHz to $f \ssim 100$ mHz \cite{2017arXiv170200786A}. This section briefly explains our calculation of the signal-to-noise for these binary mergers. 
We specify the distance of the source to the LISA detector as the  redshift $z$ using cosmological parameters from Ref.~\cite{2020A&A...641A...6P}. 

From the end of the disk migration phase, i.e. at the decoupling separation, we track the BH spin evolution using an orbit-averaged post-Newtonian formulation as implemented in the \textsc{precession} code \cite{2016PhRvD..93l4066G} down to a final, pre-merger separation. 

First, we need to ensure that the final separation $r_{\rm f}$ from our post-Newtonian evolution is consistent with the GW waveform we will utilize below. 
We obtain the final separation via
\begin{align}\label{E:SepFinal}
\begin{aligned}
r_{\rm f} = \num{2e-4} \left(\frac{f_{\rm low}}{0.1 \,{\rm mHz}}\right)^{-2/3} \left(\frac{M(1 + z)}{\Mscale}\right)^{1/3}\, {\rm pc},
\end{aligned}
\end{align} 
where $f_{\rm low} = 0.1$ mHz is the instrumental lower-frequency cut-off of the LISA detector which we take also as the waveform reference frequency $f_{\rm ref}$ (i.e. the frequency at which time-dependent quantities such as the spin directions are specified). 
For binaries with total masses $M(1 + z) > 102\,(20\,{\rm Hz}\,/f_{\rm low})\Msol$, $r_{\rm f}$ above will be too small for the post-Newtonian approximation to be valid. In those cases  we instead set 
\begin{equation}
r_{\rm f} = 10 \frac{GM}{c^2} \approx \num{9.6e-6} 
\left(\frac{M(1 + z)}{\Mscale}\right)  {\rm pc}
\end{equation} and compute the waveform reference frequency via 
\begin{equation}
f_{\rm ref} \approx 10^{-4} \left( \frac{M (1 + z) }{\Mscale} \right)^{-1}\, {\rm Hz}\,.
\end{equation}

We specify the spin directions in a  frame where the binary orbital angular momentum $L$ is along the z-axis $\mathbf{\chi}_i = ( \chi_{\rm x}, \chi_{\rm y}, \chi_{\rm z} ) = \chi_i( \sin\theta_i\cos\phi_i, \sin\theta_i\sin\phi_i, \cos\theta_i )$ where $i \in \{1,\,2\}$, $\theta_i \in [0,\,\pi]$ is the spin-orbit misalignment introduced in Sec.~\ref{sec:Meth}, and $\phi_i  \in [0,\,2\pi]$ is the in-plane angle subtended by the spin vector which we  sample uniformly at the beginning of the PN evolution phase and evolve down to $r_{\rm f}$. 

Finally, to compute the frequency domain, inspiral-merger-ringdown GW waveform $h(f)$, we use the waveform approximant \textsc{IMRPhenomXPHM}~\cite{2021PhRvD.103j4056P} through the  \textsc{pycbc} \cite{2017ApJ...849..118N} software.	 
We estimate the signal-to-noise ratio (SNR) of LISA in the long-wavelength limit as the sum of the SNRs of two L-shaped detectors $\rho^2 = \langle H_1 | H_1 \rangle + \langle H_2 | H_2 \rangle$. The SNR in each set of L-shaped arms is \cite{2015CQGra..32a5014M}, 
\begin{equation}\label{eqn:SNR}
\langle H_i | H_i \rangle =  4 \int  \frac{h_c^2(f)}{S_n(f)}  \mathrm{d} f\,, 
\end{equation} 
where $H_i = \sqrt{3} (F_{i,+}h_+ + F_{i,\times}h_\times)/2$ are the responses for the two sets of arms assuming each measures the same signal with the same detector noise, $F_{i,+\times}$ are the beam-pattern coefficients for the $h_+$ and $h_\times$ polarizations \cite{2004PhRvD..69h2005B}, and $h_c(f) = 2 f |h(f)|$ is the characteristic strain \cite{2015CQGra..32a5014M} of the GW signal and $h_n(f) = \sqrt{fS_n(f)}$ is the amplitude of the (one-sided) noise power spectral density $S_n(f)$ of LISA, which is composed of both instrumental~\citep{2021arXiv210801167B} and galactic confusion \citep{2017PhRvD..95j3012B} noises. We employ Monte Carlo integration to estimate the marginalized SNR over the binary inclination angle $\iota$ (uniform in $\cos\iota$), sky location longitude $\phi_s$ and latitude $\theta_s$ (uniform in $[0, 2\pi]$ and $\cos\theta_s$, respectively), and the GW polarization $\psi_{\rm GW}$ (uniform in $[0, \pi]$). 
We confirm the convergence of the Monte Carlo integral with a standard error of the mean $\lesssim 1$, which is sufficient for our purposes. 
In the results that follow we refer to the marginalized SNR as $\rho$. 

Nonzero BH spins that are misaligned from the binary orbital angular momentum cause relativistic spin precession during the binary inspiral and merger which modulates the waveform in amplitude, frequency, and phase by shifting the radiated power among the multipole modes of $h(f)$. In turn this affects the LISA SNR in a non-trivial manner, see e.g. Ref.~\cite{2023PhRvD.108l4045P}.

\section{Results}
\label{sec:Results}

In this section, we aim to reveal the connected aspects of binary BH disk migration, AGN jet precession, and LISA detections of binary BH mergers. This results in a large parameter space exploration of which is beyond the scope of this work. The inputs to our model are the BH-binary parameters $\{m_1, \,m_2, \,\chi_1, \,\chi_2, \,\theta_1, \,\theta_2, \,\phi_1,\, \phi_2 \}$, the disk-migration parameters $\{\alpha, \,H/R,\, f_{\rm T} \}$ and those needed  to compute the LISA SNR $\{z, \,\iota, \,\theta_s, \,\phi_s, \,\psi_{\rm GW} \}$. 
Motivated by the results of our previous study \cite{2023MNRAS.519.5031S}, we choose a fiducial set of values, 
$q = m_2/m_1 = 0.8, \,M = m_1 + m_2 = \Mscale, \,\chi_1 = \chi_2 = 0.1, \,\alpha = 0.2, \,H/R = \num{2e-3}, \,f_{\rm T} = 0.1$, and $z = 2$ (see e.g. \cite{2020MNRAS.491.2301K}). 

We sample random values of $\theta_1, \,\theta_2, \,\phi_1,$ and $\phi_2$ isotropically unless otherwise specified, and we marginalize over $\iota, \,\theta_s, \,\phi_s,$ and $\psi_{\rm GW}$ when computing the SNR $\rho$.

Figure \ref{F:singleBH} shows the dependence of $\rho$ on the Lense-Thirring jet precession timescale $t_{\rm LT}$ as the primary BH spin magnitude $\chi_1$ is varied and all else held constant for a single binary BH. 
For each value of $\chi_1$, $\theta_1$ and $\theta_2$ are equal at the beginning of the phase of disk migration, as indicated in the legend, and $\phi_1 = 1.76$ and $\phi_2 = 4.97$ are initialized at the beginning of the ensuing PN inspiral. 
We note that the modulations of $\rho$ due to the relativistic precession of the orbital plane are sensitive to these choices. 
As we assume $\chi_1$ is constant during disk migration and since $\chi_1$ remains constant throughout the PN inspiral, the dependencies of $\rho$ and $t_{\rm LT}$ on $\chi_1$ are correlated in our model, albeit from different stages in the binary evolution.
Larger values of $\chi_1$ correlate with longer $t_{\rm LT}$ and larger $\rho$ which is suppressed by larger spin-orbit misalignments, i.e., going from the red (solid) line to the blue (dashed) and to the green (dash-dot) line. 
The disks of the binaries of the green line are initialized as broken since $\theta_1 = \pi/2$, which stifles BH spin alignment and leads to large modulations in $\rho$. 
For our fiducial initial disk parameters, the binaries with initial spins aligned (red solid line) and modestly misaligned (blue dashed line) end disk migration with aligned BH spins and thus $\rho$ is without modulation. 
However, for the blue line, we instead choose to show the amount of modulation of $\rho$ in the case of inefficient spin alignment during disk migration, which can occur generally for smaller binary mass and BH spin magnitude \cite{2020MNRAS.496.3060G,2023MNRAS.519.5031S}. 

While $t_{\rm LT}$ is constant during the disk-driven binary inspiral in our model, the jet precession timescale due to the influence of the binary companion has a nearly quadratic dependence on the binary separation, $t_{\rm tid} \ssim r^{12/7}$, see Eq.~(\ref{E:Ttid2}). We explore this dependence in Fig.~\ref{F:jet_disk}, which shows the coupled evolution of the binary inspiral (left panel) and alignment of the primary BH spin  (right panel) which together determine the evolution of $t_{\rm tid}$. 
Here, we end the evolution of the binary, and hence of $t_{\rm tid}$, when the evolution of the primary BH spin $\chi_1$ ceases due to being aligned or due to encountering a broken minidisk. 
The distribution of binaries are initialized with identical parameters except that their spin orientations are isotropically distributed. 
Consistent with the results of Refs.~\cite{2020MNRAS.496.3060G,2023MNRAS.519.5031S}, this produces two subsets: binaries with $\chi_1$ aligned (dotted blue lines and circles) and binaries with $\chi_1$ misaligned due to disk breaking (dashed red lines and stars). Binaries with aligned $\chi_1$ attain the shortest $t_{\rm tid}$ unless they are initialized with nearly aligned spins in which case the dependence on $r$ dominates resulting in larger $t_{\rm tid}$. 
The longest timescales $t_{\rm tid}$ occur near the divergence $\cos\theta_1 = 0$, indicated by the peak of red stars in both panels. 
This peak is within the subset of the binaries initialized with a broken disk, where the constancy of $\cos\theta_1$ implies constant $t_{\rm tid}$ through the inspiral. For a given value of viscosity $\alpha$, the range of $\theta_1$ corresponding to initially broken disks is set by the initial value of the tidal parameter $\kappa_1 \approx 0.01$, implying that very long $t_{\rm tid}$ is generically possible via disk breaking in our model. 
We note that the fiducial set of initial parameters provide quick alignment of the primary BH relative to the binary inspiral, i.e., $\omega_1 \approx 1.45$. 
This rapid alignment is not generic since $\omega_1$ has a non-trivial dependence on the initial parameters \cite{2020MNRAS.496.3060G}.
As demonstrated in Ref.~\cite{2023MNRAS.519.5031S}, efficient spin alignment is not ubiquitous across the parameter space, implying that inefficient alignment provides $t_{\rm tid}$ inspiral evolution that is governed by the evolution of $r$. 

\begin{figure}
\centering
\includegraphics[width=0.48\textwidth]{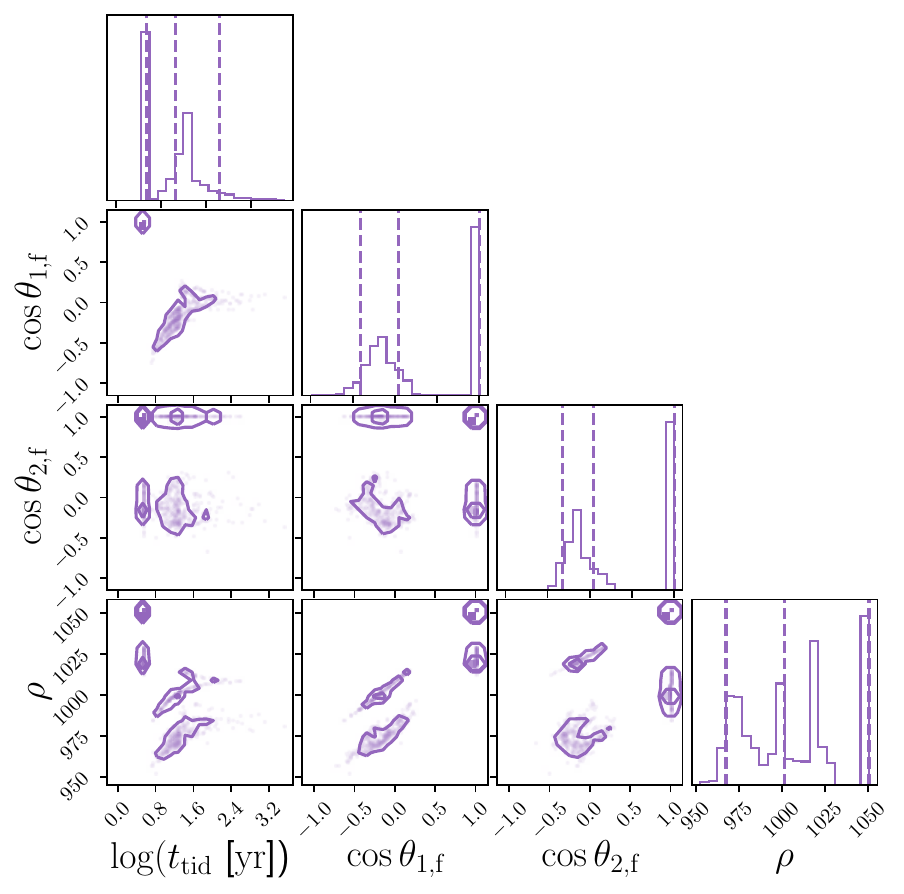}
\caption{Distributions of the tidal jet precession timescale $t_{\rm tid}$ (at $r_{\rm decoup}$), the binary BH spin orientations $\theta_1$ and $\theta_2$ (at $r_{\rm f}$), and the signal-to-noise ratio $\rho$ of the binary BH merger as seen by LISA for the set of binaries from Fig.~\ref{F:jet_disk}. The two peaks in $t_{\rm tid}$ correspond to the aligned and misaligned cases of $\theta_1$, and the four peaks in $\rho$ correspond to the four combinations of aligned and misaligned primary and secondary BH spins. 
Contours enclose 50\% and 90\% of the distributions. 
} \label{F:corner}
\end{figure}

Next, we continue evolving the distribution of binaries in Fig. \ref{F:jet_disk} down to the decoupling separation [Eq.~(\ref{E:DecoupSep})] $r_{\rm decoup} \approx 0.006$ pc to compute $t_{\rm tid}$, and then down to the pre-merger separation $r_{\rm f} \approx 10^{-5}$ pc, where the spin-orbit misalignments at $r_{\rm f}$, $\theta_{1,\rm f}$ and $\theta_{2, \rm f}$, carry the distinct imprints of efficient alignment and disk breaking that occurred during disk migration. We then compute the SNRs $\rho$ of the mergers in LISA. The resulting distributions of these parameters are presented in Fig. \ref{F:corner}. 
The tidal jet precession timescale $t_{\rm tid}$ is doubly peaked, where the steep spike arises from binaries with aligned $\chi_1$ and the broader peak at longer $t_{\rm tid}$ arises from binaries where the primary minidisk broke and hence $\chi_1$ remained misaligned. 
As $t_{\rm tid}$ is computed at $r_{\rm decoup}$ here, it is generally smaller than the values shown in Fig. \ref{F:jet_disk}, however the general trend still holds that shorter jet precession timescales $t_{\rm tid}$ require aligned $\chi_1$ shown by the sharp peak in the histogram of $t_{\rm tid}$. 

The distribution of SNRs $\rho$ has four peaks due to the four possible combinations of aligned and misaligned spins of the primary and secondary BHs: binaries in the peak at $\rho \approx 970$ have both spins misaligned (both minidisks broken), binaries in the peak at $\rho \approx 1000$ have the secondary spin aligned and primary spin misaligned (minidisk broken), binaries in the peak at $\rho \approx 1020$ have the primary aligned and secondary misaligned (minidisk broken), and binaries in the steep spike at $\rho \approx 1050$ have both spins aligned from accretion. 
While this trend is a generic outcome for the case of efficient spin alignment, the values of $\rho$ are highly sensitive to the assumed fiducial parameters. 
These binaries generally have $\theta_1\neq\theta_2$, implying that $\rho$ here can be lower than its value in the small-spin limit for the fiducial mass and redshift used in Fig.~\ref{F:singleBH}.
The distinct maxima in $\rho$ and their dependence on the binary BH spin orientations indicate that LISA will have a unique capability to probe the possible jet precession timescales $t_{\rm tid}$ of binary progenitors that evolved in gas-rich environments.

\begin{figure}
\centering
\includegraphics[width=0.48\textwidth]{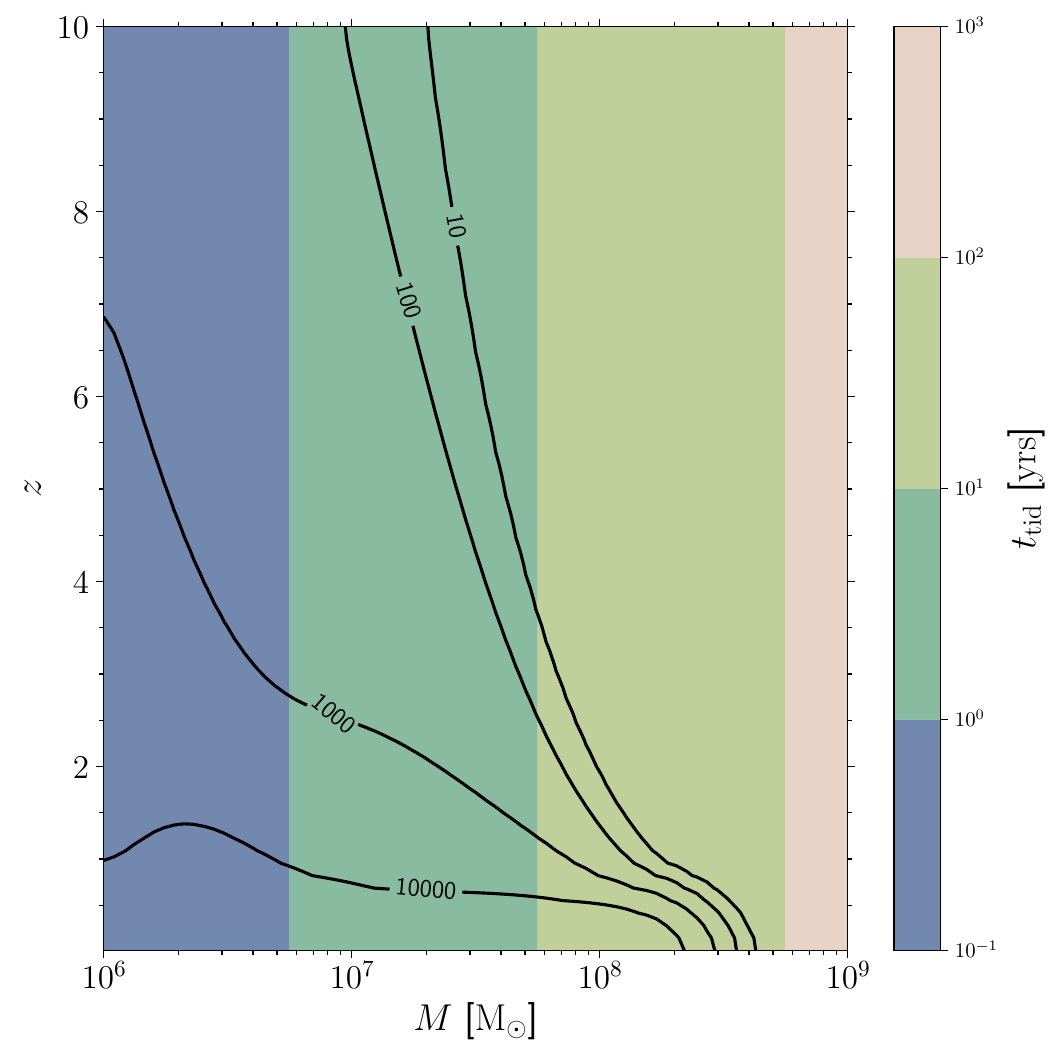}
\caption{The dependence of the tidal jet-precession timescale $t_{\rm tid}$ (colored regions) and of the LISA signal-to-noise $\rho$ (solid black lines) on the source-frame total binary mass $M$ and the redshift $z$. All other parameters are fixed to their fiducial values; the binary BH spins are aligned during disk migration. 
} \label{F:waterfall}
\end{figure}

Lastly, Fig. \ref{F:waterfall} shows how $t_{\rm tid}$ (computed at the decoupling separation as in Fig. \ref{F:corner}) and $\rho$ overlap as we vary the (source frame) total mass $M$ and redshift $z$ of a binary BH that is initialized at the start of the phase of disk migration and is terminated as a merger in the LISA detection window. 
Contours of constant $t_{\rm tid}$, depicted by the colorbar, are monotonic in $M$ and independent of $z$, and contours of constant $\rho$ are shown by the labeled, solid black lines. While $t_{\rm tid}$ is larger for higher masses, i.e., see Fig. \ref{F:timescales}, this is where $\rho$ and hence the LISA detectability quickly vanishes. 
However, here the range of values of $t_{\rm tid}$ are actually lower bounds since $t_{\rm tid}$ is larger for separations $r > r_{\rm decoup}$, i.e., see Fig. \ref{F:jet_disk}. 
As such, the largest value of $t_{\rm tid}$ in Fig. \ref{F:waterfall} reflects the fact that it is evaluated at the small separation $r = r_{\rm decoup}$.

In Fig. \ref{F:waterfall}, we assume small BH spin magnitudes $\chi_1 = \chi_2 = 0.1$ and initial spin orientations that result in aligned $\chi_1$ and $\chi_2$. The resulting range of $t_{\rm tid}$ is not very sensitive to changes in spin magnitude, except for when $\chi_1$ is misaligned due to disk breaking during disk migration, which provides for larger (smaller) $t_{\rm tid}$ at a given $M$ if $\chi_1$ is lower (higher). 
These possibilities imply correlations between the properties of precessing jets and the binary BH spin precession as seen by LISA. This situation is further complicated by the dependence of $t_{\rm tid}$ on the properties of the disk, such as $\alpha$ and $H/R$, underscoring the necessity for comprehensive astrophysical modeling in producing insightful predictions for LISA data analysis.

\section{Summary}
\label{sec:Summary}

Astrophysical jets are among the most energetic phenomena in the cosmos and in AGN they likely arise from accreting BHs. In this work, we calculated the correlations between the evolution of accreting binary BHs during their phase of disk migration and the expected SNR for the GWs emitted from their mergers as detectable by LISA. This framework simultaneously connects, for the first time, the astrophysical uncertainties of binary BH disk migration with their corresponding jet precession timescales and resultant GW signals. 
Therefore, we demonstrate that LISA can probe the precession of AGN jets by trading the necessity of coincident multi-messenger signals for the complicated astrophysical evolution of the progenitors of LISA sources. 
This work is an important step toward the development of new frameworks, possibly informed by EM constraints, to probe the properties of AGN jets through detections of GWs. 
Our main findings include:
\begin{itemize}
    \item The Lense-Thirring jet precession timescale occurs characteristically over long timescales, $t_{\rm LT} \sim$ Myr, consistent with previous studies, see e.g. \cite{1996MNRAS.282..291S,1999MNRAS.309..961N,2013MNRAS.429L..30L}. 

    \item The tidal jet precession timescale takes on a broad range of possible values, $t_{\rm tid} \ssim 1$ to $10^7$ yrs, depending strongly on the binary separation $r$ and primary BH misalignment $\theta_1$. 

    \item Accretion disk breaking, which occurs for $\theta_1 \ssim \pi/2$, is strongly correlated with long timescales $t_{\rm tid}$ as $t_{\rm tid} \ssim 1/\cos\theta_1$, implying that the dependence of the disk breaking phenomena on $\alpha$ and $\kappa$ will be important uncertainties for determining $t_{\rm tid}$. 

    \item The decoupling separation $r_{\rm decoup}$ essentially provides a lower limit for $t_{\rm tid}$ and hence sets the floor for the range of $t_{\rm tid}$ that LISA can probe. 

    \item As the SNR of LISA vanishes dramatically for higher masses $\gtrsim 10^8\Msol$, we find that LISA will be most sensitive to binaries with relatively shorter $t_{\rm tid}$ when the BH spin is aligned depending on the spin magnitude of the primary BH, suggesting possible correlations with the binary BH spin precession dynamics. 
\end{itemize}

Analytic approaches to modeling jet precession with accreting BHs, see e.g. Ref.~\cite{2007ApJ...671.1272L}, typically assume that (i) the disk is a rigid object and (ii) the jet precession is directly related to the rigid disk's precession. We have advanced assumption (i) with our semi-analytic model that self-consistently treats warp propagation through the disk. Although here we also used assumption (ii), we stress that it can be tested using astrophysical modeling frameworks similar to ours in conjunction with datasets from a GW detector such as LISA. 

We chose a modest fiducial value of the disk viscosity $\alpha = 0.2$, which is uncertain but can range from $\lesssim 0.1$ to $\ssim 0.4$ \cite{2007MNRAS.376.1740K}. While $t_{\rm LT}$ [Eq.~\ref{E:Tlt}] scales monotonically with $\alpha$, the dependence of $t_{\rm tid}$ on $\alpha$ is more complicated due to disk breaking. Less viscous disks are much more prone to breaking \cite{2020MNRAS.496.3060G}, implying that the prevalence of long $t_{\rm tid}$ timescales in the parameter space is highly sensitive to the value of $\alpha$. 
Similarly, BHs that begin binary disk migration with large values of the tidal parameter $\kappa$, i.e. with small binary separations, are also more prone to disk breaking due to strong perturbations from the companion. This effect is amplified for small values of $\alpha$ as $\kappa \propto \alpha^{-3}$. 
Therefore, we expect the longest values of $t_{\rm tid}$ to be correlated with the smallest values of $\alpha$ and with the largest values of $\kappa$.
Also, our model assumes a constant aspect ratio $H/R$ of the BH minidisks, which is likely an oversimplification 
\cite{2016MNRAS.460.1243R} and might imply that we underestimate the prevalence of disk breaking, especially considering that $\kappa \propto (H/R)^{-6}$. A more realistic prescription for $H/R$ will be important for accurately modelling the relationship between disk breaking and $t_{\rm tid}$. 
We naively expect a correlation between $t_{\rm tid}$ and the binary's merger timescale $t_{\rm merge}$ from GWs, as a larger binary separation simultaneously implies larger $t_{\rm merge}$ and $t_{\rm tid}$.

\section{Discussion}
\label{sec:Discussion}

Our conclusions have immediate implications for EM-observed precessing jets in AGN systems. Using our fiducial initial parameters, let us compare the computed tidal jet precession timescale $t_{\rm tid}$ with the observed jet precession timescale $t_{\rm obs}$ of four example AGN systems for which the distance and BH mass are known, in order of shortest to longest $t_{\rm obs}$. We assume the measured BH mass is the total mass of a binary BH. 
\begin{itemize}
    \item OJ287, the seminal blazar, has distance $z = 0.306$ \cite{1989A&AS...80..103S}, binary BH mass $M = \num{4e8}\Msol$ \cite{2002A&A...388L..48L}, and $t_{\rm obs} \approx 20$ yrs~\cite{2018MNRAS.478.3199B,2023ApJ...951..106B}, implying that a primary BH spin aligned from accretion provides $t_{\rm tid} \ssim 50$ yrs for a binary separation $r = r_{\rm decoup} \approx 0.006$ pc. 
    The corresponding LISA SNR is $\rho \approx 0$ due to the high mass and modest redshift placing it just outside the LISA horizon (see e.g., Fig.~\ref{F:waterfall}), although a smaller instrumental lower frequency cut-off would make it observable, e.g. $f_{\rm low} = 10^{-5}$ Hz as in \cite{2024arXiv240112813G}.
    
    \item NGC 1275 (3C 84) has distance $z = 0.0175$ \cite{2005MNRAS.359..755W} and binary BH mass $M = \num{3.4e8}\Msol$ \cite{1992AJ....103..691H}, and a long observed jet precession timescale $t_{\rm obs} \ssim 10^7$ yrs \cite{2006MNRAS.366..758D}, implying that a large separation $r > r_{\rm decoup}$ and an in-plane BH spin $\theta_1 = \pi/2$ from a broken accretion disk can provide $t_{\rm tid} \ssim 10^7$ yrs. 
    Compared to OJ287, for its large mass the closer distance of NGC 1275 makes it detectable by LISA with $\rho \approx 2733$. Alternatively, one could speculate that the long $t_{\rm obs}$ is from geodetic precession and the shorter VLBI-observed precession of 40 yrs \cite{2019Galax...7...72B} corresponds to $t_{\rm tid}$ which would imply a small binary separation and aligned BH spin. 

    \item 2MASXJ12032061+1319316 \cite{2017MNRAS.465.4772R} is a candidate dual AGN and has distance $z = 0.0584$, binary BH mass $M = \num{1.56e8}\Msol$, and an intermediate observed jet precession timescale $t_{\rm obs} \ssim 10^5$ yrs, implying that a large separation $r > r_{\rm decoup}$ and either an aligned (misaligned) BH spin $\theta_1 \ssim 0$ ($\theta_1 \ssim \pi/2$) from accretion disk alignment (breaking) can provide $t_{\rm tid} \ssim 10^4$ ($10^6$) yrs, and in both cases is detectable by LISA with $\rho \approx 9055$ (8045). 

    \item M81*, the nuclear core of the M81 galaxy, has distance $z < 0.001$ \cite{1994ApJ...427..628F}, binary BH mass $M = \Mscale$ \cite{2011A&A...533A.111M}, and $t_{\rm obs} \approx 7$ yrs \cite{2023A&A...672L...5V}, implying that a primary BH spin aligned from accretion provides $t_{\rm tid} \approx 5$ yrs for a binary separation $r = r_{\rm decoup} \approx 0.006$ pc. The corresponding LISA SNR is remarkably high, $\rho \gtrsim 10^6$, due to the very small redshift and modest binary mass.    
\end{itemize}
These examples show that a broad range of observed jet precession timescales can be explained without tuning the disk migration parameters. This provides a simple interpretation of observed jet precession timescales, but it also implies significant degeneracies may exist in the parameter space. 
In the four systems highlighted above, we have assumed that a binary BH exists in the unresolved sub-kinematic regime of the jet and the precession of an accretion disk around the primary BH drives the precession of the jet. This is observationally very uncertain, and in most cases one cannot definitively conclude whether the Lense-Thirring or tidal torque is responsible for the jet precession. However there is some evidence of a binary BH scenario for OJ287 \cite{2000A&AS..146..141P,2018MNRAS.478.3199B} and 2MAS \cite{2017MNRAS.465.4772R}. 
Even if a binary BH is present, the jet motion can in principle be driven by the plunging of the secondary through the primary accretion disk \cite{2016ApJ...819L..37V}. It is possible that several of these mechanisms can contribute to observed jet precession. 

If LISA is able to uncover a population of BH binaries, we can infer the prevalence of jets in their formation history and predict statistical properties of precessing-jet AGN systems in the cosmos. Considering that such LISA detections are generally expected to be dominated by mergers at redshifts greater than those of the precessing AGN jets we examined above, it might be challenging to accomplish multi-messenger (i.e. GW+EM) studies of precessing jets. 
However, radio wavelength instrumentation is improving fast, and good quality imaging is becoming available at increasingly higher redshift, see e.g. \cite{2023MNRAS.520.4427M}

While we underscore the use of GW observations and astrophysical modeling to probe precessing jets, multi-messenger studies of precessing jets are plausible. For instance, we can relate the "fossil record" provided by the jet evolution to an observed binary BH merger. 
The jet precession timescale $t_\mathrm{tid}$ can be significantly below the binary BH merger timescale (Fig.~\ref{F:timescales}), and $t_{\rm tid} \ssim 1$ yr for aligned-spin systems with masses $M \approx 10^6$ - $10^7 \Msol$ whose mergers are well observable with LISA (Fig.~\ref{F:waterfall}). 
For jets typically traveling close to light speed and often with apparent superluminal motion \citep[e.g.,][]{2008A&A...484..119B,2015ApJ...798..134H},
this implies one needs to resolve the jet structure at better than 1 pc resolution. If that could be achieved over $\ssim 100$~pc, perhaps aided by LISA revealing the host galaxy of such a binary BH merger, the fossil record of the earlier binary inspiral would be seen via an increasing frequency in the jet oscillations.
Observations of jets of this quality are indeed achievable, 
for example, Cygnus~A at redshift $z=0.056$ has been observed with sub-parsec resolution on parsec scales
\citep{2016A&A...588L...9B} and with different configurations and at different frequencies almost continuously out to the
kpc scale \citep{1998AJ....115.1295K,1998A&A...329..873K,2019ApJ...878...61N}. Other recent VLBI observations show knotty, sometimes edge-brightened parsec-scale structure that has been used to diagnose
jet precession \citep{2017A&ARv..25....4B,2017A&A...602A..29B}.

We focus on LISA detectability of AGN jet precession as we do not conduct parameter estimation of LISA sources. We can crudely approximate the errors on the binary BH spin orientations in the limit of large SNR using the results of other studies that have computed realistic errors on $\theta_1$ and $\theta_2$, which are uncertain. If we take typical uncertainties on $\theta_1$ and $\theta_2$ to be $\ssim 10^\circ$ for detectable sources \cite{2016PhRvD..93b4003K,2023PhRvD.108l4045P}, this suggests that our binaries have uncertainties $\ssim 10/1000 = 0.01^\circ$. This would be sufficient for measuring the distinct peaks in $\theta_1$ and $\theta_2$ shown in Fig.~\ref{F:corner} and for probing the complicated correlations between $t_{\rm tid}$ and $\rho$. Future studies such as ours will open the door to understanding the jet precession of high redshift AGN.

Our conclusions suggest correlations between binaries that underwent disk migration and emitted precessing jets with sources detectable by LISA, depending on the spins of the massive binary BH population. 
These possibilities can be imprinted on their LISA signal. 
For instance, a population of binaries with broken disks, i.e., those with $\cos\theta \ssim 0$, will exhibit longer $t_{\rm tid}$ and enter the LISA band experiencing significant binary BH spin precession. Binaries with aligned spins from accretion during disk migration will have short $t_{\rm tid}$ and can enter the LISA band with large precession frequency of the binary BH orbital angular momentum. Also, LISA measurements of binaries with large nutation amplitudes of the binary orbital angular momentum might be indicative of progenitors that encountered double disk-breaking during disk migration and complicated jet precession timescales. 

EM observations of OJ287 have revealed a jet motion that occurs within a precession cycle, referred to as jet nutation \cite{2018MNRAS.478.3199B}. 
It seems unlikely for this jet nutation to be caused by the nutation of the orbital angular momentum of a binary BH which is a spin-spin coupling effect that is likely negligible at the large distances where these jets exist during disk migration. Instead, the jet nutation could be linked to accretion disk breaking, e.g., if the disk breaks and BH alignment is suppressed then the inner and outer portions of the broken disk can precess at different orientations such that the outer portion of the disk precesses at a rate comparable to the binary orbital period while the inner portion continues precessing normally.  

A few caveats of our analysis are worth discussing. Importantly, while our results depend on the assumptions of our model of binary BH evolution and accretion, most notably adopting a Shakura-Sunyaev disk prescription compared to performing full GRMHD simulations, this work is a first step toward future studies with better accounting of astrophysical and general relativistic processes. More specifically, we assume that disk migration and BH spin alignment occur in a quasi-adiabatic inspiral, where the mass and spin magnitude of the accreting BH are constant. Investigating this assumption is an active topic of research and a potential area for improving our semi-analytic model and hence our predictions for $t_{\rm tid}$ and LISA. We also assume that the disk radius $R_{\rm d}$ in Eq.~(\ref{E:Ttid}) is given by $R_{\rm tid}$, which is a model-dependent choice, implying our predictions for $t_{\rm tid}$ may be quantitatively different than those from other studies. We do not consider other scenarios that might explain AGN jet precession timescales, but it will be important to include many possibilities when attempting to find likely formation histories of these precessing-jet systems. Future studies will open the door to testing our assumptions about the connection between jet precession and binary BH evolution in gas-rich AGN. 

Ultimately, our work motivates many future studies: (i)
using multi-messenger frameworks to constrain the population of AGN jets with future GW and EM datasets in conjunction with astrophysical modeling; (ii) connecting cosmological simulations of galactic evolution with observable properties of jets via a subgrid process, see e.g. \cite{2023arXiv231117144B}, that accounts for accretion disk precession and breaking; (iii) using Bayesian inference and LISA datasets to predict statistical properties of jets possibly arising from binary BH disk migration; (iv) studying the possible correlations between the recoil velocity of a binary BH merger and the jet observables for the post-merger BH remnant if in a gas-rich environment; (v) broadband-GW constraints on the population of AGNs with precessing jets by leveraging the complementarity between pular-timing array and LISA observations~\cite{2017ogw..book...43C}; and (vi) using a similar modeling approach to ours but for accreting stellar-mass binaries that produce precessing jets and later emit GW chirps detectable by ground-based detectors such as LIGO. 

\acknowledgements
The authors would like to thank Samar Safi-Harb, Matt Nicholl, Alberto Vecchio, Hannah Middleton, Christopher J. Moore, Clement Bonnerot, and Tyrone Woods for helpful discussions and comments. 
N.S. is supported by the Natural Sciences and Engineering Research Council of Canada (NSERC) through the Canada Research Chairs and Discovery Grants programs.
N.S. and D.G. are supported by Leverhulme Trust Grant No. RPG-2019-350. 
D.G. is supported by ERC Starting Grant No.~945155--GWmining, 
Cariplo Foundation Grant No.~2021-0555, 
MUR PRIN Grant No.~2022-Z9X4XS,
MSCA Fellowships No.~101064542--StochRewind and No.~101149270--ProtoBH, 
and the ICSC National Research Centre funded by NextGenerationEU. 
Computational work was performed at CINECA with allocations 
through INFN and Bicocca.

\newpage

\bibliography{jetLISA}

\begin{thebibliography}{111}%
\makeatletter
\providecommand \@ifxundefined [1]{%
 \@ifx{#1\undefined}
}%
\providecommand \@ifnum [1]{%
 \ifnum #1\expandafter \@firstoftwo
 \else \expandafter \@secondoftwo
 \fi
}%
\providecommand \@ifx [1]{%
 \ifx #1\expandafter \@firstoftwo
 \else \expandafter \@secondoftwo
 \fi
}%
\providecommand \natexlab [1]{#1}%
\providecommand \enquote  [1]{``#1''}%
\providecommand \bibnamefont  [1]{#1}%
\providecommand \bibfnamefont [1]{#1}%
\providecommand \citenamefont [1]{#1}%
\providecommand \href@noop [0]{\@secondoftwo}%
\providecommand \href [0]{\begingroup \@sanitize@url \@href}%
\providecommand \@href[1]{\@@startlink{#1}\@@href}%
\providecommand \@@href[1]{\endgroup#1\@@endlink}%
\providecommand \@sanitize@url [0]{\catcode `\\12\catcode `\$12\catcode `\&12\catcode `\#12\catcode `\^12\catcode `\_12\catcode `\%12\relax}%
\providecommand \@@startlink[1]{}%
\providecommand \@@endlink[0]{}%
\providecommand \url  [0]{\begingroup\@sanitize@url \@url }%
\providecommand \@url [1]{\endgroup\@href {#1}{\urlprefix }}%
\providecommand \urlprefix  [0]{URL }%
\providecommand \Eprint [0]{\href }%
\providecommand \doibase [0]{https://doi.org/}%
\providecommand \selectlanguage [0]{\@gobble}%
\providecommand \bibinfo  [0]{\@secondoftwo}%
\providecommand \bibfield  [0]{\@secondoftwo}%
\providecommand \translation [1]{[#1]}%
\providecommand \BibitemOpen [0]{}%
\providecommand \bibitemStop [0]{}%
\providecommand \bibitemNoStop [0]{.\EOS\space}%
\providecommand \EOS [0]{\spacefactor3000\relax}%
\providecommand \BibitemShut  [1]{\csname bibitem#1\endcsname}%
\let\auto@bib@innerbib\@empty
\bibitem [{\citenamefont {{de Gouveia Dal Pino}}(2005)}]{2005AdSpR..35..908D}%
  \BibitemOpen
  \bibfield  {author} {\bibinfo {author} {\bibfnamefont {E.~M.}\ \bibnamefont {{de Gouveia Dal Pino}}},\ }\href {https://doi.org/10.1016/j.asr.2005.03.145} {\bibfield  {journal} {\bibinfo  {journal} {\advss}\ }\textbf {\bibinfo {volume} {35}},\ \bibinfo {pages} {908} (\bibinfo {year} {2005})},\ \Eprint {https://arxiv.org/abs/astro-ph/0406319} {arXiv:astro-ph/0406319 [astro-ph]} \BibitemShut {NoStop}%
\bibitem [{\citenamefont {{Maiolino}}\ \emph {et~al.}(2017)\citenamefont {{Maiolino}}, \citenamefont {{Russell}}, \citenamefont {{Fabian}}, \citenamefont {{Carniani}}, \citenamefont {{Gallagher}}, \citenamefont {{Cazzoli}}, \citenamefont {{Arribas}}, \citenamefont {{Belfiore}}, \citenamefont {{Bellocchi}}, \citenamefont {{Colina}}, \citenamefont {{Cresci}}, \citenamefont {{Ishibashi}}, \citenamefont {{Marconi}}, \citenamefont {{Mannucci}}, \citenamefont {{Oliva}},\ and\ \citenamefont {{Sturm}}}]{2017Natur.544..202M}%
  \BibitemOpen
  \bibfield  {author} {\bibinfo {author} {\bibfnamefont {R.}~\bibnamefont {{Maiolino}}}, \bibinfo {author} {\bibfnamefont {H.~R.}\ \bibnamefont {{Russell}}}, \bibinfo {author} {\bibfnamefont {A.~C.}\ \bibnamefont {{Fabian}}}, \bibinfo {author} {\bibfnamefont {S.}~\bibnamefont {{Carniani}}}, \bibinfo {author} {\bibfnamefont {R.}~\bibnamefont {{Gallagher}}}, \bibinfo {author} {\bibfnamefont {S.}~\bibnamefont {{Cazzoli}}}, \bibinfo {author} {\bibfnamefont {S.}~\bibnamefont {{Arribas}}}, \bibinfo {author} {\bibfnamefont {F.}~\bibnamefont {{Belfiore}}}, \bibinfo {author} {\bibfnamefont {E.}~\bibnamefont {{Bellocchi}}}, \bibinfo {author} {\bibfnamefont {L.}~\bibnamefont {{Colina}}}, \bibinfo {author} {\bibfnamefont {G.}~\bibnamefont {{Cresci}}}, \bibinfo {author} {\bibfnamefont {W.}~\bibnamefont {{Ishibashi}}}, \bibinfo {author} {\bibfnamefont {A.}~\bibnamefont {{Marconi}}}, \bibinfo {author} {\bibfnamefont {F.}~\bibnamefont {{Mannucci}}}, \bibinfo {author} {\bibfnamefont {E.}~\bibnamefont {{Oliva}}},\ and\
  \bibinfo {author} {\bibfnamefont {E.}~\bibnamefont {{Sturm}}},\ }\href {https://doi.org/10.1038/nature21677} {\bibfield  {journal} {\bibinfo  {journal} {Nature}\ }\textbf {\bibinfo {volume} {544}},\ \bibinfo {pages} {202} (\bibinfo {year} {2017})},\ \Eprint {https://arxiv.org/abs/1703.08587} {arXiv:1703.08587 [astro-ph.GA]} \BibitemShut {NoStop}%
\bibitem [{\citenamefont {{Hardcastle}}\ and\ \citenamefont {{Croston}}(2020)}]{2020NewAR..8801539H}%
  \BibitemOpen
  \bibfield  {author} {\bibinfo {author} {\bibfnamefont {M.~J.}\ \bibnamefont {{Hardcastle}}}\ and\ \bibinfo {author} {\bibfnamefont {J.~H.}\ \bibnamefont {{Croston}}},\ }\href {https://doi.org/10.1016/j.newar.2020.101539} {\bibfield  {journal} {\bibinfo  {journal} {New Astron. Rev.}\ }\textbf {\bibinfo {volume} {88}},\ \bibinfo {eid} {101539} (\bibinfo {year} {2020})},\ \Eprint {https://arxiv.org/abs/2003.06137} {arXiv:2003.06137 [astro-ph.HE]} \BibitemShut {NoStop}%
\bibitem [{\citenamefont {{Baldi}}(2023)}]{2023A&ARv..31....3B}%
  \BibitemOpen
  \bibfield  {author} {\bibinfo {author} {\bibfnamefont {R.~D.}\ \bibnamefont {{Baldi}}},\ }\href {https://doi.org/10.1007/s00159-023-00148-3} {\bibfield  {journal} {\bibinfo  {journal} {Astron. Astrophys. Rev.}\ }\textbf {\bibinfo {volume} {31}},\ \bibinfo {eid} {3} (\bibinfo {year} {2023})},\ \Eprint {https://arxiv.org/abs/2307.08379} {arXiv:2307.08379 [astro-ph.GA]} \BibitemShut {NoStop}%
\bibitem [{\citenamefont {{Sabater}}\ \emph {et~al.}(2019)\citenamefont {{Sabater}}, \citenamefont {{Best}}, \citenamefont {{Hardcastle}}, \citenamefont {{Shimwell}}, \citenamefont {{Tasse}}, \citenamefont {{Williams}}, \citenamefont {{Br{\"u}ggen}}, \citenamefont {{Cochrane}}, \citenamefont {{Croston}}, \citenamefont {{de Gasperin}}, \citenamefont {{Duncan}}, \citenamefont {{G{\"u}rkan}}, \citenamefont {{Mechev}}, \citenamefont {{Morabito}}, \citenamefont {{Prandoni}}, \citenamefont {{R{\"o}ttgering}}, \citenamefont {{Smith}}, \citenamefont {{Harwood}}, \citenamefont {{Mingo}}, \citenamefont {{Mooney}},\ and\ \citenamefont {{Saxena}}}]{2019A&A...622A..17S}%
  \BibitemOpen
  \bibfield  {author} {\bibinfo {author} {\bibfnamefont {J.}~\bibnamefont {{Sabater}}}, \bibinfo {author} {\bibfnamefont {P.~N.}\ \bibnamefont {{Best}}}, \bibinfo {author} {\bibfnamefont {M.~J.}\ \bibnamefont {{Hardcastle}}}, \bibinfo {author} {\bibfnamefont {T.~W.}\ \bibnamefont {{Shimwell}}}, \bibinfo {author} {\bibfnamefont {C.}~\bibnamefont {{Tasse}}}, \bibinfo {author} {\bibfnamefont {W.~L.}\ \bibnamefont {{Williams}}}, \bibinfo {author} {\bibfnamefont {M.}~\bibnamefont {{Br{\"u}ggen}}}, \bibinfo {author} {\bibfnamefont {R.~K.}\ \bibnamefont {{Cochrane}}}, \bibinfo {author} {\bibfnamefont {J.~H.}\ \bibnamefont {{Croston}}}, \bibinfo {author} {\bibfnamefont {F.}~\bibnamefont {{de Gasperin}}}, \bibinfo {author} {\bibfnamefont {K.~J.}\ \bibnamefont {{Duncan}}}, \bibinfo {author} {\bibfnamefont {G.}~\bibnamefont {{G{\"u}rkan}}}, \bibinfo {author} {\bibfnamefont {A.~P.}\ \bibnamefont {{Mechev}}}, \bibinfo {author} {\bibfnamefont {L.~K.}\ \bibnamefont {{Morabito}}}, \bibinfo {author} {\bibfnamefont
  {I.}~\bibnamefont {{Prandoni}}}, \bibinfo {author} {\bibfnamefont {H.~J.~A.}\ \bibnamefont {{R{\"o}ttgering}}}, \bibinfo {author} {\bibfnamefont {D.~J.~B.}\ \bibnamefont {{Smith}}}, \bibinfo {author} {\bibfnamefont {J.~J.}\ \bibnamefont {{Harwood}}}, \bibinfo {author} {\bibfnamefont {B.}~\bibnamefont {{Mingo}}}, \bibinfo {author} {\bibfnamefont {S.}~\bibnamefont {{Mooney}}},\ and\ \bibinfo {author} {\bibfnamefont {A.}~\bibnamefont {{Saxena}}},\ }\href {https://doi.org/10.1051/0004-6361/201833883} {\bibfield  {journal} {\bibinfo  {journal} {Astron. Astrophys.}\ }\textbf {\bibinfo {volume} {622}},\ \bibinfo {eid} {A17} (\bibinfo {year} {2019})},\ \Eprint {https://arxiv.org/abs/1811.05528} {arXiv:1811.05528 [astro-ph.GA]} \BibitemShut {NoStop}%
\bibitem [{\citenamefont {{Purser}}\ \emph {et~al.}(2021)\citenamefont {{Purser}}, \citenamefont {{Lumsden}}, \citenamefont {{Hoare}},\ and\ \citenamefont {{Kurtz}}}]{2021MNRAS.504..338P}%
  \BibitemOpen
  \bibfield  {author} {\bibinfo {author} {\bibfnamefont {S.~J.~D.}\ \bibnamefont {{Purser}}}, \bibinfo {author} {\bibfnamefont {S.~L.}\ \bibnamefont {{Lumsden}}}, \bibinfo {author} {\bibfnamefont {M.~G.}\ \bibnamefont {{Hoare}}},\ and\ \bibinfo {author} {\bibfnamefont {S.}~\bibnamefont {{Kurtz}}},\ }\href {https://doi.org/10.1093/mnras/stab747} {\bibfield  {journal} {\bibinfo  {journal} {Mon. Not. R. Astron. Soc.}\ }\textbf {\bibinfo {volume} {504}},\ \bibinfo {pages} {338} (\bibinfo {year} {2021})},\ \Eprint {https://arxiv.org/abs/2103.08990} {arXiv:2103.08990 [astro-ph.GA]} \BibitemShut {NoStop}%
\bibitem [{\citenamefont {{Reipurth}}\ and\ \citenamefont {{Bally}}(2001)}]{2001ARA&A..39..403R}%
  \BibitemOpen
  \bibfield  {author} {\bibinfo {author} {\bibfnamefont {B.}~\bibnamefont {{Reipurth}}}\ and\ \bibinfo {author} {\bibfnamefont {J.}~\bibnamefont {{Bally}}},\ }\href {https://doi.org/10.1146/annurev.astro.39.1.403} {\bibfield  {journal} {\bibinfo  {journal} {Annu. Rev. Astron. Astrophys.}\ }\textbf {\bibinfo {volume} {39}},\ \bibinfo {pages} {403} (\bibinfo {year} {2001})}\BibitemShut {NoStop}%
\bibitem [{\citenamefont {{Lee}}(2020)}]{2020A&ARv..28....1L}%
  \BibitemOpen
  \bibfield  {author} {\bibinfo {author} {\bibfnamefont {C.-F.}\ \bibnamefont {{Lee}}},\ }\href {https://doi.org/10.1007/s00159-020-0123-7} {\bibfield  {journal} {\bibinfo  {journal} {Astron. Astrophys. Rev.}\ }\textbf {\bibinfo {volume} {28}},\ \bibinfo {eid} {1} (\bibinfo {year} {2020})},\ \Eprint {https://arxiv.org/abs/2002.05823} {arXiv:2002.05823 [astro-ph.GA]} \BibitemShut {NoStop}%
\bibitem [{\citenamefont {{Mirabel}}\ and\ \citenamefont {{Rodr{\'\i}guez}}(1998)}]{1998Natur.392..673M}%
  \BibitemOpen
  \bibfield  {author} {\bibinfo {author} {\bibfnamefont {I.~F.}\ \bibnamefont {{Mirabel}}}\ and\ \bibinfo {author} {\bibfnamefont {L.~F.}\ \bibnamefont {{Rodr{\'\i}guez}}},\ }\href {https://doi.org/10.1038/33603} {\bibfield  {journal} {\bibinfo  {journal} {Nature}\ }\textbf {\bibinfo {volume} {392}},\ \bibinfo {pages} {673} (\bibinfo {year} {1998})}\BibitemShut {NoStop}%
\bibitem [{\citenamefont {{Fender}}\ \emph {et~al.}(2023)\citenamefont {{Fender}}, \citenamefont {{Mooley}}, \citenamefont {{Motta}}, \citenamefont {{Bright}}, \citenamefont {{Williams}}, \citenamefont {{Rushton}}, \citenamefont {{Beswick}}, \citenamefont {{Miller-Jones}}, \citenamefont {{Kimura}}, \citenamefont {{Isogai}},\ and\ \citenamefont {{Kato}}}]{2023MNRAS.518.1243F}%
  \BibitemOpen
  \bibfield  {author} {\bibinfo {author} {\bibfnamefont {R.~P.}\ \bibnamefont {{Fender}}}, \bibinfo {author} {\bibfnamefont {K.~P.}\ \bibnamefont {{Mooley}}}, \bibinfo {author} {\bibfnamefont {S.~E.}\ \bibnamefont {{Motta}}}, \bibinfo {author} {\bibfnamefont {J.~S.}\ \bibnamefont {{Bright}}}, \bibinfo {author} {\bibfnamefont {D.~R.~A.}\ \bibnamefont {{Williams}}}, \bibinfo {author} {\bibfnamefont {A.~P.}\ \bibnamefont {{Rushton}}}, \bibinfo {author} {\bibfnamefont {R.~J.}\ \bibnamefont {{Beswick}}}, \bibinfo {author} {\bibfnamefont {J.~C.~A.}\ \bibnamefont {{Miller-Jones}}}, \bibinfo {author} {\bibfnamefont {M.}~\bibnamefont {{Kimura}}}, \bibinfo {author} {\bibfnamefont {K.}~\bibnamefont {{Isogai}}},\ and\ \bibinfo {author} {\bibfnamefont {T.}~\bibnamefont {{Kato}}},\ }\href {https://doi.org/10.1093/mnras/stac1836} {\bibfield  {journal} {\bibinfo  {journal} {Mon. Not. R. Astron. Soc.}\ }\textbf {\bibinfo {volume} {518}},\ \bibinfo {pages} {1243} (\bibinfo {year} {2023})},\ \Eprint
  {https://arxiv.org/abs/2206.09831} {arXiv:2206.09831 [astro-ph.HE]} \BibitemShut {NoStop}%
\bibitem [{\citenamefont {{Fender}}\ and\ \citenamefont {{Gallo}}(2014)}]{2014SSRv..183..323F}%
  \BibitemOpen
  \bibfield  {author} {\bibinfo {author} {\bibfnamefont {R.}~\bibnamefont {{Fender}}}\ and\ \bibinfo {author} {\bibfnamefont {E.}~\bibnamefont {{Gallo}}},\ }\href {https://doi.org/10.1007/s11214-014-0069-z} {\bibfield  {journal} {\bibinfo  {journal} {Space Sci. Rev.}\ }\textbf {\bibinfo {volume} {183}},\ \bibinfo {pages} {323} (\bibinfo {year} {2014})},\ \Eprint {https://arxiv.org/abs/1407.3674} {arXiv:1407.3674 [astro-ph.HE]} \BibitemShut {NoStop}%
\bibitem [{\citenamefont {{Ghisellini}}\ \emph {et~al.}(2014)\citenamefont {{Ghisellini}}, \citenamefont {{Tavecchio}}, \citenamefont {{Maraschi}}, \citenamefont {{Celotti}},\ and\ \citenamefont {{Sbarrato}}}]{2014Natur.515..376G}%
  \BibitemOpen
  \bibfield  {author} {\bibinfo {author} {\bibfnamefont {G.}~\bibnamefont {{Ghisellini}}}, \bibinfo {author} {\bibfnamefont {F.}~\bibnamefont {{Tavecchio}}}, \bibinfo {author} {\bibfnamefont {L.}~\bibnamefont {{Maraschi}}}, \bibinfo {author} {\bibfnamefont {A.}~\bibnamefont {{Celotti}}},\ and\ \bibinfo {author} {\bibfnamefont {T.}~\bibnamefont {{Sbarrato}}},\ }\href {https://doi.org/10.1038/nature13856} {\bibfield  {journal} {\bibinfo  {journal} {Nature}\ }\textbf {\bibinfo {volume} {515}},\ \bibinfo {pages} {376} (\bibinfo {year} {2014})},\ \Eprint {https://arxiv.org/abs/1411.5368} {arXiv:1411.5368 [astro-ph.HE]} \BibitemShut {NoStop}%
\bibitem [{\citenamefont {{Lister}}\ \emph {et~al.}(2016)\citenamefont {{Lister}} \emph {et~al.}}]{2016Galax...4...29L}%
  \BibitemOpen
  \bibfield  {author} {\bibinfo {author} {\bibfnamefont {M.}~\bibnamefont {{Lister}}} \emph {et~al.},\ }\href {https://doi.org/10.3390/galaxies4030029} {\bibfield  {journal} {\bibinfo  {journal} {Galaxies}\ }\textbf {\bibinfo {volume} {4}},\ \bibinfo {eid} {29} (\bibinfo {year} {2016})}\BibitemShut {NoStop}%
\bibitem [{\citenamefont {{van den Eijnden}}\ \emph {et~al.}(2018)\citenamefont {{van den Eijnden}}, \citenamefont {{Degenaar}}, \citenamefont {{Russell}}, \citenamefont {{Wijnands}}, \citenamefont {{Miller-Jones}}, \citenamefont {{Sivakoff}},\ and\ \citenamefont {{Hern{\'a}ndez Santisteban}}}]{2018Natur.562..233V}%
  \BibitemOpen
  \bibfield  {author} {\bibinfo {author} {\bibfnamefont {J.}~\bibnamefont {{van den Eijnden}}}, \bibinfo {author} {\bibfnamefont {N.}~\bibnamefont {{Degenaar}}}, \bibinfo {author} {\bibfnamefont {T.~D.}\ \bibnamefont {{Russell}}}, \bibinfo {author} {\bibfnamefont {R.}~\bibnamefont {{Wijnands}}}, \bibinfo {author} {\bibfnamefont {J.~C.~A.}\ \bibnamefont {{Miller-Jones}}}, \bibinfo {author} {\bibfnamefont {G.~R.}\ \bibnamefont {{Sivakoff}}},\ and\ \bibinfo {author} {\bibfnamefont {J.~V.}\ \bibnamefont {{Hern{\'a}ndez Santisteban}}},\ }\href {https://doi.org/10.1038/s41586-018-0524-1} {\bibfield  {journal} {\bibinfo  {journal} {Nature}\ }\textbf {\bibinfo {volume} {562}},\ \bibinfo {pages} {233} (\bibinfo {year} {2018})},\ \Eprint {https://arxiv.org/abs/1809.10204} {arXiv:1809.10204 [astro-ph.HE]} \BibitemShut {NoStop}%
\bibitem [{\citenamefont {{Gallo}}\ \emph {et~al.}(2018)\citenamefont {{Gallo}}, \citenamefont {{Degenaar}},\ and\ \citenamefont {{van den Eijnden}}}]{2018MNRAS.478L.132G}%
  \BibitemOpen
  \bibfield  {author} {\bibinfo {author} {\bibfnamefont {E.}~\bibnamefont {{Gallo}}}, \bibinfo {author} {\bibfnamefont {N.}~\bibnamefont {{Degenaar}}},\ and\ \bibinfo {author} {\bibfnamefont {J.}~\bibnamefont {{van den Eijnden}}},\ }\href {https://doi.org/10.1093/mnrasl/sly083} {\bibfield  {journal} {\bibinfo  {journal} {Mon. Not. R. Astron. Soc.}\ }\textbf {\bibinfo {volume} {478}},\ \bibinfo {pages} {L132} (\bibinfo {year} {2018})},\ \Eprint {https://arxiv.org/abs/1805.01905} {arXiv:1805.01905 [astro-ph.HE]} \BibitemShut {NoStop}%
\bibitem [{\citenamefont {{Blandford}}\ \emph {et~al.}(2019)\citenamefont {{Blandford}}, \citenamefont {{Meier}},\ and\ \citenamefont {{Readhead}}}]{2019ARA&A..57..467B}%
  \BibitemOpen
  \bibfield  {author} {\bibinfo {author} {\bibfnamefont {R.}~\bibnamefont {{Blandford}}}, \bibinfo {author} {\bibfnamefont {D.}~\bibnamefont {{Meier}}},\ and\ \bibinfo {author} {\bibfnamefont {A.}~\bibnamefont {{Readhead}}},\ }\href {https://doi.org/10.1146/annurev-astro-081817-051948} {\bibfield  {journal} {\bibinfo  {journal} {Annu. Rev. Astron. Astrophys.}\ }\textbf {\bibinfo {volume} {57}},\ \bibinfo {pages} {467} (\bibinfo {year} {2019})},\ \Eprint {https://arxiv.org/abs/1812.06025} {arXiv:1812.06025 [astro-ph.HE]} \BibitemShut {NoStop}%
\bibitem [{\citenamefont {{Hovatta}}\ and\ \citenamefont {{Lindfors}}(2019)}]{2019NewAR..8701541H}%
  \BibitemOpen
  \bibfield  {author} {\bibinfo {author} {\bibfnamefont {T.}~\bibnamefont {{Hovatta}}}\ and\ \bibinfo {author} {\bibfnamefont {E.}~\bibnamefont {{Lindfors}}},\ }\href {https://doi.org/10.1016/j.newar.2020.101541} {\bibfield  {journal} {\bibinfo  {journal} {New Astron. Rev.}\ }\textbf {\bibinfo {volume} {87}},\ \bibinfo {eid} {101541} (\bibinfo {year} {2019})},\ \Eprint {https://arxiv.org/abs/2003.06322} {arXiv:2003.06322 [astro-ph.HE]} \BibitemShut {NoStop}%
\bibitem [{\citenamefont {{Lister}}\ \emph {et~al.}(2013)\citenamefont {{Lister}}, \citenamefont {{Aller}}, \citenamefont {{Aller}}, \citenamefont {{Homan}}, \citenamefont {{Kellermann}}, \citenamefont {{Kovalev}}, \citenamefont {{Pushkarev}}, \citenamefont {{Richards}}, \citenamefont {{Ros}},\ and\ \citenamefont {{Savolainen}}}]{2013AJ....146..120L}%
  \BibitemOpen
  \bibfield  {author} {\bibinfo {author} {\bibfnamefont {M.~L.}\ \bibnamefont {{Lister}}}, \bibinfo {author} {\bibfnamefont {M.~F.}\ \bibnamefont {{Aller}}}, \bibinfo {author} {\bibfnamefont {H.~D.}\ \bibnamefont {{Aller}}}, \bibinfo {author} {\bibfnamefont {D.~C.}\ \bibnamefont {{Homan}}}, \bibinfo {author} {\bibfnamefont {K.~I.}\ \bibnamefont {{Kellermann}}}, \bibinfo {author} {\bibfnamefont {Y.~Y.}\ \bibnamefont {{Kovalev}}}, \bibinfo {author} {\bibfnamefont {A.~B.}\ \bibnamefont {{Pushkarev}}}, \bibinfo {author} {\bibfnamefont {J.~L.}\ \bibnamefont {{Richards}}}, \bibinfo {author} {\bibfnamefont {E.}~\bibnamefont {{Ros}}},\ and\ \bibinfo {author} {\bibfnamefont {T.}~\bibnamefont {{Savolainen}}},\ }\href {https://doi.org/10.1088/0004-6256/146/5/120} {\bibfield  {journal} {\bibinfo  {journal} {Astron. J.}\ }\textbf {\bibinfo {volume} {146}},\ \bibinfo {eid} {120} (\bibinfo {year} {2013})},\ \Eprint {https://arxiv.org/abs/1308.2713} {arXiv:1308.2713 [astro-ph.CO]} \BibitemShut {NoStop}%
\bibitem [{\citenamefont {{Krause}}\ \emph {et~al.}(2019)\citenamefont {{Krause}}, \citenamefont {{Shabala}}, \citenamefont {{Hardcastle}}, \citenamefont {{Bicknell}}, \citenamefont {{B{\"o}hringer}}, \citenamefont {{Chon}}, \citenamefont {{Nawaz}}, \citenamefont {{Sarzi}},\ and\ \citenamefont {{Wagner}}}]{2019MNRAS.482..240K}%
  \BibitemOpen
  \bibfield  {author} {\bibinfo {author} {\bibfnamefont {M.~G.~H.}\ \bibnamefont {{Krause}}}, \bibinfo {author} {\bibfnamefont {S.~S.}\ \bibnamefont {{Shabala}}}, \bibinfo {author} {\bibfnamefont {M.~J.}\ \bibnamefont {{Hardcastle}}}, \bibinfo {author} {\bibfnamefont {G.~V.}\ \bibnamefont {{Bicknell}}}, \bibinfo {author} {\bibfnamefont {H.}~\bibnamefont {{B{\"o}hringer}}}, \bibinfo {author} {\bibfnamefont {G.}~\bibnamefont {{Chon}}}, \bibinfo {author} {\bibfnamefont {M.~A.}\ \bibnamefont {{Nawaz}}}, \bibinfo {author} {\bibfnamefont {M.}~\bibnamefont {{Sarzi}}},\ and\ \bibinfo {author} {\bibfnamefont {A.~Y.}\ \bibnamefont {{Wagner}}},\ }\href {https://doi.org/10.1093/mnras/sty2558} {\bibfield  {journal} {\bibinfo  {journal} {Mon. Not. R. Astron. Soc.}\ }\textbf {\bibinfo {volume} {482}},\ \bibinfo {pages} {240} (\bibinfo {year} {2019})},\ \Eprint {https://arxiv.org/abs/1809.04050} {arXiv:1809.04050 [astro-ph.HE]} \BibitemShut {NoStop}%
\bibitem [{\citenamefont {{Mingo}}\ \emph {et~al.}(2022)\citenamefont {{Mingo}}, \citenamefont {{Croston}}, \citenamefont {{Best}}, \citenamefont {{Duncan}}, \citenamefont {{Hardcastle}}, \citenamefont {{Kondapally}}, \citenamefont {{Prandoni}}, \citenamefont {{Sabater}}, \citenamefont {{Shimwell}}, \citenamefont {{Williams}}, \citenamefont {{Baldi}}, \citenamefont {{Bonato}}, \citenamefont {{Bondi}}, \citenamefont {{Dabhade}}, \citenamefont {{G{\"u}rkan}}, \citenamefont {{Ineson}}, \citenamefont {{Magliocchetti}}, \citenamefont {{Miley}}, \citenamefont {{Pierce}},\ and\ \citenamefont {{R{\"o}ttgering}}}]{2022MNRAS.511.3250M}%
  \BibitemOpen
  \bibfield  {author} {\bibinfo {author} {\bibfnamefont {B.}~\bibnamefont {{Mingo}}}, \bibinfo {author} {\bibfnamefont {J.~H.}\ \bibnamefont {{Croston}}}, \bibinfo {author} {\bibfnamefont {P.~N.}\ \bibnamefont {{Best}}}, \bibinfo {author} {\bibfnamefont {K.~J.}\ \bibnamefont {{Duncan}}}, \bibinfo {author} {\bibfnamefont {M.~J.}\ \bibnamefont {{Hardcastle}}}, \bibinfo {author} {\bibfnamefont {R.}~\bibnamefont {{Kondapally}}}, \bibinfo {author} {\bibfnamefont {I.}~\bibnamefont {{Prandoni}}}, \bibinfo {author} {\bibfnamefont {J.}~\bibnamefont {{Sabater}}}, \bibinfo {author} {\bibfnamefont {T.~W.}\ \bibnamefont {{Shimwell}}}, \bibinfo {author} {\bibfnamefont {W.~L.}\ \bibnamefont {{Williams}}}, \bibinfo {author} {\bibfnamefont {R.~D.}\ \bibnamefont {{Baldi}}}, \bibinfo {author} {\bibfnamefont {M.}~\bibnamefont {{Bonato}}}, \bibinfo {author} {\bibfnamefont {M.}~\bibnamefont {{Bondi}}}, \bibinfo {author} {\bibfnamefont {P.}~\bibnamefont {{Dabhade}}}, \bibinfo {author} {\bibfnamefont {G.}~\bibnamefont
  {{G{\"u}rkan}}}, \bibinfo {author} {\bibfnamefont {J.}~\bibnamefont {{Ineson}}}, \bibinfo {author} {\bibfnamefont {M.}~\bibnamefont {{Magliocchetti}}}, \bibinfo {author} {\bibfnamefont {G.}~\bibnamefont {{Miley}}}, \bibinfo {author} {\bibfnamefont {J.~C.~S.}\ \bibnamefont {{Pierce}}},\ and\ \bibinfo {author} {\bibfnamefont {H.~J.~A.}\ \bibnamefont {{R{\"o}ttgering}}},\ }\href {https://doi.org/10.1093/mnras/stac140} {\bibfield  {journal} {\bibinfo  {journal} {Mon. Not. R. Astron. Soc.}\ }\textbf {\bibinfo {volume} {511}},\ \bibinfo {pages} {3250} (\bibinfo {year} {2022})},\ \Eprint {https://arxiv.org/abs/2201.04433} {arXiv:2201.04433 [astro-ph.GA]} \BibitemShut {NoStop}%
\bibitem [{\citenamefont {{Rieger}}(2019)}]{2019Galax...7...28R}%
  \BibitemOpen
  \bibfield  {author} {\bibinfo {author} {\bibfnamefont {F.~M.}\ \bibnamefont {{Rieger}}},\ }\href {https://doi.org/10.3390/galaxies7010028} {\bibfield  {journal} {\bibinfo  {journal} {Galaxies}\ }\textbf {\bibinfo {volume} {7}},\ \bibinfo {eid} {28} (\bibinfo {year} {2019})},\ \Eprint {https://arxiv.org/abs/1901.10216} {arXiv:1901.10216 [astro-ph.HE]} \BibitemShut {NoStop}%
\bibitem [{\citenamefont {{Shabala}}\ \emph {et~al.}(2020)\citenamefont {{Shabala}}, \citenamefont {{Jurlin}}, \citenamefont {{Morganti}}, \citenamefont {{Brienza}}, \citenamefont {{Hardcastle}}, \citenamefont {{Godfrey}}, \citenamefont {{Krause}},\ and\ \citenamefont {{Turner}}}]{2020MNRAS.496.1706S}%
  \BibitemOpen
  \bibfield  {author} {\bibinfo {author} {\bibfnamefont {S.~S.}\ \bibnamefont {{Shabala}}}, \bibinfo {author} {\bibfnamefont {N.}~\bibnamefont {{Jurlin}}}, \bibinfo {author} {\bibfnamefont {R.}~\bibnamefont {{Morganti}}}, \bibinfo {author} {\bibfnamefont {M.}~\bibnamefont {{Brienza}}}, \bibinfo {author} {\bibfnamefont {M.~J.}\ \bibnamefont {{Hardcastle}}}, \bibinfo {author} {\bibfnamefont {L.~E.~H.}\ \bibnamefont {{Godfrey}}}, \bibinfo {author} {\bibfnamefont {M.~G.~H.}\ \bibnamefont {{Krause}}},\ and\ \bibinfo {author} {\bibfnamefont {R.~J.}\ \bibnamefont {{Turner}}},\ }\href {https://doi.org/10.1093/mnras/staa1172} {\bibfield  {journal} {\bibinfo  {journal} {Mon. Not. R. Astron. Soc.}\ }\textbf {\bibinfo {volume} {496}},\ \bibinfo {pages} {1706} (\bibinfo {year} {2020})},\ \Eprint {https://arxiv.org/abs/2004.08979} {arXiv:2004.08979 [astro-ph.GA]} \BibitemShut {NoStop}%
\bibitem [{\citenamefont {{Oei}}\ \emph {et~al.}(2023)\citenamefont {{Oei}}, \citenamefont {{van Weeren}}, \citenamefont {{Gast}}, \citenamefont {{Botteon}}, \citenamefont {{Hardcastle}}, \citenamefont {{Dabhade}}, \citenamefont {{Shimwell}}, \citenamefont {{R{\"o}ttgering}},\ and\ \citenamefont {{Drabent}}}]{2023A&A...672A.163O}%
  \BibitemOpen
  \bibfield  {author} {\bibinfo {author} {\bibfnamefont {M.~S.~S.~L.}\ \bibnamefont {{Oei}}}, \bibinfo {author} {\bibfnamefont {R.~J.}\ \bibnamefont {{van Weeren}}}, \bibinfo {author} {\bibfnamefont {A.~R.~D.~J.~G.~I.~B.}\ \bibnamefont {{Gast}}}, \bibinfo {author} {\bibfnamefont {A.}~\bibnamefont {{Botteon}}}, \bibinfo {author} {\bibfnamefont {M.~J.}\ \bibnamefont {{Hardcastle}}}, \bibinfo {author} {\bibfnamefont {P.}~\bibnamefont {{Dabhade}}}, \bibinfo {author} {\bibfnamefont {T.~W.}\ \bibnamefont {{Shimwell}}}, \bibinfo {author} {\bibfnamefont {H.~J.~A.}\ \bibnamefont {{R{\"o}ttgering}}},\ and\ \bibinfo {author} {\bibfnamefont {A.}~\bibnamefont {{Drabent}}},\ }\href {https://doi.org/10.1051/0004-6361/202243572} {\bibfield  {journal} {\bibinfo  {journal} {Astron. Astrophys.}\ }\textbf {\bibinfo {volume} {672}},\ \bibinfo {eid} {A163} (\bibinfo {year} {2023})},\ \Eprint {https://arxiv.org/abs/2210.10234} {arXiv:2210.10234 [astro-ph.GA]} \BibitemShut {NoStop}%
\bibitem [{\citenamefont {{Nandi}}\ \emph {et~al.}(2021)\citenamefont {{Nandi}}, \citenamefont {{Caproni}}, \citenamefont {{Kharb}}, \citenamefont {{Sebastian}},\ and\ \citenamefont {{Roy}}}]{2021ApJ...908..178N}%
  \BibitemOpen
  \bibfield  {author} {\bibinfo {author} {\bibfnamefont {S.}~\bibnamefont {{Nandi}}}, \bibinfo {author} {\bibfnamefont {A.}~\bibnamefont {{Caproni}}}, \bibinfo {author} {\bibfnamefont {P.}~\bibnamefont {{Kharb}}}, \bibinfo {author} {\bibfnamefont {B.}~\bibnamefont {{Sebastian}}},\ and\ \bibinfo {author} {\bibfnamefont {R.}~\bibnamefont {{Roy}}},\ }\href {https://doi.org/10.3847/1538-4357/abd2ba} {\bibfield  {journal} {\bibinfo  {journal} {Astrophys. J.}\ }\textbf {\bibinfo {volume} {908}},\ \bibinfo {eid} {178} (\bibinfo {year} {2021})},\ \Eprint {https://arxiv.org/abs/2012.06290} {arXiv:2012.06290 [astro-ph.HE]} \BibitemShut {NoStop}%
\bibitem [{\citenamefont {{Cui}}\ \emph {et~al.}(2023)\citenamefont {{Cui}}, \citenamefont {{Hada}}, \citenamefont {{Kawashima}}, \citenamefont {{Kino}}, \citenamefont {{Lin}}, \citenamefont {{Mizuno}}, \citenamefont {{Ro}}, \citenamefont {{Honma}}, \citenamefont {{Yi}}, \citenamefont {{Yu}}, \citenamefont {{Park}}, \citenamefont {{Jiang}}, \citenamefont {{Shen}}, \citenamefont {{Kravchenko}} \emph {et~al.}}]{2023Natur.621..711C}%
  \BibitemOpen
  \bibfield  {author} {\bibinfo {author} {\bibfnamefont {Y.}~\bibnamefont {{Cui}}}, \bibinfo {author} {\bibfnamefont {K.}~\bibnamefont {{Hada}}}, \bibinfo {author} {\bibfnamefont {T.}~\bibnamefont {{Kawashima}}}, \bibinfo {author} {\bibfnamefont {M.}~\bibnamefont {{Kino}}}, \bibinfo {author} {\bibfnamefont {W.}~\bibnamefont {{Lin}}}, \bibinfo {author} {\bibfnamefont {Y.}~\bibnamefont {{Mizuno}}}, \bibinfo {author} {\bibfnamefont {H.}~\bibnamefont {{Ro}}}, \bibinfo {author} {\bibfnamefont {M.}~\bibnamefont {{Honma}}}, \bibinfo {author} {\bibfnamefont {K.}~\bibnamefont {{Yi}}}, \bibinfo {author} {\bibfnamefont {J.}~\bibnamefont {{Yu}}}, \bibinfo {author} {\bibfnamefont {J.}~\bibnamefont {{Park}}}, \bibinfo {author} {\bibfnamefont {W.}~\bibnamefont {{Jiang}}}, \bibinfo {author} {\bibfnamefont {Z.}~\bibnamefont {{Shen}}}, \bibinfo {author} {\bibfnamefont {E.}~\bibnamefont {{Kravchenko}}}, \emph {et~al.},\ }\href {https://doi.org/10.1038/s41586-023-06479-6} {\bibfield  {journal} {\bibinfo  {journal} {Nature}\
  }\textbf {\bibinfo {volume} {621}},\ \bibinfo {pages} {711} (\bibinfo {year} {2023})},\ \Eprint {https://arxiv.org/abs/2310.09015} {arXiv:2310.09015 [astro-ph.HE]} \BibitemShut {NoStop}%
\bibitem [{\citenamefont {{Britzen}}\ \emph {et~al.}(2023)\citenamefont {{Britzen}}, \citenamefont {{Zaja{\v{c}}ek}}, \citenamefont {{Gopal-Krishna}}, \citenamefont {{Fendt}}, \citenamefont {{Kun}}, \citenamefont {{Jaron}}, \citenamefont {{Sillanp{\"a}{\"a}}},\ and\ \citenamefont {{Eckart}}}]{2023ApJ...951..106B}%
  \BibitemOpen
  \bibfield  {author} {\bibinfo {author} {\bibfnamefont {S.}~\bibnamefont {{Britzen}}}, \bibinfo {author} {\bibfnamefont {M.}~\bibnamefont {{Zaja{\v{c}}ek}}}, \bibinfo {author} {\bibnamefont {{Gopal-Krishna}}}, \bibinfo {author} {\bibfnamefont {C.}~\bibnamefont {{Fendt}}}, \bibinfo {author} {\bibfnamefont {E.}~\bibnamefont {{Kun}}}, \bibinfo {author} {\bibfnamefont {F.}~\bibnamefont {{Jaron}}}, \bibinfo {author} {\bibfnamefont {A.}~\bibnamefont {{Sillanp{\"a}{\"a}}}},\ and\ \bibinfo {author} {\bibfnamefont {A.}~\bibnamefont {{Eckart}}},\ }\href {https://doi.org/10.3847/1538-4357/accbbc} {\bibfield  {journal} {\bibinfo  {journal} {Astrophys. J.}\ }\textbf {\bibinfo {volume} {951}},\ \bibinfo {eid} {106} (\bibinfo {year} {2023})},\ \Eprint {https://arxiv.org/abs/2307.05838} {arXiv:2307.05838 [astro-ph.HE]} \BibitemShut {NoStop}%
\bibitem [{\citenamefont {{Fendt}}\ and\ \citenamefont {{Yardimci}}(2022)}]{2022ApJ...933...71F}%
  \BibitemOpen
  \bibfield  {author} {\bibinfo {author} {\bibfnamefont {C.}~\bibnamefont {{Fendt}}}\ and\ \bibinfo {author} {\bibfnamefont {M.}~\bibnamefont {{Yardimci}}},\ }\href {https://doi.org/10.3847/1538-4357/ac7145} {\bibfield  {journal} {\bibinfo  {journal} {Astrophys. J.}\ }\textbf {\bibinfo {volume} {933}},\ \bibinfo {eid} {71} (\bibinfo {year} {2022})},\ \Eprint {https://arxiv.org/abs/2205.08498} {arXiv:2205.08498 [astro-ph.HE]} \BibitemShut {NoStop}%
\bibitem [{\citenamefont {{Britzen}}\ \emph {et~al.}(2018)\citenamefont {{Britzen}}, \citenamefont {{Fendt}}, \citenamefont {{Witzel}}, \citenamefont {{Qian}}, \citenamefont {{Pashchenko}}, \citenamefont {{Kurtanidze}}, \citenamefont {{Zajacek}}, \citenamefont {{Martinez}}, \citenamefont {{Karas}}, \citenamefont {{Aller}}, \citenamefont {{Aller}}, \citenamefont {{Eckart}}, \citenamefont {{Nilsson}}, \citenamefont {{Ar{\'e}valo}}, \citenamefont {{Cuadra}}, \citenamefont {{Subroweit}},\ and\ \citenamefont {{Witzel}}}]{2018MNRAS.478.3199B}%
  \BibitemOpen
  \bibfield  {author} {\bibinfo {author} {\bibfnamefont {S.}~\bibnamefont {{Britzen}}}, \bibinfo {author} {\bibfnamefont {C.}~\bibnamefont {{Fendt}}}, \bibinfo {author} {\bibfnamefont {G.}~\bibnamefont {{Witzel}}}, \bibinfo {author} {\bibfnamefont {S.~J.}\ \bibnamefont {{Qian}}}, \bibinfo {author} {\bibfnamefont {I.~N.}\ \bibnamefont {{Pashchenko}}}, \bibinfo {author} {\bibfnamefont {O.}~\bibnamefont {{Kurtanidze}}}, \bibinfo {author} {\bibfnamefont {M.}~\bibnamefont {{Zajacek}}}, \bibinfo {author} {\bibfnamefont {G.}~\bibnamefont {{Martinez}}}, \bibinfo {author} {\bibfnamefont {V.}~\bibnamefont {{Karas}}}, \bibinfo {author} {\bibfnamefont {M.}~\bibnamefont {{Aller}}}, \bibinfo {author} {\bibfnamefont {H.}~\bibnamefont {{Aller}}}, \bibinfo {author} {\bibfnamefont {A.}~\bibnamefont {{Eckart}}}, \bibinfo {author} {\bibfnamefont {K.}~\bibnamefont {{Nilsson}}}, \bibinfo {author} {\bibfnamefont {P.}~\bibnamefont {{Ar{\'e}valo}}}, \bibinfo {author} {\bibfnamefont {J.}~\bibnamefont {{Cuadra}}}, \bibinfo {author}
  {\bibfnamefont {M.}~\bibnamefont {{Subroweit}}},\ and\ \bibinfo {author} {\bibfnamefont {A.}~\bibnamefont {{Witzel}}},\ }\href {https://doi.org/10.1093/mnras/sty1026} {\bibfield  {journal} {\bibinfo  {journal} {Mon. Not. R. Astron. Soc.}\ }\textbf {\bibinfo {volume} {478}},\ \bibinfo {pages} {3199} (\bibinfo {year} {2018})}\BibitemShut {NoStop}%
\bibitem [{\citenamefont {{Dey}}\ \emph {et~al.}(2018)\citenamefont {{Dey}}, \citenamefont {{Valtonen}}, \citenamefont {{Gopakumar}}, \citenamefont {{Zola}}, \citenamefont {{Hudec}}, \citenamefont {{Pihajoki}}, \citenamefont {{Ciprini}}, \citenamefont {{Matsumoto}}, \citenamefont {{Sadakane}}, \citenamefont {{Kidger}}, \citenamefont {{Nilsson}}, \citenamefont {{Mikkola}}, \citenamefont {{Sillanp{\"a}{\"a}}}, \citenamefont {{Takalo}}, \citenamefont {{Lehto}}, \citenamefont {{Berdyugin}}, \citenamefont {{Piirola}}, \citenamefont {{Jermak}}, \citenamefont {{Baliyan}}, \citenamefont {{Pursimo}}, \citenamefont {{Caton}} \emph {et~al.}}]{2018ApJ...866...11D}%
  \BibitemOpen
  \bibfield  {author} {\bibinfo {author} {\bibfnamefont {L.}~\bibnamefont {{Dey}}}, \bibinfo {author} {\bibfnamefont {M.~J.}\ \bibnamefont {{Valtonen}}}, \bibinfo {author} {\bibfnamefont {A.}~\bibnamefont {{Gopakumar}}}, \bibinfo {author} {\bibfnamefont {S.}~\bibnamefont {{Zola}}}, \bibinfo {author} {\bibfnamefont {R.}~\bibnamefont {{Hudec}}}, \bibinfo {author} {\bibfnamefont {P.}~\bibnamefont {{Pihajoki}}}, \bibinfo {author} {\bibfnamefont {S.}~\bibnamefont {{Ciprini}}}, \bibinfo {author} {\bibfnamefont {K.}~\bibnamefont {{Matsumoto}}}, \bibinfo {author} {\bibfnamefont {K.}~\bibnamefont {{Sadakane}}}, \bibinfo {author} {\bibfnamefont {M.}~\bibnamefont {{Kidger}}}, \bibinfo {author} {\bibfnamefont {K.}~\bibnamefont {{Nilsson}}}, \bibinfo {author} {\bibfnamefont {S.}~\bibnamefont {{Mikkola}}}, \bibinfo {author} {\bibfnamefont {A.}~\bibnamefont {{Sillanp{\"a}{\"a}}}}, \bibinfo {author} {\bibfnamefont {L.~O.}\ \bibnamefont {{Takalo}}}, \bibinfo {author} {\bibfnamefont {H.~J.}\ \bibnamefont {{Lehto}}}, \bibinfo
  {author} {\bibfnamefont {A.}~\bibnamefont {{Berdyugin}}}, \bibinfo {author} {\bibfnamefont {V.}~\bibnamefont {{Piirola}}}, \bibinfo {author} {\bibfnamefont {H.}~\bibnamefont {{Jermak}}}, \bibinfo {author} {\bibfnamefont {K.~S.}\ \bibnamefont {{Baliyan}}}, \bibinfo {author} {\bibfnamefont {T.}~\bibnamefont {{Pursimo}}}, \bibinfo {author} {\bibfnamefont {D.~B.}\ \bibnamefont {{Caton}}}, \emph {et~al.},\ }\href {https://doi.org/10.3847/1538-4357/aadd95} {\bibfield  {journal} {\bibinfo  {journal} {Astrophys. J.}\ }\textbf {\bibinfo {volume} {866}},\ \bibinfo {eid} {11} (\bibinfo {year} {2018})},\ \Eprint {https://arxiv.org/abs/1808.09309} {arXiv:1808.09309 [astro-ph.HE]} \BibitemShut {NoStop}%
\bibitem [{\citenamefont {{Rozgonyi}}\ and\ \citenamefont {{Frey}}(2016)}]{2016Galax...4...10R}%
  \BibitemOpen
  \bibfield  {author} {\bibinfo {author} {\bibfnamefont {K.}~\bibnamefont {{Rozgonyi}}}\ and\ \bibinfo {author} {\bibfnamefont {S.}~\bibnamefont {{Frey}}},\ }\href {https://doi.org/10.3390/galaxies4030010} {\bibfield  {journal} {\bibinfo  {journal} {Galaxies}\ }\textbf {\bibinfo {volume} {4}},\ \bibinfo {eid} {10} (\bibinfo {year} {2016})},\ \Eprint {https://arxiv.org/abs/1608.06219} {arXiv:1608.06219 [astro-ph.GA]} \BibitemShut {NoStop}%
\bibitem [{\citenamefont {{Mart{\'\i}-Vidal}}\ \emph {et~al.}(2011)\citenamefont {{Mart{\'\i}-Vidal}}, \citenamefont {{Marcaide}}, \citenamefont {{Alberdi}}, \citenamefont {{P{\'e}rez-Torres}}, \citenamefont {{Ros}},\ and\ \citenamefont {{Guirado}}}]{2011A&A...533A.111M}%
  \BibitemOpen
  \bibfield  {author} {\bibinfo {author} {\bibfnamefont {I.}~\bibnamefont {{Mart{\'\i}-Vidal}}}, \bibinfo {author} {\bibfnamefont {J.~M.}\ \bibnamefont {{Marcaide}}}, \bibinfo {author} {\bibfnamefont {A.}~\bibnamefont {{Alberdi}}}, \bibinfo {author} {\bibfnamefont {M.~A.}\ \bibnamefont {{P{\'e}rez-Torres}}}, \bibinfo {author} {\bibfnamefont {E.}~\bibnamefont {{Ros}}},\ and\ \bibinfo {author} {\bibfnamefont {J.~C.}\ \bibnamefont {{Guirado}}},\ }\href {https://doi.org/10.1051/0004-6361/201117211} {\bibfield  {journal} {\bibinfo  {journal} {Astron. Astrophys.}\ }\textbf {\bibinfo {volume} {533}},\ \bibinfo {eid} {A111} (\bibinfo {year} {2011})},\ \Eprint {https://arxiv.org/abs/1107.0704} {arXiv:1107.0704 [astro-ph.HE]} \BibitemShut {NoStop}%
\bibitem [{\citenamefont {{von Fellenberg}}\ \emph {et~al.}(2023)\citenamefont {{von Fellenberg}}, \citenamefont {{Janssen}}, \citenamefont {{Davelaar}}, \citenamefont {{Zaja{\v{c}}ek}}, \citenamefont {{Britzen}}, \citenamefont {{Falcke}}, \citenamefont {{K{\"o}rding}},\ and\ \citenamefont {{Ros}}}]{2023A&A...672L...5V}%
  \BibitemOpen
  \bibfield  {author} {\bibinfo {author} {\bibfnamefont {S.~D.}\ \bibnamefont {{von Fellenberg}}}, \bibinfo {author} {\bibfnamefont {M.}~\bibnamefont {{Janssen}}}, \bibinfo {author} {\bibfnamefont {J.}~\bibnamefont {{Davelaar}}}, \bibinfo {author} {\bibfnamefont {M.}~\bibnamefont {{Zaja{\v{c}}ek}}}, \bibinfo {author} {\bibfnamefont {S.}~\bibnamefont {{Britzen}}}, \bibinfo {author} {\bibfnamefont {H.}~\bibnamefont {{Falcke}}}, \bibinfo {author} {\bibfnamefont {E.}~\bibnamefont {{K{\"o}rding}}},\ and\ \bibinfo {author} {\bibfnamefont {E.}~\bibnamefont {{Ros}}},\ }\href {https://doi.org/10.1051/0004-6361/202245506} {\bibfield  {journal} {\bibinfo  {journal} {Astron. Astrophys.}\ }\textbf {\bibinfo {volume} {672}},\ \bibinfo {eid} {L5} (\bibinfo {year} {2023})},\ \Eprint {https://arxiv.org/abs/2303.00603} {arXiv:2303.00603 [astro-ph.HE]} \BibitemShut {NoStop}%
\bibitem [{\citenamefont {{Nawaz}}\ \emph {et~al.}(2016)\citenamefont {{Nawaz}}, \citenamefont {{Bicknell}}, \citenamefont {{Wagner}}, \citenamefont {{Sutherland}},\ and\ \citenamefont {{McNamara}}}]{2016MNRAS.458..802N}%
  \BibitemOpen
  \bibfield  {author} {\bibinfo {author} {\bibfnamefont {M.~A.}\ \bibnamefont {{Nawaz}}}, \bibinfo {author} {\bibfnamefont {G.~V.}\ \bibnamefont {{Bicknell}}}, \bibinfo {author} {\bibfnamefont {A.~Y.}\ \bibnamefont {{Wagner}}}, \bibinfo {author} {\bibfnamefont {R.~S.}\ \bibnamefont {{Sutherland}}},\ and\ \bibinfo {author} {\bibfnamefont {B.~R.}\ \bibnamefont {{McNamara}}},\ }\href {https://doi.org/10.1093/mnras/stw330} {\bibfield  {journal} {\bibinfo  {journal} {Mon. Not. R. Astron. Soc.}\ }\textbf {\bibinfo {volume} {458}},\ \bibinfo {pages} {802} (\bibinfo {year} {2016})},\ \Eprint {https://arxiv.org/abs/1602.02969} {arXiv:1602.02969 [astro-ph.GA]} \BibitemShut {NoStop}%
\bibitem [{\citenamefont {{Falceta-Gon{\c{c}}alves}}\ \emph {et~al.}(2010)\citenamefont {{Falceta-Gon{\c{c}}alves}}, \citenamefont {{Caproni}}, \citenamefont {{Abraham}}, \citenamefont {{Teixeira}},\ and\ \citenamefont {{de Gouveia Dal Pino}}}]{2010ApJ...713L..74F}%
  \BibitemOpen
  \bibfield  {author} {\bibinfo {author} {\bibfnamefont {D.}~\bibnamefont {{Falceta-Gon{\c{c}}alves}}}, \bibinfo {author} {\bibfnamefont {A.}~\bibnamefont {{Caproni}}}, \bibinfo {author} {\bibfnamefont {Z.}~\bibnamefont {{Abraham}}}, \bibinfo {author} {\bibfnamefont {D.~M.}\ \bibnamefont {{Teixeira}}},\ and\ \bibinfo {author} {\bibfnamefont {E.~M.}\ \bibnamefont {{de Gouveia Dal Pino}}},\ }\href {https://doi.org/10.1088/2041-8205/713/1/L74} {\bibfield  {journal} {\bibinfo  {journal} {Astrophys. J. Lett.}\ }\textbf {\bibinfo {volume} {713}},\ \bibinfo {pages} {L74} (\bibinfo {year} {2010})},\ \Eprint {https://arxiv.org/abs/1003.2406} {arXiv:1003.2406 [astro-ph.CO]} \BibitemShut {NoStop}%
\bibitem [{\citenamefont {{Britzen}}\ \emph {et~al.}(2019)\citenamefont {{Britzen}}, \citenamefont {{Fendt}}, \citenamefont {{Zaja{\v{c}}ek}}, \citenamefont {{Jaron}}, \citenamefont {{Pashchenko}}, \citenamefont {{Aller}},\ and\ \citenamefont {{Aller}}}]{2019Galax...7...72B}%
  \BibitemOpen
  \bibfield  {author} {\bibinfo {author} {\bibfnamefont {S.}~\bibnamefont {{Britzen}}}, \bibinfo {author} {\bibfnamefont {C.}~\bibnamefont {{Fendt}}}, \bibinfo {author} {\bibfnamefont {M.}~\bibnamefont {{Zaja{\v{c}}ek}}}, \bibinfo {author} {\bibfnamefont {F.}~\bibnamefont {{Jaron}}}, \bibinfo {author} {\bibfnamefont {I.}~\bibnamefont {{Pashchenko}}}, \bibinfo {author} {\bibfnamefont {M.~F.}\ \bibnamefont {{Aller}}},\ and\ \bibinfo {author} {\bibfnamefont {H.~D.}\ \bibnamefont {{Aller}}},\ }\href {https://doi.org/10.3390/galaxies7030072} {\bibfield  {journal} {\bibinfo  {journal} {Galaxies}\ }\textbf {\bibinfo {volume} {7}},\ \bibinfo {eid} {72} (\bibinfo {year} {2019})}\BibitemShut {NoStop}%
\bibitem [{\citenamefont {{Dunn}}\ \emph {et~al.}(2006)\citenamefont {{Dunn}}, \citenamefont {{Fabian}},\ and\ \citenamefont {{Sanders}}}]{2006MNRAS.366..758D}%
  \BibitemOpen
  \bibfield  {author} {\bibinfo {author} {\bibfnamefont {R.~J.~H.}\ \bibnamefont {{Dunn}}}, \bibinfo {author} {\bibfnamefont {A.~C.}\ \bibnamefont {{Fabian}}},\ and\ \bibinfo {author} {\bibfnamefont {J.~S.}\ \bibnamefont {{Sanders}}},\ }\href {https://doi.org/10.1111/j.1365-2966.2005.09928.x} {\bibfield  {journal} {\bibinfo  {journal} {Mon. Not. R. Astron. Soc.}\ }\textbf {\bibinfo {volume} {366}},\ \bibinfo {pages} {758} (\bibinfo {year} {2006})},\ \Eprint {https://arxiv.org/abs/astro-ph/0512022} {arXiv:astro-ph/0512022 [astro-ph]} \BibitemShut {NoStop}%
\bibitem [{\citenamefont {{Zaninetti}}\ and\ \citenamefont {{van Horn}}(1988)}]{1988A&A...189...45Z}%
  \BibitemOpen
  \bibfield  {author} {\bibinfo {author} {\bibfnamefont {L.}~\bibnamefont {{Zaninetti}}}\ and\ \bibinfo {author} {\bibfnamefont {H.~M.}\ \bibnamefont {{van Horn}}},\ }\href@noop {} {\bibfield  {journal} {\bibinfo  {journal} {\aap}\ }\textbf {\bibinfo {volume} {189}},\ \bibinfo {pages} {45} (\bibinfo {year} {1988})}\BibitemShut {NoStop}%
\bibitem [{\citenamefont {{Makeev}}\ \emph {et~al.}(2023)\citenamefont {{Makeev}}, \citenamefont {{Kovalev}},\ and\ \citenamefont {{Pushkarev}}}]{2023arXiv230107751M}%
  \BibitemOpen
  \bibfield  {author} {\bibinfo {author} {\bibfnamefont {V.~A.}\ \bibnamefont {{Makeev}}}, \bibinfo {author} {\bibfnamefont {Y.~Y.}\ \bibnamefont {{Kovalev}}},\ and\ \bibinfo {author} {\bibfnamefont {A.~B.}\ \bibnamefont {{Pushkarev}}},\ }\href@noop {} {\bibfield  {journal} {\bibinfo  {journal} {{}}\ } (\bibinfo {year} {2023})},\ \Eprint {https://arxiv.org/abs/2301.07751} {arXiv:2301.07751 [astro-ph.HE]} \BibitemShut {NoStop}%
\bibitem [{\citenamefont {{Horton}}\ \emph {et~al.}(2020)\citenamefont {{Horton}}, \citenamefont {{Krause}},\ and\ \citenamefont {{Hardcastle}}}]{2020MNRAS.499.5765H}%
  \BibitemOpen
  \bibfield  {author} {\bibinfo {author} {\bibfnamefont {M.~A.}\ \bibnamefont {{Horton}}}, \bibinfo {author} {\bibfnamefont {M.~G.~H.}\ \bibnamefont {{Krause}}},\ and\ \bibinfo {author} {\bibfnamefont {M.~J.}\ \bibnamefont {{Hardcastle}}},\ }\href {https://doi.org/10.1093/mnras/staa3020} {\bibfield  {journal} {\bibinfo  {journal} {Mon. Not. R. Astron. Soc.}\ }\textbf {\bibinfo {volume} {499}},\ \bibinfo {pages} {5765} (\bibinfo {year} {2020})},\ \Eprint {https://arxiv.org/abs/2010.00480} {arXiv:2010.00480 [astro-ph.GA]} \BibitemShut {NoStop}%
\bibitem [{\citenamefont {{Falcke}}\ and\ \citenamefont {{Biermann}}(1996)}]{1996A&A...308..321F}%
  \BibitemOpen
  \bibfield  {author} {\bibinfo {author} {\bibfnamefont {H.}~\bibnamefont {{Falcke}}}\ and\ \bibinfo {author} {\bibfnamefont {P.~L.}\ \bibnamefont {{Biermann}}},\ }\href {https://doi.org/10.48550/arXiv.astro-ph/9506138} {\bibfield  {journal} {\bibinfo  {journal} {Astron. Astrophys.}\ }\textbf {\bibinfo {volume} {308}},\ \bibinfo {pages} {321} (\bibinfo {year} {1996})},\ \Eprint {https://arxiv.org/abs/astro-ph/9506138} {arXiv:astro-ph/9506138 [astro-ph]} \BibitemShut {NoStop}%
\bibitem [{\citenamefont {{Liska}}\ \emph {et~al.}(2018)\citenamefont {{Liska}}, \citenamefont {{Hesp}}, \citenamefont {{Tchekhovskoy}}, \citenamefont {{Ingram}}, \citenamefont {{van der Klis}},\ and\ \citenamefont {{Markoff}}}]{2018MNRAS.474L..81L}%
  \BibitemOpen
  \bibfield  {author} {\bibinfo {author} {\bibfnamefont {M.}~\bibnamefont {{Liska}}}, \bibinfo {author} {\bibfnamefont {C.}~\bibnamefont {{Hesp}}}, \bibinfo {author} {\bibfnamefont {A.}~\bibnamefont {{Tchekhovskoy}}}, \bibinfo {author} {\bibfnamefont {A.}~\bibnamefont {{Ingram}}}, \bibinfo {author} {\bibfnamefont {M.}~\bibnamefont {{van der Klis}}},\ and\ \bibinfo {author} {\bibfnamefont {S.}~\bibnamefont {{Markoff}}},\ }\href {https://doi.org/10.1093/mnrasl/slx174} {\bibfield  {journal} {\bibinfo  {journal} {Mon. Not. R. Astron. Soc.}\ }\textbf {\bibinfo {volume} {474}},\ \bibinfo {pages} {L81} (\bibinfo {year} {2018})},\ \Eprint {https://arxiv.org/abs/1707.06619} {arXiv:1707.06619 [astro-ph.HE]} \BibitemShut {NoStop}%
\bibitem [{\citenamefont {{Blackman}}\ and\ \citenamefont {{Lebedev}}(2022)}]{2022NewAR..9501661B}%
  \BibitemOpen
  \bibfield  {author} {\bibinfo {author} {\bibfnamefont {E.~G.}\ \bibnamefont {{Blackman}}}\ and\ \bibinfo {author} {\bibfnamefont {S.~V.}\ \bibnamefont {{Lebedev}}},\ }\href {https://doi.org/10.1016/j.newar.2022.101661} {\bibfield  {journal} {\bibinfo  {journal} {New Astron. Rev.}\ }\textbf {\bibinfo {volume} {95}},\ \bibinfo {eid} {101661} (\bibinfo {year} {2022})},\ \Eprint {https://arxiv.org/abs/2009.08057} {arXiv:2009.08057 [astro-ph.HE]} \BibitemShut {NoStop}%
\bibitem [{\citenamefont {{Amaro-Seoane}}\ \emph {et~al.}(2017)\citenamefont {{Amaro-Seoane}} \emph {et~al.}}]{2017arXiv170200786A}%
  \BibitemOpen
  \bibfield  {author} {\bibinfo {author} {\bibfnamefont {P.}~\bibnamefont {{Amaro-Seoane}}} \emph {et~al.},\ }\href@noop {} {\bibfield  {journal} {\bibinfo  {journal} {{}}\ } (\bibinfo {year} {2017})},\ \Eprint {https://arxiv.org/abs/1702.00786} {arXiv:1702.00786 [astro-ph.IM]} \BibitemShut {NoStop}%
\bibitem [{\citenamefont {{Colpi}}\ \emph {et~al.}(2024)\citenamefont {{Colpi}}, \citenamefont {{Danzmann}}, \citenamefont {{Hewitson}} \emph {et~al.}}]{2024arXiv240207571C}%
  \BibitemOpen
  \bibfield  {author} {\bibinfo {author} {\bibfnamefont {M.}~\bibnamefont {{Colpi}}}, \bibinfo {author} {\bibfnamefont {K.}~\bibnamefont {{Danzmann}}}, \bibinfo {author} {\bibfnamefont {M.}~\bibnamefont {{Hewitson}}}, \emph {et~al.},\ }\href@noop {} {\bibfield  {journal} {\bibinfo  {journal} {{}}\ } (\bibinfo {year} {2024})},\ \Eprint {https://arxiv.org/abs/2402.07571} {arXiv:2402.07571 [astro-ph.CO]} \BibitemShut {NoStop}%
\bibitem [{\citenamefont {{Barausse}}(2012)}]{2012MNRAS.423.2533B}%
  \BibitemOpen
  \bibfield  {author} {\bibinfo {author} {\bibfnamefont {E.}~\bibnamefont {{Barausse}}},\ }\href {https://doi.org/10.1111/j.1365-2966.2012.21057.x} {\bibfield  {journal} {\bibinfo  {journal} {Mon. Not. R. Astron. Soc.}\ }\textbf {\bibinfo {volume} {423}},\ \bibinfo {pages} {2533} (\bibinfo {year} {2012})},\ \Eprint {https://arxiv.org/abs/1201.5888} {arXiv:1201.5888 [astro-ph.CO]} \BibitemShut {NoStop}%
\bibitem [{\citenamefont {{Mayer}}(2013)}]{2013CQGra..30x4008M}%
  \BibitemOpen
  \bibfield  {author} {\bibinfo {author} {\bibfnamefont {L.}~\bibnamefont {{Mayer}}},\ }\href {https://doi.org/10.1088/0264-9381/30/24/244008} {\bibfield  {journal} {\bibinfo  {journal} {\cqg}\ }\textbf {\bibinfo {volume} {30}},\ \bibinfo {eid} {244008} (\bibinfo {year} {2013})},\ \Eprint {https://arxiv.org/abs/1308.0431} {arXiv:1308.0431 [astro-ph.CO]} \BibitemShut {NoStop}%
\bibitem [{\citenamefont {{Li}}\ \emph {et~al.}(2022)\citenamefont {{Li}}, \citenamefont {{Bogdanovi{\'c}}}, \citenamefont {{Ballantyne}},\ and\ \citenamefont {{Bonetti}}}]{2022ApJ...933..104L}%
  \BibitemOpen
  \bibfield  {author} {\bibinfo {author} {\bibfnamefont {K.}~\bibnamefont {{Li}}}, \bibinfo {author} {\bibfnamefont {T.}~\bibnamefont {{Bogdanovi{\'c}}}}, \bibinfo {author} {\bibfnamefont {D.~R.}\ \bibnamefont {{Ballantyne}}},\ and\ \bibinfo {author} {\bibfnamefont {M.}~\bibnamefont {{Bonetti}}},\ }\href {https://doi.org/10.3847/1538-4357/ac74b5} {\bibfield  {journal} {\bibinfo  {journal} {Astrophys. J.}\ }\textbf {\bibinfo {volume} {933}},\ \bibinfo {eid} {104} (\bibinfo {year} {2022})},\ \Eprint {https://arxiv.org/abs/2201.11088} {arXiv:2201.11088 [astro-ph.GA]} \BibitemShut {NoStop}%
\bibitem [{\citenamefont {{Colpi}}(2014)}]{2014SSRv..183..189C}%
  \BibitemOpen
  \bibfield  {author} {\bibinfo {author} {\bibfnamefont {M.}~\bibnamefont {{Colpi}}},\ }\href {https://doi.org/10.1007/s11214-014-0067-1} {\bibfield  {journal} {\bibinfo  {journal} {Space Sci. Rev.}\ }\textbf {\bibinfo {volume} {183}},\ \bibinfo {pages} {189} (\bibinfo {year} {2014})},\ \Eprint {https://arxiv.org/abs/1407.3102} {arXiv:1407.3102 [astro-ph.GA]} \BibitemShut {NoStop}%
\bibitem [{\citenamefont {{Volonteri}}\ \emph {et~al.}(2021)\citenamefont {{Volonteri}}, \citenamefont {{Habouzit}},\ and\ \citenamefont {{Colpi}}}]{2021NatRP...3..732V}%
  \BibitemOpen
  \bibfield  {author} {\bibinfo {author} {\bibfnamefont {M.}~\bibnamefont {{Volonteri}}}, \bibinfo {author} {\bibfnamefont {M.}~\bibnamefont {{Habouzit}}},\ and\ \bibinfo {author} {\bibfnamefont {M.}~\bibnamefont {{Colpi}}},\ }\href {https://doi.org/10.1038/s42254-021-00364-9} {\bibfield  {journal} {\bibinfo  {journal} {\natrp}\ }\textbf {\bibinfo {volume} {3}},\ \bibinfo {pages} {732} (\bibinfo {year} {2021})},\ \Eprint {https://arxiv.org/abs/2110.10175} {arXiv:2110.10175 [astro-ph.GA]} \BibitemShut {NoStop}%
\bibitem [{\citenamefont {{Bourne}}\ \emph {et~al.}(2023)\citenamefont {{Bourne}}, \citenamefont {{Fiacconi}}, \citenamefont {{Sijacki}}, \citenamefont {{Piotrowska}},\ and\ \citenamefont {{Koudmani}}}]{2023arXiv231117144B}%
  \BibitemOpen
  \bibfield  {author} {\bibinfo {author} {\bibfnamefont {M.~A.}\ \bibnamefont {{Bourne}}}, \bibinfo {author} {\bibfnamefont {D.}~\bibnamefont {{Fiacconi}}}, \bibinfo {author} {\bibfnamefont {D.}~\bibnamefont {{Sijacki}}}, \bibinfo {author} {\bibfnamefont {J.~M.}\ \bibnamefont {{Piotrowska}}},\ and\ \bibinfo {author} {\bibfnamefont {S.}~\bibnamefont {{Koudmani}}},\ }\href@noop {} {\bibfield  {journal} {\bibinfo  {journal} {{}}\ } (\bibinfo {year} {2023})},\ \Eprint {https://arxiv.org/abs/2311.17144} {arXiv:2311.17144 [astro-ph.HE]} \BibitemShut {NoStop}%
\bibitem [{\citenamefont {{Begelman}}\ \emph {et~al.}(1980)\citenamefont {{Begelman}}, \citenamefont {{Blandford}},\ and\ \citenamefont {{Rees}}}]{1980Natur.287..307B}%
  \BibitemOpen
  \bibfield  {author} {\bibinfo {author} {\bibfnamefont {M.~C.}\ \bibnamefont {{Begelman}}}, \bibinfo {author} {\bibfnamefont {R.~D.}\ \bibnamefont {{Blandford}}},\ and\ \bibinfo {author} {\bibfnamefont {M.~J.}\ \bibnamefont {{Rees}}},\ }\href {https://doi.org/10.1038/287307a0} {\bibfield  {journal} {\bibinfo  {journal} {Nature}\ }\textbf {\bibinfo {volume} {287}},\ \bibinfo {pages} {307} (\bibinfo {year} {1980})}\BibitemShut {NoStop}%
\bibitem [{\citenamefont {{Mangiagli}}\ \emph {et~al.}(2022)\citenamefont {{Mangiagli}}, \citenamefont {{Caprini}}, \citenamefont {{Volonteri}}, \citenamefont {{Marsat}}, \citenamefont {{Vergani}}, \citenamefont {{Tamanini}},\ and\ \citenamefont {{Inchausp{\'e}}}}]{2022PhRvD.106j3017M}%
  \BibitemOpen
  \bibfield  {author} {\bibinfo {author} {\bibfnamefont {A.}~\bibnamefont {{Mangiagli}}}, \bibinfo {author} {\bibfnamefont {C.}~\bibnamefont {{Caprini}}}, \bibinfo {author} {\bibfnamefont {M.}~\bibnamefont {{Volonteri}}}, \bibinfo {author} {\bibfnamefont {S.}~\bibnamefont {{Marsat}}}, \bibinfo {author} {\bibfnamefont {S.}~\bibnamefont {{Vergani}}}, \bibinfo {author} {\bibfnamefont {N.}~\bibnamefont {{Tamanini}}},\ and\ \bibinfo {author} {\bibfnamefont {H.}~\bibnamefont {{Inchausp{\'e}}}},\ }\href {https://doi.org/10.1103/PhysRevD.106.103017} {\bibfield  {journal} {\bibinfo  {journal} {Phys. Rev. D}\ }\textbf {\bibinfo {volume} {106}},\ \bibinfo {eid} {103017} (\bibinfo {year} {2022})},\ \Eprint {https://arxiv.org/abs/2207.10678} {arXiv:2207.10678 [astro-ph.HE]} \BibitemShut {NoStop}%
\bibitem [{\citenamefont {{Cattorini}}\ and\ \citenamefont {{Giacomazzo}}(2024)}]{2024APh...15402892C}%
  \BibitemOpen
  \bibfield  {author} {\bibinfo {author} {\bibfnamefont {F.}~\bibnamefont {{Cattorini}}}\ and\ \bibinfo {author} {\bibfnamefont {B.}~\bibnamefont {{Giacomazzo}}},\ }\href {https://doi.org/10.1016/j.astropartphys.2023.102892} {\bibfield  {journal} {\bibinfo  {journal} {\app}\ }\textbf {\bibinfo {volume} {154}},\ \bibinfo {eid} {102892} (\bibinfo {year} {2024})},\ \Eprint {https://arxiv.org/abs/2401.02521} {arXiv:2401.02521 [astro-ph.HE]} \BibitemShut {NoStop}%
\bibitem [{\citenamefont {{Gerosa}}\ \emph {et~al.}(2020)\citenamefont {{Gerosa}}, \citenamefont {{Rosotti}},\ and\ \citenamefont {{Barbieri}}}]{2020MNRAS.496.3060G}%
  \BibitemOpen
  \bibfield  {author} {\bibinfo {author} {\bibfnamefont {D.}~\bibnamefont {{Gerosa}}}, \bibinfo {author} {\bibfnamefont {G.}~\bibnamefont {{Rosotti}}},\ and\ \bibinfo {author} {\bibfnamefont {R.}~\bibnamefont {{Barbieri}}},\ }\href {https://doi.org/10.1093/mnras/staa1693} {\bibfield  {journal} {\bibinfo  {journal} {Mon. Not. R. Astron. Soc.}\ }\textbf {\bibinfo {volume} {496}},\ \bibinfo {pages} {3060} (\bibinfo {year} {2020})},\ \Eprint {https://arxiv.org/abs/2004.02894} {arXiv:2004.02894 [astro-ph.GA]} \BibitemShut {NoStop}%
\bibitem [{\citenamefont {{Bowen}}\ \emph {et~al.}(2018)\citenamefont {{Bowen}}, \citenamefont {{Mewes}}, \citenamefont {{Campanelli}}, \citenamefont {{Noble}}, \citenamefont {{Krolik}},\ and\ \citenamefont {{Zilh{\~a}o}}}]{2018ApJ...853L..17B}%
  \BibitemOpen
  \bibfield  {author} {\bibinfo {author} {\bibfnamefont {D.~B.}\ \bibnamefont {{Bowen}}}, \bibinfo {author} {\bibfnamefont {V.}~\bibnamefont {{Mewes}}}, \bibinfo {author} {\bibfnamefont {M.}~\bibnamefont {{Campanelli}}}, \bibinfo {author} {\bibfnamefont {S.~C.}\ \bibnamefont {{Noble}}}, \bibinfo {author} {\bibfnamefont {J.~H.}\ \bibnamefont {{Krolik}}},\ and\ \bibinfo {author} {\bibfnamefont {M.}~\bibnamefont {{Zilh{\~a}o}}},\ }\href {https://doi.org/10.3847/2041-8213/aaa756} {\bibfield  {journal} {\bibinfo  {journal} {Astrophys. J. Lett.}\ }\textbf {\bibinfo {volume} {853}},\ \bibinfo {eid} {L17} (\bibinfo {year} {2018})},\ \Eprint {https://arxiv.org/abs/1712.05451} {arXiv:1712.05451 [astro-ph.HE]} \BibitemShut {NoStop}%
\bibitem [{\citenamefont {{Papaloizou}}\ and\ \citenamefont {{Pringle}}(1983)}]{1983MNRAS.202.1181P}%
  \BibitemOpen
  \bibfield  {author} {\bibinfo {author} {\bibfnamefont {J.~C.~B.}\ \bibnamefont {{Papaloizou}}}\ and\ \bibinfo {author} {\bibfnamefont {J.~E.}\ \bibnamefont {{Pringle}}},\ }\href {https://doi.org/10.1093/mnras/202.4.1181} {\bibfield  {journal} {\bibinfo  {journal} {Mon. Not. R. Astron. Soc.}\ }\textbf {\bibinfo {volume} {202}},\ \bibinfo {pages} {1181} (\bibinfo {year} {1983})}\BibitemShut {NoStop}%
\bibitem [{\citenamefont {{Scheuer}}\ and\ \citenamefont {{Feiler}}(1996)}]{1996MNRAS.282..291S}%
  \BibitemOpen
  \bibfield  {author} {\bibinfo {author} {\bibfnamefont {P.~A.~G.}\ \bibnamefont {{Scheuer}}}\ and\ \bibinfo {author} {\bibfnamefont {R.}~\bibnamefont {{Feiler}}},\ }\href {https://doi.org/10.1093/mnras/282.1.291} {\bibfield  {journal} {\bibinfo  {journal} {Mon. Not. R. Astron. Soc.}\ }\textbf {\bibinfo {volume} {282}},\ \bibinfo {pages} {291} (\bibinfo {year} {1996})}\BibitemShut {NoStop}%
\bibitem [{\citenamefont {{Tremaine}}\ and\ \citenamefont {{Davis}}(2014)}]{2014MNRAS.441.1408T}%
  \BibitemOpen
  \bibfield  {author} {\bibinfo {author} {\bibfnamefont {S.}~\bibnamefont {{Tremaine}}}\ and\ \bibinfo {author} {\bibfnamefont {S.~W.}\ \bibnamefont {{Davis}}},\ }\href {https://doi.org/10.1093/mnras/stu663} {\bibfield  {journal} {\bibinfo  {journal} {Mon. Not. R. Astron. Soc.}\ }\textbf {\bibinfo {volume} {441}},\ \bibinfo {pages} {1408} (\bibinfo {year} {2014})},\ \Eprint {https://arxiv.org/abs/1308.1964} {arXiv:1308.1964 [astro-ph.HE]} \BibitemShut {NoStop}%
\bibitem [{\citenamefont {{Bardeen}}\ and\ \citenamefont {{Petterson}}(1975)}]{1975ApJ...195L..65B}%
  \BibitemOpen
  \bibfield  {author} {\bibinfo {author} {\bibfnamefont {J.~M.}\ \bibnamefont {{Bardeen}}}\ and\ \bibinfo {author} {\bibfnamefont {J.~A.}\ \bibnamefont {{Petterson}}},\ }\href {https://doi.org/10.1086/181711} {\bibfield  {journal} {\bibinfo  {journal} {Astrophys. J. Lett.}\ }\textbf {\bibinfo {volume} {195}},\ \bibinfo {pages} {L65} (\bibinfo {year} {1975})}\BibitemShut {NoStop}%
\bibitem [{\citenamefont {{Rees}}(1978)}]{1978Natur.275..516R}%
  \BibitemOpen
  \bibfield  {author} {\bibinfo {author} {\bibfnamefont {M.~J.}\ \bibnamefont {{Rees}}},\ }\href {https://doi.org/10.1038/275516a0} {\bibfield  {journal} {\bibinfo  {journal} {Nature}\ }\textbf {\bibinfo {volume} {275}},\ \bibinfo {pages} {516} (\bibinfo {year} {1978})}\BibitemShut {NoStop}%
\bibitem [{\citenamefont {{Kumar}}\ and\ \citenamefont {{Pringle}}(1985)}]{1985MNRAS.213..435K}%
  \BibitemOpen
  \bibfield  {author} {\bibinfo {author} {\bibfnamefont {S.}~\bibnamefont {{Kumar}}}\ and\ \bibinfo {author} {\bibfnamefont {J.~E.}\ \bibnamefont {{Pringle}}},\ }\href {https://doi.org/10.1093/mnras/213.3.435} {\bibfield  {journal} {\bibinfo  {journal} {Mon. Not. R. Astron. Soc.}\ }\textbf {\bibinfo {volume} {213}},\ \bibinfo {pages} {435} (\bibinfo {year} {1985})}\BibitemShut {NoStop}%
\bibitem [{\citenamefont {{Steinle}}\ and\ \citenamefont {{Gerosa}}(2023)}]{2023MNRAS.519.5031S}%
  \BibitemOpen
  \bibfield  {author} {\bibinfo {author} {\bibfnamefont {N.}~\bibnamefont {{Steinle}}}\ and\ \bibinfo {author} {\bibfnamefont {D.}~\bibnamefont {{Gerosa}}},\ }\href {https://doi.org/10.1093/mnras/stac3821} {\bibfield  {journal} {\bibinfo  {journal} {Mon. Not. R. Astron. Soc.}\ }\textbf {\bibinfo {volume} {519}},\ \bibinfo {pages} {5031} (\bibinfo {year} {2023})},\ \Eprint {https://arxiv.org/abs/2211.00044} {arXiv:2211.00044 [astro-ph.HE]} \BibitemShut {NoStop}%
\bibitem [{\citenamefont {{Ivanov}}\ \emph {et~al.}(1999)\citenamefont {{Ivanov}}, \citenamefont {{Papaloizou}},\ and\ \citenamefont {{Polnarev}}}]{1999MNRAS.307...79I}%
  \BibitemOpen
  \bibfield  {author} {\bibinfo {author} {\bibfnamefont {P.~B.}\ \bibnamefont {{Ivanov}}}, \bibinfo {author} {\bibfnamefont {J.~C.~B.}\ \bibnamefont {{Papaloizou}}},\ and\ \bibinfo {author} {\bibfnamefont {A.~G.}\ \bibnamefont {{Polnarev}}},\ }\href {https://doi.org/10.1046/j.1365-8711.1999.02623.x} {\bibfield  {journal} {\bibinfo  {journal} {Mon. Not. R. Astron. Soc.}\ }\textbf {\bibinfo {volume} {307}},\ \bibinfo {pages} {79} (\bibinfo {year} {1999})},\ \Eprint {https://arxiv.org/abs/astro-ph/9812198} {arXiv:astro-ph/9812198 [astro-ph]} \BibitemShut {NoStop}%
\bibitem [{\citenamefont {{Ogilvie}}\ and\ \citenamefont {{Latter}}(2013)}]{2013MNRAS.433.2403O}%
  \BibitemOpen
  \bibfield  {author} {\bibinfo {author} {\bibfnamefont {G.~I.}\ \bibnamefont {{Ogilvie}}}\ and\ \bibinfo {author} {\bibfnamefont {H.~N.}\ \bibnamefont {{Latter}}},\ }\href {https://doi.org/10.1093/mnras/stt916} {\bibfield  {journal} {\bibinfo  {journal} {Mon. Not. R. Astron. Soc.}\ }\textbf {\bibinfo {volume} {433}},\ \bibinfo {pages} {2403} (\bibinfo {year} {2013})},\ \Eprint {https://arxiv.org/abs/1303.0263} {arXiv:1303.0263 [astro-ph.SR]} \BibitemShut {NoStop}%
\bibitem [{\citenamefont {{Shakura}}\ and\ \citenamefont {{Sunyaev}}(1973)}]{1973A&A....24..337S}%
  \BibitemOpen
  \bibfield  {author} {\bibinfo {author} {\bibfnamefont {N.~I.}\ \bibnamefont {{Shakura}}}\ and\ \bibinfo {author} {\bibfnamefont {R.~A.}\ \bibnamefont {{Sunyaev}}},\ }\href@noop {} {\bibfield  {journal} {\bibinfo  {journal} {Astron. Astrophys.}\ }\textbf {\bibinfo {volume} {24}},\ \bibinfo {pages} {337} (\bibinfo {year} {1973})}\BibitemShut {NoStop}%
\bibitem [{\citenamefont {{Gerosa}}\ \emph {et~al.}(2015)\citenamefont {{Gerosa}}, \citenamefont {{Veronesi}}, \citenamefont {{Lodato}},\ and\ \citenamefont {{Rosotti}}}]{2015MNRAS.451.3941G}%
  \BibitemOpen
  \bibfield  {author} {\bibinfo {author} {\bibfnamefont {D.}~\bibnamefont {{Gerosa}}}, \bibinfo {author} {\bibfnamefont {B.}~\bibnamefont {{Veronesi}}}, \bibinfo {author} {\bibfnamefont {G.}~\bibnamefont {{Lodato}}},\ and\ \bibinfo {author} {\bibfnamefont {G.}~\bibnamefont {{Rosotti}}},\ }\href {https://doi.org/10.1093/mnras/stv1214} {\bibfield  {journal} {\bibinfo  {journal} {Mon. Not. R. Astron. Soc.}\ }\textbf {\bibinfo {volume} {451}},\ \bibinfo {pages} {3941} (\bibinfo {year} {2015})},\ \Eprint {https://arxiv.org/abs/1503.06807} {arXiv:1503.06807 [astro-ph.GA]} \BibitemShut {NoStop}%
\bibitem [{\citenamefont {{Martin}}\ \emph {et~al.}(2007)\citenamefont {{Martin}}, \citenamefont {{Pringle}},\ and\ \citenamefont {{Tout}}}]{2007MNRAS.381.1617M}%
  \BibitemOpen
  \bibfield  {author} {\bibinfo {author} {\bibfnamefont {R.~G.}\ \bibnamefont {{Martin}}}, \bibinfo {author} {\bibfnamefont {J.~E.}\ \bibnamefont {{Pringle}}},\ and\ \bibinfo {author} {\bibfnamefont {C.~A.}\ \bibnamefont {{Tout}}},\ }\href {https://doi.org/10.1111/j.1365-2966.2007.12349.x} {\bibfield  {journal} {\bibinfo  {journal} {Mon. Not. R. Astron. Soc.}\ }\textbf {\bibinfo {volume} {381}},\ \bibinfo {pages} {1617} (\bibinfo {year} {2007})},\ \Eprint {https://arxiv.org/abs/0708.2034} {arXiv:0708.2034 [astro-ph]} \BibitemShut {NoStop}%
\bibitem [{\citenamefont {{Martin}}\ \emph {et~al.}(2009)\citenamefont {{Martin}}, \citenamefont {{Pringle}},\ and\ \citenamefont {{Tout}}}]{2009MNRAS.400..383M}%
  \BibitemOpen
  \bibfield  {author} {\bibinfo {author} {\bibfnamefont {R.~G.}\ \bibnamefont {{Martin}}}, \bibinfo {author} {\bibfnamefont {J.~E.}\ \bibnamefont {{Pringle}}},\ and\ \bibinfo {author} {\bibfnamefont {C.~A.}\ \bibnamefont {{Tout}}},\ }\href {https://doi.org/10.1111/j.1365-2966.2009.15465.x} {\bibfield  {journal} {\bibinfo  {journal} {Mon. Not. R. Astron. Soc.}\ }\textbf {\bibinfo {volume} {400}},\ \bibinfo {pages} {383} (\bibinfo {year} {2009})},\ \Eprint {https://arxiv.org/abs/0907.5142} {arXiv:0907.5142 [astro-ph.HE]} \BibitemShut {NoStop}%
\bibitem [{\citenamefont {{Ogilvie}}(1999)}]{1999MNRAS.304..557O}%
  \BibitemOpen
  \bibfield  {author} {\bibinfo {author} {\bibfnamefont {G.~I.}\ \bibnamefont {{Ogilvie}}},\ }\href {https://doi.org/10.1046/j.1365-8711.1999.02340.x} {\bibfield  {journal} {\bibinfo  {journal} {Mon. Not. R. Astron. Soc.}\ }\textbf {\bibinfo {volume} {304}},\ \bibinfo {pages} {557} (\bibinfo {year} {1999})},\ \Eprint {https://arxiv.org/abs/astro-ph/9812073} {arXiv:astro-ph/9812073 [astro-ph]} \BibitemShut {NoStop}%
\bibitem [{\citenamefont {{Nelson}}\ and\ \citenamefont {{Papaloizou}}(2000)}]{2000MNRAS.315..570N}%
  \BibitemOpen
  \bibfield  {author} {\bibinfo {author} {\bibfnamefont {R.~P.}\ \bibnamefont {{Nelson}}}\ and\ \bibinfo {author} {\bibfnamefont {J.~C.~B.}\ \bibnamefont {{Papaloizou}}},\ }\href {https://doi.org/10.1046/j.1365-8711.2000.03478.x} {\bibfield  {journal} {\bibinfo  {journal} {Mon. Not. R. Astron. Soc.}\ }\textbf {\bibinfo {volume} {315}},\ \bibinfo {pages} {570} (\bibinfo {year} {2000})},\ \Eprint {https://arxiv.org/abs/astro-ph/0001439} {arXiv:astro-ph/0001439 [astro-ph]} \BibitemShut {NoStop}%
\bibitem [{\citenamefont {{Nealon}}\ \emph {et~al.}(2022)\citenamefont {{Nealon}}, \citenamefont {{Ragusa}}, \citenamefont {{Gerosa}}, \citenamefont {{Rosotti}},\ and\ \citenamefont {{Barbieri}}}]{2022MNRAS.509.5608N}%
  \BibitemOpen
  \bibfield  {author} {\bibinfo {author} {\bibfnamefont {R.}~\bibnamefont {{Nealon}}}, \bibinfo {author} {\bibfnamefont {E.}~\bibnamefont {{Ragusa}}}, \bibinfo {author} {\bibfnamefont {D.}~\bibnamefont {{Gerosa}}}, \bibinfo {author} {\bibfnamefont {G.}~\bibnamefont {{Rosotti}}},\ and\ \bibinfo {author} {\bibfnamefont {R.}~\bibnamefont {{Barbieri}}},\ }\href {https://doi.org/10.1093/mnras/stab3328} {\bibfield  {journal} {\bibinfo  {journal} {Mon. Not. R. Astron. Soc.}\ }\textbf {\bibinfo {volume} {509}},\ \bibinfo {pages} {5608} (\bibinfo {year} {2022})},\ \Eprint {https://arxiv.org/abs/2111.08065} {arXiv:2111.08065 [astro-ph.HE]} \BibitemShut {NoStop}%
\bibitem [{\citenamefont {{Roos}}(1988)}]{1988ApJ...334...95R}%
  \BibitemOpen
  \bibfield  {author} {\bibinfo {author} {\bibfnamefont {N.}~\bibnamefont {{Roos}}},\ }\href {https://doi.org/10.1086/166820} {\bibfield  {journal} {\bibinfo  {journal} {Astrophys. J.}\ }\textbf {\bibinfo {volume} {334}},\ \bibinfo {pages} {95} (\bibinfo {year} {1988})}\BibitemShut {NoStop}%
\bibitem [{\citenamefont {{Liu}}\ and\ \citenamefont {{Chen}}(2007)}]{2007ApJ...671.1272L}%
  \BibitemOpen
  \bibfield  {author} {\bibinfo {author} {\bibfnamefont {F.~K.}\ \bibnamefont {{Liu}}}\ and\ \bibinfo {author} {\bibfnamefont {X.}~\bibnamefont {{Chen}}},\ }\href {https://doi.org/10.1086/522910} {\bibfield  {journal} {\bibinfo  {journal} {Astrophys. J.}\ }\textbf {\bibinfo {volume} {671}},\ \bibinfo {pages} {1272} (\bibinfo {year} {2007})},\ \Eprint {https://arxiv.org/abs/0705.1077} {arXiv:0705.1077 [astro-ph]} \BibitemShut {NoStop}%
\bibitem [{\citenamefont {{Natarajan}}\ and\ \citenamefont {{Pringle}}(1998)}]{1998ApJ...506L..97N}%
  \BibitemOpen
  \bibfield  {author} {\bibinfo {author} {\bibfnamefont {P.}~\bibnamefont {{Natarajan}}}\ and\ \bibinfo {author} {\bibfnamefont {J.~E.}\ \bibnamefont {{Pringle}}},\ }\href {https://doi.org/10.1086/311658} {\bibfield  {journal} {\bibinfo  {journal} {Astrophys. J. Lett.}\ }\textbf {\bibinfo {volume} {506}},\ \bibinfo {pages} {L97} (\bibinfo {year} {1998})},\ \Eprint {https://arxiv.org/abs/astro-ph/9808187} {arXiv:astro-ph/9808187 [astro-ph]} \BibitemShut {NoStop}%
\bibitem [{\citenamefont {{Natarajan}}\ and\ \citenamefont {{Armitage}}(1999)}]{1999MNRAS.309..961N}%
  \BibitemOpen
  \bibfield  {author} {\bibinfo {author} {\bibfnamefont {P.}~\bibnamefont {{Natarajan}}}\ and\ \bibinfo {author} {\bibfnamefont {P.~J.}\ \bibnamefont {{Armitage}}},\ }\href {https://doi.org/10.1046/j.1365-8711.1999.02917.x} {\bibfield  {journal} {\bibinfo  {journal} {Mon. Not. R. Astron. Soc.}\ }\textbf {\bibinfo {volume} {309}},\ \bibinfo {pages} {961} (\bibinfo {year} {1999})},\ \Eprint {https://arxiv.org/abs/astro-ph/9812001} {arXiv:astro-ph/9812001 [astro-ph]} \BibitemShut {NoStop}%
\bibitem [{\citenamefont {{Lodato}}\ and\ \citenamefont {{Gerosa}}(2013)}]{2013MNRAS.429L..30L}%
  \BibitemOpen
  \bibfield  {author} {\bibinfo {author} {\bibfnamefont {G.}~\bibnamefont {{Lodato}}}\ and\ \bibinfo {author} {\bibfnamefont {D.}~\bibnamefont {{Gerosa}}},\ }\href {https://doi.org/10.1093/mnrasl/sls018} {\bibfield  {journal} {\bibinfo  {journal} {Mon. Not. R. Astron. Soc.}\ }\textbf {\bibinfo {volume} {429}},\ \bibinfo {pages} {L30} (\bibinfo {year} {2013})},\ \Eprint {https://arxiv.org/abs/1211.0284} {arXiv:1211.0284 [astro-ph.CO]} \BibitemShut {NoStop}%
\bibitem [{\citenamefont {{Katz}}(1997)}]{1997ApJ...478..527K}%
  \BibitemOpen
  \bibfield  {author} {\bibinfo {author} {\bibfnamefont {J.~I.}\ \bibnamefont {{Katz}}},\ }\href {https://doi.org/10.1086/303811} {\bibfield  {journal} {\bibinfo  {journal} {Astrophys. J.}\ }\textbf {\bibinfo {volume} {478}},\ \bibinfo {pages} {527} (\bibinfo {year} {1997})}\BibitemShut {NoStop}%
\bibitem [{\citenamefont {{Farris}}\ \emph {et~al.}(2012)\citenamefont {{Farris}}, \citenamefont {{Gold}}, \citenamefont {{Paschalidis}}, \citenamefont {{Etienne}},\ and\ \citenamefont {{Shapiro}}}]{2012PhRvL.109v1102F}%
  \BibitemOpen
  \bibfield  {author} {\bibinfo {author} {\bibfnamefont {B.~D.}\ \bibnamefont {{Farris}}}, \bibinfo {author} {\bibfnamefont {R.}~\bibnamefont {{Gold}}}, \bibinfo {author} {\bibfnamefont {V.}~\bibnamefont {{Paschalidis}}}, \bibinfo {author} {\bibfnamefont {Z.~B.}\ \bibnamefont {{Etienne}}},\ and\ \bibinfo {author} {\bibfnamefont {S.~L.}\ \bibnamefont {{Shapiro}}},\ }\href {https://doi.org/10.1103/PhysRevLett.109.221102} {\bibfield  {journal} {\bibinfo  {journal} {Phys. Rev. Lett.}\ }\textbf {\bibinfo {volume} {109}},\ \bibinfo {eid} {221102} (\bibinfo {year} {2012})},\ \Eprint {https://arxiv.org/abs/1207.3354} {arXiv:1207.3354 [astro-ph.HE]} \BibitemShut {NoStop}%
\bibitem [{\citenamefont {{Peters}}\ and\ \citenamefont {{Mathews}}(1963)}]{1963PhRv..131..435P}%
  \BibitemOpen
  \bibfield  {author} {\bibinfo {author} {\bibfnamefont {P.~C.}\ \bibnamefont {{Peters}}}\ and\ \bibinfo {author} {\bibfnamefont {J.}~\bibnamefont {{Mathews}}},\ }\href {https://doi.org/10.1103/PhysRev.131.435} {\bibfield  {journal} {\bibinfo  {journal} {\pr}\ }\textbf {\bibinfo {volume} {131}},\ \bibinfo {pages} {435} (\bibinfo {year} {1963})}\BibitemShut {NoStop}%
\bibitem [{\citenamefont {{Aghanim}}\ \emph {et~al.}(2020)\citenamefont {{Aghanim}} \emph {et~al.}}]{2020A&A...641A...6P}%
  \BibitemOpen
  \bibfield  {author} {\bibinfo {author} {\bibfnamefont {N.}~\bibnamefont {{Aghanim}}} \emph {et~al.},\ }\href {https://doi.org/10.1051/0004-6361/201833910} {\bibfield  {journal} {\bibinfo  {journal} {Astron. Astrophys.}\ }\textbf {\bibinfo {volume} {641}},\ \bibinfo {eid} {A6} (\bibinfo {year} {2020})},\ \Eprint {https://arxiv.org/abs/1807.06209} {arXiv:1807.06209 [astro-ph.CO]} \BibitemShut {NoStop}%
\bibitem [{\citenamefont {{Gerosa}}\ and\ \citenamefont {{Kesden}}(2016)}]{2016PhRvD..93l4066G}%
  \BibitemOpen
  \bibfield  {author} {\bibinfo {author} {\bibfnamefont {D.}~\bibnamefont {{Gerosa}}}\ and\ \bibinfo {author} {\bibfnamefont {M.}~\bibnamefont {{Kesden}}},\ }\href {https://doi.org/10.1103/PhysRevD.93.124066} {\bibfield  {journal} {\bibinfo  {journal} {Phys. Rev. D}\ }\textbf {\bibinfo {volume} {93}},\ \bibinfo {eid} {124066} (\bibinfo {year} {2016})},\ \Eprint {https://arxiv.org/abs/1605.01067} {arXiv:1605.01067 [astro-ph.HE]} \BibitemShut {NoStop}%
\bibitem [{\citenamefont {{Pratten}}\ \emph {et~al.}(2021)\citenamefont {{Pratten}}, \citenamefont {{Garc{\'\i}a-Quir{\'o}s}}, \citenamefont {{Colleoni}}, \citenamefont {{Ramos-Buades}}, \citenamefont {{Estell{\'e}s}}, \citenamefont {{Mateu-Lucena}}, \citenamefont {{Jaume}}, \citenamefont {{Haney}}, \citenamefont {{Keitel}}, \citenamefont {{Thompson}},\ and\ \citenamefont {{Husa}}}]{2021PhRvD.103j4056P}%
  \BibitemOpen
  \bibfield  {author} {\bibinfo {author} {\bibfnamefont {G.}~\bibnamefont {{Pratten}}}, \bibinfo {author} {\bibfnamefont {C.}~\bibnamefont {{Garc{\'\i}a-Quir{\'o}s}}}, \bibinfo {author} {\bibfnamefont {M.}~\bibnamefont {{Colleoni}}}, \bibinfo {author} {\bibfnamefont {A.}~\bibnamefont {{Ramos-Buades}}}, \bibinfo {author} {\bibfnamefont {H.}~\bibnamefont {{Estell{\'e}s}}}, \bibinfo {author} {\bibfnamefont {M.}~\bibnamefont {{Mateu-Lucena}}}, \bibinfo {author} {\bibfnamefont {R.}~\bibnamefont {{Jaume}}}, \bibinfo {author} {\bibfnamefont {M.}~\bibnamefont {{Haney}}}, \bibinfo {author} {\bibfnamefont {D.}~\bibnamefont {{Keitel}}}, \bibinfo {author} {\bibfnamefont {J.~E.}\ \bibnamefont {{Thompson}}},\ and\ \bibinfo {author} {\bibfnamefont {S.}~\bibnamefont {{Husa}}},\ }\href {https://doi.org/10.1103/PhysRevD.103.104056} {\bibfield  {journal} {\bibinfo  {journal} {Phys. Rev. D}\ }\textbf {\bibinfo {volume} {103}},\ \bibinfo {eid} {104056} (\bibinfo {year} {2021})},\ \Eprint {https://arxiv.org/abs/2004.06503}
  {arXiv:2004.06503 [gr-qc]} \BibitemShut {NoStop}%
\bibitem [{\citenamefont {{Nitz}}\ \emph {et~al.}(2017)\citenamefont {{Nitz}}, \citenamefont {{Dent}}, \citenamefont {{Dal Canton}}, \citenamefont {{Fairhurst}},\ and\ \citenamefont {{Brown}}}]{2017ApJ...849..118N}%
  \BibitemOpen
  \bibfield  {author} {\bibinfo {author} {\bibfnamefont {A.~H.}\ \bibnamefont {{Nitz}}}, \bibinfo {author} {\bibfnamefont {T.}~\bibnamefont {{Dent}}}, \bibinfo {author} {\bibfnamefont {T.}~\bibnamefont {{Dal Canton}}}, \bibinfo {author} {\bibfnamefont {S.}~\bibnamefont {{Fairhurst}}},\ and\ \bibinfo {author} {\bibfnamefont {D.~A.}\ \bibnamefont {{Brown}}},\ }\href {https://doi.org/10.3847/1538-4357/aa8f50} {\bibfield  {journal} {\bibinfo  {journal} {Astrophys. J.}\ }\textbf {\bibinfo {volume} {849}},\ \bibinfo {eid} {118} (\bibinfo {year} {2017})},\ \Eprint {https://arxiv.org/abs/1705.01513} {arXiv:1705.01513 [gr-qc]} \BibitemShut {NoStop}%
\bibitem [{\citenamefont {{Moore}}\ \emph {et~al.}(2015)\citenamefont {{Moore}}, \citenamefont {{Cole}},\ and\ \citenamefont {{Berry}}}]{2015CQGra..32a5014M}%
  \BibitemOpen
  \bibfield  {author} {\bibinfo {author} {\bibfnamefont {C.~J.}\ \bibnamefont {{Moore}}}, \bibinfo {author} {\bibfnamefont {R.~H.}\ \bibnamefont {{Cole}}},\ and\ \bibinfo {author} {\bibfnamefont {C.~P.~L.}\ \bibnamefont {{Berry}}},\ }\href {https://doi.org/10.1088/0264-9381/32/1/015014} {\bibfield  {journal} {\bibinfo  {journal} {\cqg}\ }\textbf {\bibinfo {volume} {32}},\ \bibinfo {eid} {015014} (\bibinfo {year} {2015})},\ \Eprint {https://arxiv.org/abs/1408.0740} {arXiv:1408.0740 [gr-qc]} \BibitemShut {NoStop}%
\bibitem [{\citenamefont {{Barack}}\ and\ \citenamefont {{Cutler}}(2004)}]{2004PhRvD..69h2005B}%
  \BibitemOpen
  \bibfield  {author} {\bibinfo {author} {\bibfnamefont {L.}~\bibnamefont {{Barack}}}\ and\ \bibinfo {author} {\bibfnamefont {C.}~\bibnamefont {{Cutler}}},\ }\href {https://doi.org/10.1103/PhysRevD.69.082005} {\bibfield  {journal} {\bibinfo  {journal} {Phys. Rev. D}\ }\textbf {\bibinfo {volume} {69}},\ \bibinfo {eid} {082005} (\bibinfo {year} {2004})},\ \Eprint {https://arxiv.org/abs/gr-qc/0310125} {arXiv:gr-qc/0310125 [gr-qc]} \BibitemShut {NoStop}%
\bibitem [{\citenamefont {{Babak}}\ \emph {et~al.}(2021)\citenamefont {{Babak}}, \citenamefont {{Hewitson}},\ and\ \citenamefont {{Petiteau}}}]{2021arXiv210801167B}%
  \BibitemOpen
  \bibfield  {author} {\bibinfo {author} {\bibfnamefont {S.}~\bibnamefont {{Babak}}}, \bibinfo {author} {\bibfnamefont {M.}~\bibnamefont {{Hewitson}}},\ and\ \bibinfo {author} {\bibfnamefont {A.}~\bibnamefont {{Petiteau}}},\ }\href@noop {} {\bibfield  {journal} {\bibinfo  {journal} {{}}\ } (\bibinfo {year} {2021})},\ \Eprint {https://arxiv.org/abs/2108.01167} {arXiv:2108.01167 [astro-ph.IM]} \BibitemShut {NoStop}%
\bibitem [{\citenamefont {{Babak}}\ \emph {et~al.}(2017)\citenamefont {{Babak}}, \citenamefont {{Gair}}, \citenamefont {{Sesana}}, \citenamefont {{Barausse}}, \citenamefont {{Sopuerta}}, \citenamefont {{Berry}}, \citenamefont {{Berti}}, \citenamefont {{Amaro-Seoane}}, \citenamefont {{Petiteau}},\ and\ \citenamefont {{Klein}}}]{2017PhRvD..95j3012B}%
  \BibitemOpen
  \bibfield  {author} {\bibinfo {author} {\bibfnamefont {S.}~\bibnamefont {{Babak}}}, \bibinfo {author} {\bibfnamefont {J.}~\bibnamefont {{Gair}}}, \bibinfo {author} {\bibfnamefont {A.}~\bibnamefont {{Sesana}}}, \bibinfo {author} {\bibfnamefont {E.}~\bibnamefont {{Barausse}}}, \bibinfo {author} {\bibfnamefont {C.~F.}\ \bibnamefont {{Sopuerta}}}, \bibinfo {author} {\bibfnamefont {C.~P.~L.}\ \bibnamefont {{Berry}}}, \bibinfo {author} {\bibfnamefont {E.}~\bibnamefont {{Berti}}}, \bibinfo {author} {\bibfnamefont {P.}~\bibnamefont {{Amaro-Seoane}}}, \bibinfo {author} {\bibfnamefont {A.}~\bibnamefont {{Petiteau}}},\ and\ \bibinfo {author} {\bibfnamefont {A.}~\bibnamefont {{Klein}}},\ }\href {https://doi.org/10.1103/PhysRevD.95.103012} {\bibfield  {journal} {\bibinfo  {journal} {Phys. Rev. D}\ }\textbf {\bibinfo {volume} {95}},\ \bibinfo {eid} {103012} (\bibinfo {year} {2017})},\ \Eprint {https://arxiv.org/abs/1703.09722} {arXiv:1703.09722 [gr-qc]} \BibitemShut {NoStop}%
\bibitem [{\citenamefont {{Pratten}}\ \emph {et~al.}(2023)\citenamefont {{Pratten}}, \citenamefont {{Schmidt}}, \citenamefont {{Middleton}},\ and\ \citenamefont {{Vecchio}}}]{2023PhRvD.108l4045P}%
  \BibitemOpen
  \bibfield  {author} {\bibinfo {author} {\bibfnamefont {G.}~\bibnamefont {{Pratten}}}, \bibinfo {author} {\bibfnamefont {P.}~\bibnamefont {{Schmidt}}}, \bibinfo {author} {\bibfnamefont {H.}~\bibnamefont {{Middleton}}},\ and\ \bibinfo {author} {\bibfnamefont {A.}~\bibnamefont {{Vecchio}}},\ }\href {https://doi.org/10.1103/PhysRevD.108.124045} {\bibfield  {journal} {\bibinfo  {journal} {Phys. Rev. D}\ }\textbf {\bibinfo {volume} {108}},\ \bibinfo {eid} {124045} (\bibinfo {year} {2023})},\ \Eprint {https://arxiv.org/abs/2307.13026} {arXiv:2307.13026 [gr-qc]} \BibitemShut {NoStop}%
\bibitem [{\citenamefont {{Katz}}\ \emph {et~al.}(2020)\citenamefont {{Katz}}, \citenamefont {{Kelley}}, \citenamefont {{Dosopoulou}}, \citenamefont {{Berry}}, \citenamefont {{Blecha}},\ and\ \citenamefont {{Larson}}}]{2020MNRAS.491.2301K}%
  \BibitemOpen
  \bibfield  {author} {\bibinfo {author} {\bibfnamefont {M.~L.}\ \bibnamefont {{Katz}}}, \bibinfo {author} {\bibfnamefont {L.~Z.}\ \bibnamefont {{Kelley}}}, \bibinfo {author} {\bibfnamefont {F.}~\bibnamefont {{Dosopoulou}}}, \bibinfo {author} {\bibfnamefont {S.}~\bibnamefont {{Berry}}}, \bibinfo {author} {\bibfnamefont {L.}~\bibnamefont {{Blecha}}},\ and\ \bibinfo {author} {\bibfnamefont {S.~L.}\ \bibnamefont {{Larson}}},\ }\href {https://doi.org/10.1093/mnras/stz3102} {\bibfield  {journal} {\bibinfo  {journal} {Mon. Not. R. Astron. Soc.}\ }\textbf {\bibinfo {volume} {491}},\ \bibinfo {pages} {2301} (\bibinfo {year} {2020})},\ \Eprint {https://arxiv.org/abs/1908.05779} {arXiv:1908.05779 [astro-ph.HE]} \BibitemShut {NoStop}%
\bibitem [{\citenamefont {{King}}\ \emph {et~al.}(2007)\citenamefont {{King}}, \citenamefont {{Pringle}},\ and\ \citenamefont {{Livio}}}]{2007MNRAS.376.1740K}%
  \BibitemOpen
  \bibfield  {author} {\bibinfo {author} {\bibfnamefont {A.~R.}\ \bibnamefont {{King}}}, \bibinfo {author} {\bibfnamefont {J.~E.}\ \bibnamefont {{Pringle}}},\ and\ \bibinfo {author} {\bibfnamefont {M.}~\bibnamefont {{Livio}}},\ }\href {https://doi.org/10.1111/j.1365-2966.2007.11556.x} {\bibfield  {journal} {\bibinfo  {journal} {Mon. Not. R. Astron. Soc.}\ }\textbf {\bibinfo {volume} {376}},\ \bibinfo {pages} {1740} (\bibinfo {year} {2007})},\ \Eprint {https://arxiv.org/abs/astro-ph/0701803} {arXiv:astro-ph/0701803 [astro-ph]} \BibitemShut {NoStop}%
\bibitem [{\citenamefont {{Ragusa}}\ \emph {et~al.}(2016)\citenamefont {{Ragusa}}, \citenamefont {{Lodato}},\ and\ \citenamefont {{Price}}}]{2016MNRAS.460.1243R}%
  \BibitemOpen
  \bibfield  {author} {\bibinfo {author} {\bibfnamefont {E.}~\bibnamefont {{Ragusa}}}, \bibinfo {author} {\bibfnamefont {G.}~\bibnamefont {{Lodato}}},\ and\ \bibinfo {author} {\bibfnamefont {D.~J.}\ \bibnamefont {{Price}}},\ }\href {https://doi.org/10.1093/mnras/stw1081} {\bibfield  {journal} {\bibinfo  {journal} {Mon. Not. R. Astron. Soc.}\ }\textbf {\bibinfo {volume} {460}},\ \bibinfo {pages} {1243} (\bibinfo {year} {2016})},\ \Eprint {https://arxiv.org/abs/1605.01730} {arXiv:1605.01730 [astro-ph.HE]} \BibitemShut {NoStop}%
\bibitem [{\citenamefont {{Stickel}}\ \emph {et~al.}(1989)\citenamefont {{Stickel}}, \citenamefont {{Fried}},\ and\ \citenamefont {{Kuehr}}}]{1989A&AS...80..103S}%
  \BibitemOpen
  \bibfield  {author} {\bibinfo {author} {\bibfnamefont {M.}~\bibnamefont {{Stickel}}}, \bibinfo {author} {\bibfnamefont {J.~W.}\ \bibnamefont {{Fried}}},\ and\ \bibinfo {author} {\bibfnamefont {H.}~\bibnamefont {{Kuehr}}},\ }\href@noop {} {\bibfield  {journal} {\bibinfo  {journal} {Astron. Astrophys. Sup.}\ }\textbf {\bibinfo {volume} {80}},\ \bibinfo {pages} {103} (\bibinfo {year} {1989})}\BibitemShut {NoStop}%
\bibitem [{\citenamefont {{Liu}}\ and\ \citenamefont {{Wu}}(2002)}]{2002A&A...388L..48L}%
  \BibitemOpen
  \bibfield  {author} {\bibinfo {author} {\bibfnamefont {F.~K.}\ \bibnamefont {{Liu}}}\ and\ \bibinfo {author} {\bibfnamefont {X.~B.}\ \bibnamefont {{Wu}}},\ }\href {https://doi.org/10.1051/0004-6361:20020566} {\bibfield  {journal} {\bibinfo  {journal} {Astron. Astrophys.}\ }\textbf {\bibinfo {volume} {388}},\ \bibinfo {pages} {L48} (\bibinfo {year} {2002})},\ \Eprint {https://arxiv.org/abs/astro-ph/0212475} {arXiv:astro-ph/0212475 [astro-ph]} \BibitemShut {NoStop}%
\bibitem [{\citenamefont {{Gao}}\ \emph {et~al.}(2024)\citenamefont {{Gao}}, \citenamefont {{Hu}}, \citenamefont {{Li}}, \citenamefont {{Zhang}},\ and\ \citenamefont {{Mei}}}]{2024arXiv240112813G}%
  \BibitemOpen
  \bibfield  {author} {\bibinfo {author} {\bibfnamefont {J.}~\bibnamefont {{Gao}}}, \bibinfo {author} {\bibfnamefont {Y.-M.}\ \bibnamefont {{Hu}}}, \bibinfo {author} {\bibfnamefont {E.-K.}\ \bibnamefont {{Li}}}, \bibinfo {author} {\bibfnamefont {J.-d.}\ \bibnamefont {{Zhang}}},\ and\ \bibinfo {author} {\bibfnamefont {J.}~\bibnamefont {{Mei}}},\ }\href@noop {} {\bibfield  {journal} {\bibinfo  {journal} {{}}\ } (\bibinfo {year} {2024})},\ \Eprint {https://arxiv.org/abs/2401.12813} {arXiv:2401.12813 [astro-ph.GA]} \BibitemShut {NoStop}%
\bibitem [{\citenamefont {{Wilman}}\ \emph {et~al.}(2005)\citenamefont {{Wilman}}, \citenamefont {{Edge}},\ and\ \citenamefont {{Johnstone}}}]{2005MNRAS.359..755W}%
  \BibitemOpen
  \bibfield  {author} {\bibinfo {author} {\bibfnamefont {R.~J.}\ \bibnamefont {{Wilman}}}, \bibinfo {author} {\bibfnamefont {A.~C.}\ \bibnamefont {{Edge}}},\ and\ \bibinfo {author} {\bibfnamefont {R.~M.}\ \bibnamefont {{Johnstone}}},\ }\href {https://doi.org/10.1111/j.1365-2966.2005.08956.x} {\bibfield  {journal} {\bibinfo  {journal} {Mon. Not. R. Astron. Soc.}\ }\textbf {\bibinfo {volume} {359}},\ \bibinfo {pages} {755} (\bibinfo {year} {2005})},\ \Eprint {https://arxiv.org/abs/astro-ph/0502537} {arXiv:astro-ph/0502537 [astro-ph]} \BibitemShut {NoStop}%
\bibitem [{\citenamefont {{Holtzman}}\ \emph {et~al.}(1992)\citenamefont {{Holtzman}}, \citenamefont {{Faber}}, \citenamefont {{Shaya}}, \citenamefont {{Lauer}}, \citenamefont {{Groth}}, \citenamefont {{Hunter}}, \citenamefont {{Baum}}, \citenamefont {{Ewald}}, \citenamefont {{Hester}}, \citenamefont {{Light}}, \citenamefont {{Lynds}}, \citenamefont {{O'Neil}},\ and\ \citenamefont {{Westphal}}}]{1992AJ....103..691H}%
  \BibitemOpen
  \bibfield  {author} {\bibinfo {author} {\bibfnamefont {J.~A.}\ \bibnamefont {{Holtzman}}}, \bibinfo {author} {\bibfnamefont {S.~M.}\ \bibnamefont {{Faber}}}, \bibinfo {author} {\bibfnamefont {E.~J.}\ \bibnamefont {{Shaya}}}, \bibinfo {author} {\bibfnamefont {T.~R.}\ \bibnamefont {{Lauer}}}, \bibinfo {author} {\bibfnamefont {J.}~\bibnamefont {{Groth}}}, \bibinfo {author} {\bibfnamefont {D.~A.}\ \bibnamefont {{Hunter}}}, \bibinfo {author} {\bibfnamefont {W.~A.}\ \bibnamefont {{Baum}}}, \bibinfo {author} {\bibfnamefont {S.~P.}\ \bibnamefont {{Ewald}}}, \bibinfo {author} {\bibfnamefont {J.~J.}\ \bibnamefont {{Hester}}}, \bibinfo {author} {\bibfnamefont {R.~M.}\ \bibnamefont {{Light}}}, \bibinfo {author} {\bibfnamefont {C.~R.}\ \bibnamefont {{Lynds}}}, \bibinfo {author} {\bibfnamefont {J.}~\bibnamefont {{O'Neil}}, \bibfnamefont {E.~J.}},\ and\ \bibinfo {author} {\bibfnamefont {J.~A.}\ \bibnamefont {{Westphal}}},\ }\href {https://doi.org/10.1086/116094} {\bibfield  {journal} {\bibinfo  {journal} {Astron. J.}\
  }\textbf {\bibinfo {volume} {103}},\ \bibinfo {pages} {691} (\bibinfo {year} {1992})}\BibitemShut {NoStop}%
\bibitem [{\citenamefont {{Rubinur}}\ \emph {et~al.}(2017)\citenamefont {{Rubinur}}, \citenamefont {{Das}}, \citenamefont {{Kharb}},\ and\ \citenamefont {{Honey}}}]{2017MNRAS.465.4772R}%
  \BibitemOpen
  \bibfield  {author} {\bibinfo {author} {\bibfnamefont {K.}~\bibnamefont {{Rubinur}}}, \bibinfo {author} {\bibfnamefont {M.}~\bibnamefont {{Das}}}, \bibinfo {author} {\bibfnamefont {P.}~\bibnamefont {{Kharb}}},\ and\ \bibinfo {author} {\bibfnamefont {M.}~\bibnamefont {{Honey}}},\ }\href {https://doi.org/10.1093/mnras/stw2981} {\bibfield  {journal} {\bibinfo  {journal} {Mon. Not. R. Astron. Soc.}\ }\textbf {\bibinfo {volume} {465}},\ \bibinfo {pages} {4772} (\bibinfo {year} {2017})}\BibitemShut {NoStop}%
\bibitem [{\citenamefont {{Freedman}}\ \emph {et~al.}(1994)\citenamefont {{Freedman}}, \citenamefont {{Hughes}}, \citenamefont {{Madore}}, \citenamefont {{Mould}}, \citenamefont {{Lee}}, \citenamefont {{Stetson}}, \citenamefont {{Kennicutt}}, \citenamefont {{Turner}}, \citenamefont {{Ferrarese}}, \citenamefont {{Ford}}, \citenamefont {{Graham}}, \citenamefont {{Hill}}, \citenamefont {{Hoessel}}, \citenamefont {{Huchra}},\ and\ \citenamefont {{Illingworth}}}]{1994ApJ...427..628F}%
  \BibitemOpen
  \bibfield  {author} {\bibinfo {author} {\bibfnamefont {W.~L.}\ \bibnamefont {{Freedman}}}, \bibinfo {author} {\bibfnamefont {S.~M.}\ \bibnamefont {{Hughes}}}, \bibinfo {author} {\bibfnamefont {B.~F.}\ \bibnamefont {{Madore}}}, \bibinfo {author} {\bibfnamefont {J.~R.}\ \bibnamefont {{Mould}}}, \bibinfo {author} {\bibfnamefont {M.~G.}\ \bibnamefont {{Lee}}}, \bibinfo {author} {\bibfnamefont {P.}~\bibnamefont {{Stetson}}}, \bibinfo {author} {\bibfnamefont {R.~C.}\ \bibnamefont {{Kennicutt}}}, \bibinfo {author} {\bibfnamefont {A.}~\bibnamefont {{Turner}}}, \bibinfo {author} {\bibfnamefont {L.}~\bibnamefont {{Ferrarese}}}, \bibinfo {author} {\bibfnamefont {H.}~\bibnamefont {{Ford}}}, \bibinfo {author} {\bibfnamefont {J.~A.}\ \bibnamefont {{Graham}}}, \bibinfo {author} {\bibfnamefont {R.}~\bibnamefont {{Hill}}}, \bibinfo {author} {\bibfnamefont {J.~G.}\ \bibnamefont {{Hoessel}}}, \bibinfo {author} {\bibfnamefont {J.}~\bibnamefont {{Huchra}}},\ and\ \bibinfo {author} {\bibfnamefont {G.~D.}\ \bibnamefont
  {{Illingworth}}},\ }\href {https://doi.org/10.1086/174172} {\bibfield  {journal} {\bibinfo  {journal} {Astrophys. J.}\ }\textbf {\bibinfo {volume} {427}},\ \bibinfo {pages} {628} (\bibinfo {year} {1994})}\BibitemShut {NoStop}%
\bibitem [{\citenamefont {{Pursimo}}\ \emph {et~al.}(2000)\citenamefont {{Pursimo}}, \citenamefont {{Takalo}}, \citenamefont {{Sillanp{\"a}{\"a}}}, \citenamefont {{Kidger}}, \citenamefont {{Lehto}}, \citenamefont {{Heidt}}, \citenamefont {{Charles}} \emph {et~al.}}]{2000A&AS..146..141P}%
  \BibitemOpen
  \bibfield  {author} {\bibinfo {author} {\bibfnamefont {T.}~\bibnamefont {{Pursimo}}}, \bibinfo {author} {\bibfnamefont {L.~O.}\ \bibnamefont {{Takalo}}}, \bibinfo {author} {\bibfnamefont {A.}~\bibnamefont {{Sillanp{\"a}{\"a}}}}, \bibinfo {author} {\bibfnamefont {M.}~\bibnamefont {{Kidger}}}, \bibinfo {author} {\bibfnamefont {H.~J.}\ \bibnamefont {{Lehto}}}, \bibinfo {author} {\bibfnamefont {J.}~\bibnamefont {{Heidt}}}, \bibinfo {author} {\bibfnamefont {P.~A.}\ \bibnamefont {{Charles}}}, \emph {et~al.},\ }\href {https://doi.org/10.1051/aas:2000264} {\bibfield  {journal} {\bibinfo  {journal} {Astron. Astrophys. Sup.}\ }\textbf {\bibinfo {volume} {146}},\ \bibinfo {pages} {141} (\bibinfo {year} {2000})}\BibitemShut {NoStop}%
\bibitem [{\citenamefont {{Valtonen}}\ \emph {et~al.}(2016)\citenamefont {{Valtonen}}, \citenamefont {{Zola}}, \citenamefont {{Ciprini}}, \citenamefont {{Gopakumar}}, \citenamefont {{Matsumoto}}, \citenamefont {{Sadakane}}, \citenamefont {{Kidger}}, \citenamefont {{Gazeas}}, \citenamefont {{Nilsson}}, \citenamefont {{Berdyugin}}, \citenamefont {{Piirola}}, \citenamefont {{Jermak}}, \citenamefont {{Baliyan}} \emph {et~al.}}]{2016ApJ...819L..37V}%
  \BibitemOpen
  \bibfield  {author} {\bibinfo {author} {\bibfnamefont {M.~J.}\ \bibnamefont {{Valtonen}}}, \bibinfo {author} {\bibfnamefont {S.}~\bibnamefont {{Zola}}}, \bibinfo {author} {\bibfnamefont {S.}~\bibnamefont {{Ciprini}}}, \bibinfo {author} {\bibfnamefont {A.}~\bibnamefont {{Gopakumar}}}, \bibinfo {author} {\bibfnamefont {K.}~\bibnamefont {{Matsumoto}}}, \bibinfo {author} {\bibfnamefont {K.}~\bibnamefont {{Sadakane}}}, \bibinfo {author} {\bibfnamefont {M.}~\bibnamefont {{Kidger}}}, \bibinfo {author} {\bibfnamefont {K.}~\bibnamefont {{Gazeas}}}, \bibinfo {author} {\bibfnamefont {K.}~\bibnamefont {{Nilsson}}}, \bibinfo {author} {\bibfnamefont {A.}~\bibnamefont {{Berdyugin}}}, \bibinfo {author} {\bibfnamefont {V.}~\bibnamefont {{Piirola}}}, \bibinfo {author} {\bibfnamefont {H.}~\bibnamefont {{Jermak}}}, \bibinfo {author} {\bibfnamefont {K.~S.}\ \bibnamefont {{Baliyan}}}, \emph {et~al.},\ }\href {https://doi.org/10.3847/2041-8205/819/2/L37} {\bibfield  {journal} {\bibinfo  {journal} {Astrophys. J. Lett.}\ }\textbf
  {\bibinfo {volume} {819}},\ \bibinfo {eid} {L37} (\bibinfo {year} {2016})},\ \Eprint {https://arxiv.org/abs/1603.04171} {arXiv:1603.04171 [astro-ph.HE]} \BibitemShut {NoStop}%
\bibitem [{\citenamefont {{Mahatma}}\ \emph {et~al.}(2023)\citenamefont {{Mahatma}}, \citenamefont {{Basu}}, \citenamefont {{Hardcastle}}, \citenamefont {{Morabito}},\ and\ \citenamefont {{van Weeren}}}]{2023MNRAS.520.4427M}%
  \BibitemOpen
  \bibfield  {author} {\bibinfo {author} {\bibfnamefont {V.~H.}\ \bibnamefont {{Mahatma}}}, \bibinfo {author} {\bibfnamefont {A.}~\bibnamefont {{Basu}}}, \bibinfo {author} {\bibfnamefont {M.~J.}\ \bibnamefont {{Hardcastle}}}, \bibinfo {author} {\bibfnamefont {L.~K.}\ \bibnamefont {{Morabito}}},\ and\ \bibinfo {author} {\bibfnamefont {R.~J.}\ \bibnamefont {{van Weeren}}},\ }\href {https://doi.org/10.1093/mnras/stad395} {\bibfield  {journal} {\bibinfo  {journal} {\mnras}\ }\textbf {\bibinfo {volume} {520}},\ \bibinfo {pages} {4427} (\bibinfo {year} {2023})},\ \Eprint {https://arxiv.org/abs/2302.01357} {arXiv:2302.01357 [astro-ph.GA]} \BibitemShut {NoStop}%
\bibitem [{\citenamefont {{Britzen}}\ \emph {et~al.}(2008)\citenamefont {{Britzen}}, \citenamefont {{Vermeulen}}, \citenamefont {{Campbell}}, \citenamefont {{Taylor}}, \citenamefont {{Pearson}}, \citenamefont {{Readhead}}, \citenamefont {{Xu}}, \citenamefont {{Browne}}, \citenamefont {{Henstock}},\ and\ \citenamefont {{Wilkinson}}}]{2008A&A...484..119B}%
  \BibitemOpen
  \bibfield  {author} {\bibinfo {author} {\bibfnamefont {S.}~\bibnamefont {{Britzen}}}, \bibinfo {author} {\bibfnamefont {R.~C.}\ \bibnamefont {{Vermeulen}}}, \bibinfo {author} {\bibfnamefont {R.~M.}\ \bibnamefont {{Campbell}}}, \bibinfo {author} {\bibfnamefont {G.~B.}\ \bibnamefont {{Taylor}}}, \bibinfo {author} {\bibfnamefont {T.~J.}\ \bibnamefont {{Pearson}}}, \bibinfo {author} {\bibfnamefont {A.~C.~S.}\ \bibnamefont {{Readhead}}}, \bibinfo {author} {\bibfnamefont {W.}~\bibnamefont {{Xu}}}, \bibinfo {author} {\bibfnamefont {I.~W.}\ \bibnamefont {{Browne}}}, \bibinfo {author} {\bibfnamefont {D.~R.}\ \bibnamefont {{Henstock}}},\ and\ \bibinfo {author} {\bibfnamefont {P.}~\bibnamefont {{Wilkinson}}},\ }\href {https://doi.org/10.1051/0004-6361:20077717} {\bibfield  {journal} {\bibinfo  {journal} {Astron. Astrophys.}\ }\textbf {\bibinfo {volume} {484}},\ \bibinfo {pages} {119} (\bibinfo {year} {2008})},\ \Eprint {https://arxiv.org/abs/0802.4148} {arXiv:0802.4148 [astro-ph]} \BibitemShut {NoStop}%
\bibitem [{\citenamefont {{Homan}}\ \emph {et~al.}(2015)\citenamefont {{Homan}}, \citenamefont {{Lister}}, \citenamefont {{Kovalev}}, \citenamefont {{Pushkarev}}, \citenamefont {{Savolainen}}, \citenamefont {{Kellermann}}, \citenamefont {{Richards}},\ and\ \citenamefont {{Ros}}}]{2015ApJ...798..134H}%
  \BibitemOpen
  \bibfield  {author} {\bibinfo {author} {\bibfnamefont {D.~C.}\ \bibnamefont {{Homan}}}, \bibinfo {author} {\bibfnamefont {M.~L.}\ \bibnamefont {{Lister}}}, \bibinfo {author} {\bibfnamefont {Y.~Y.}\ \bibnamefont {{Kovalev}}}, \bibinfo {author} {\bibfnamefont {A.~B.}\ \bibnamefont {{Pushkarev}}}, \bibinfo {author} {\bibfnamefont {T.}~\bibnamefont {{Savolainen}}}, \bibinfo {author} {\bibfnamefont {K.~I.}\ \bibnamefont {{Kellermann}}}, \bibinfo {author} {\bibfnamefont {J.~L.}\ \bibnamefont {{Richards}}},\ and\ \bibinfo {author} {\bibfnamefont {E.}~\bibnamefont {{Ros}}},\ }\href {https://doi.org/10.1088/0004-637X/798/2/134} {\bibfield  {journal} {\bibinfo  {journal} {Astrophys. J.}\ }\textbf {\bibinfo {volume} {798}},\ \bibinfo {eid} {134} (\bibinfo {year} {2015})},\ \Eprint {https://arxiv.org/abs/1410.8502} {arXiv:1410.8502 [astro-ph.HE]} \BibitemShut {NoStop}%
\bibitem [{\citenamefont {{Boccardi}}\ \emph {et~al.}(2016)\citenamefont {{Boccardi}}, \citenamefont {{Krichbaum}}, \citenamefont {{Bach}}, \citenamefont {{Bremer}},\ and\ \citenamefont {{Zensus}}}]{2016A&A...588L...9B}%
  \BibitemOpen
  \bibfield  {author} {\bibinfo {author} {\bibfnamefont {B.}~\bibnamefont {{Boccardi}}}, \bibinfo {author} {\bibfnamefont {T.~P.}\ \bibnamefont {{Krichbaum}}}, \bibinfo {author} {\bibfnamefont {U.}~\bibnamefont {{Bach}}}, \bibinfo {author} {\bibfnamefont {M.}~\bibnamefont {{Bremer}}},\ and\ \bibinfo {author} {\bibfnamefont {J.~A.}\ \bibnamefont {{Zensus}}},\ }\href {https://doi.org/10.1051/0004-6361/201628412} {\bibfield  {journal} {\bibinfo  {journal} {Astron. Astrophys.}\ }\textbf {\bibinfo {volume} {588}},\ \bibinfo {eid} {L9} (\bibinfo {year} {2016})},\ \Eprint {https://arxiv.org/abs/1603.04221} {arXiv:1603.04221 [astro-ph.HE]} \BibitemShut {NoStop}%
\bibitem [{\citenamefont {{Kellermann}}\ \emph {et~al.}(1998)\citenamefont {{Kellermann}}, \citenamefont {{Vermeulen}}, \citenamefont {{Zensus}},\ and\ \citenamefont {{Cohen}}}]{1998AJ....115.1295K}%
  \BibitemOpen
  \bibfield  {author} {\bibinfo {author} {\bibfnamefont {K.~I.}\ \bibnamefont {{Kellermann}}}, \bibinfo {author} {\bibfnamefont {R.~C.}\ \bibnamefont {{Vermeulen}}}, \bibinfo {author} {\bibfnamefont {J.~A.}\ \bibnamefont {{Zensus}}},\ and\ \bibinfo {author} {\bibfnamefont {M.~H.}\ \bibnamefont {{Cohen}}},\ }\href {https://doi.org/10.1086/300308} {\bibfield  {journal} {\bibinfo  {journal} {Astron. J.}\ }\textbf {\bibinfo {volume} {115}},\ \bibinfo {pages} {1295} (\bibinfo {year} {1998})},\ \Eprint {https://arxiv.org/abs/astro-ph/9801010} {arXiv:astro-ph/9801010 [astro-ph]} \BibitemShut {NoStop}%
\bibitem [{\citenamefont {{Krichbaum}}\ \emph {et~al.}(1998)\citenamefont {{Krichbaum}}, \citenamefont {{Alef}}, \citenamefont {{Witzel}}, \citenamefont {{Zensus}}, \citenamefont {{Booth}}, \citenamefont {{Greve}},\ and\ \citenamefont {{Rogers}}}]{1998A&A...329..873K}%
  \BibitemOpen
  \bibfield  {author} {\bibinfo {author} {\bibfnamefont {T.~P.}\ \bibnamefont {{Krichbaum}}}, \bibinfo {author} {\bibfnamefont {W.}~\bibnamefont {{Alef}}}, \bibinfo {author} {\bibfnamefont {A.}~\bibnamefont {{Witzel}}}, \bibinfo {author} {\bibfnamefont {J.~A.}\ \bibnamefont {{Zensus}}}, \bibinfo {author} {\bibfnamefont {R.~S.}\ \bibnamefont {{Booth}}}, \bibinfo {author} {\bibfnamefont {A.}~\bibnamefont {{Greve}}},\ and\ \bibinfo {author} {\bibfnamefont {A.~E.~E.}\ \bibnamefont {{Rogers}}},\ }\href@noop {} {\bibfield  {journal} {\bibinfo  {journal} {\aap}\ }\textbf {\bibinfo {volume} {329}},\ \bibinfo {pages} {873} (\bibinfo {year} {1998})}\BibitemShut {NoStop}%
\bibitem [{\citenamefont {{Nakahara}}\ \emph {et~al.}(2019)\citenamefont {{Nakahara}}, \citenamefont {{Doi}}, \citenamefont {{Murata}}, \citenamefont {{Nakamura}}, \citenamefont {{Hada}},\ and\ \citenamefont {{Asada}}}]{2019ApJ...878...61N}%
  \BibitemOpen
  \bibfield  {author} {\bibinfo {author} {\bibfnamefont {S.}~\bibnamefont {{Nakahara}}}, \bibinfo {author} {\bibfnamefont {A.}~\bibnamefont {{Doi}}}, \bibinfo {author} {\bibfnamefont {Y.}~\bibnamefont {{Murata}}}, \bibinfo {author} {\bibfnamefont {M.}~\bibnamefont {{Nakamura}}}, \bibinfo {author} {\bibfnamefont {K.}~\bibnamefont {{Hada}}},\ and\ \bibinfo {author} {\bibfnamefont {K.}~\bibnamefont {{Asada}}},\ }\href {https://doi.org/10.3847/1538-4357/ab1b0e} {\bibfield  {journal} {\bibinfo  {journal} {Astrophys. J.}\ }\textbf {\bibinfo {volume} {878}},\ \bibinfo {eid} {61} (\bibinfo {year} {2019})}\BibitemShut {NoStop}%
\bibitem [{\citenamefont {{Boccardi}}\ \emph {et~al.}(2017)\citenamefont {{Boccardi}}, \citenamefont {{Krichbaum}}, \citenamefont {{Ros}},\ and\ \citenamefont {{Zensus}}}]{2017A&ARv..25....4B}%
  \BibitemOpen
  \bibfield  {author} {\bibinfo {author} {\bibfnamefont {B.}~\bibnamefont {{Boccardi}}}, \bibinfo {author} {\bibfnamefont {T.~P.}\ \bibnamefont {{Krichbaum}}}, \bibinfo {author} {\bibfnamefont {E.}~\bibnamefont {{Ros}}},\ and\ \bibinfo {author} {\bibfnamefont {J.~A.}\ \bibnamefont {{Zensus}}},\ }\href {https://doi.org/10.1007/s00159-017-0105-6} {\bibfield  {journal} {\bibinfo  {journal} {Astron. Astrophys. Rev.}\ }\textbf {\bibinfo {volume} {25}},\ \bibinfo {eid} {4} (\bibinfo {year} {2017})},\ \Eprint {https://arxiv.org/abs/1711.07548} {arXiv:1711.07548 [astro-ph.HE]} \BibitemShut {NoStop}%
\bibitem [{\citenamefont {{Britzen}}\ \emph {et~al.}(2017)\citenamefont {{Britzen}}, \citenamefont {{Qian}}, \citenamefont {{Steffen}}, \citenamefont {{Kun}}, \citenamefont {{Karouzos}}, \citenamefont {{Gergely}}, \citenamefont {{Schmidt}}, \citenamefont {{Aller}}, \citenamefont {{Aller}}, \citenamefont {{Krause}}, \citenamefont {{Fendt}}, \citenamefont {{B{\"o}ttcher}}, \citenamefont {{Witzel}}, \citenamefont {{Eckart}},\ and\ \citenamefont {{Moser}}}]{2017A&A...602A..29B}%
  \BibitemOpen
  \bibfield  {author} {\bibinfo {author} {\bibfnamefont {S.}~\bibnamefont {{Britzen}}}, \bibinfo {author} {\bibfnamefont {S.~J.}\ \bibnamefont {{Qian}}}, \bibinfo {author} {\bibfnamefont {W.}~\bibnamefont {{Steffen}}}, \bibinfo {author} {\bibfnamefont {E.}~\bibnamefont {{Kun}}}, \bibinfo {author} {\bibfnamefont {M.}~\bibnamefont {{Karouzos}}}, \bibinfo {author} {\bibfnamefont {L.}~\bibnamefont {{Gergely}}}, \bibinfo {author} {\bibfnamefont {J.}~\bibnamefont {{Schmidt}}}, \bibinfo {author} {\bibfnamefont {M.}~\bibnamefont {{Aller}}}, \bibinfo {author} {\bibfnamefont {H.}~\bibnamefont {{Aller}}}, \bibinfo {author} {\bibfnamefont {M.}~\bibnamefont {{Krause}}}, \bibinfo {author} {\bibfnamefont {C.}~\bibnamefont {{Fendt}}}, \bibinfo {author} {\bibfnamefont {M.}~\bibnamefont {{B{\"o}ttcher}}}, \bibinfo {author} {\bibfnamefont {A.}~\bibnamefont {{Witzel}}}, \bibinfo {author} {\bibfnamefont {A.}~\bibnamefont {{Eckart}}},\ and\ \bibinfo {author} {\bibfnamefont {L.}~\bibnamefont {{Moser}}},\ }\href
  {https://doi.org/10.1051/0004-6361/201629999} {\bibfield  {journal} {\bibinfo  {journal} {Astron. Astrophys.}\ }\textbf {\bibinfo {volume} {602}},\ \bibinfo {eid} {A29} (\bibinfo {year} {2017})}\BibitemShut {NoStop}%
\bibitem [{\citenamefont {{Klein}}\ \emph {et~al.}(2016)\citenamefont {{Klein}}, \citenamefont {{Barausse}}, \citenamefont {{Sesana}}, \citenamefont {{Petiteau}}, \citenamefont {{Berti}}, \citenamefont {{Babak}}, \citenamefont {{Gair}}, \citenamefont {{Aoudia}}, \citenamefont {{Hinder}}, \citenamefont {{Ohme}},\ and\ \citenamefont {{Wardell}}}]{2016PhRvD..93b4003K}%
  \BibitemOpen
  \bibfield  {author} {\bibinfo {author} {\bibfnamefont {A.}~\bibnamefont {{Klein}}}, \bibinfo {author} {\bibfnamefont {E.}~\bibnamefont {{Barausse}}}, \bibinfo {author} {\bibfnamefont {A.}~\bibnamefont {{Sesana}}}, \bibinfo {author} {\bibfnamefont {A.}~\bibnamefont {{Petiteau}}}, \bibinfo {author} {\bibfnamefont {E.}~\bibnamefont {{Berti}}}, \bibinfo {author} {\bibfnamefont {S.}~\bibnamefont {{Babak}}}, \bibinfo {author} {\bibfnamefont {J.}~\bibnamefont {{Gair}}}, \bibinfo {author} {\bibfnamefont {S.}~\bibnamefont {{Aoudia}}}, \bibinfo {author} {\bibfnamefont {I.}~\bibnamefont {{Hinder}}}, \bibinfo {author} {\bibfnamefont {F.}~\bibnamefont {{Ohme}}},\ and\ \bibinfo {author} {\bibfnamefont {B.}~\bibnamefont {{Wardell}}},\ }\href {https://doi.org/10.1103/PhysRevD.93.024003} {\bibfield  {journal} {\bibinfo  {journal} {Phys. Rev. D}\ }\textbf {\bibinfo {volume} {93}},\ \bibinfo {eid} {024003} (\bibinfo {year} {2016})},\ \Eprint {https://arxiv.org/abs/1511.05581} {arXiv:1511.05581 [gr-qc]} \BibitemShut {NoStop}%
\bibitem [{\citenamefont {{Colpi}}\ and\ \citenamefont {{Sesana}}(2017)}]{2017ogw..book...43C}%
  \BibitemOpen
  \bibfield  {author} {\bibinfo {author} {\bibfnamefont {M.}~\bibnamefont {{Colpi}}}\ and\ \bibinfo {author} {\bibfnamefont {A.}~\bibnamefont {{Sesana}}},\ }in\ \href {https://doi.org/10.1142/9789813141766_0002} {\emph {\bibinfo {booktitle} {An Overview of Gravitational Waves: Theory}}}\ (\bibinfo  {publisher} {World Scientific Publishing},\ \bibinfo {year} {2017})\ pp.\ \bibinfo {pages} {43--140}\BibitemShut {NoStop}%
\end{thebibliography}%
\end{document}